\PassOptionsToPackage{unicode}{hyperref}
\PassOptionsToPackage{hyphens}{url}
\PassOptionsToPackage{dvipsnames,svgnames,x11names}{xcolor}
\documentclass[
  letterpaper,
  DIV=11,
  numbers=noendperiod]{scrartcl}

\usepackage{amsmath,amssymb}
\usepackage[T1]{fontenc}
\usepackage[utf8]{inputenc}
\usepackage{textcomp}
\usepackage{lmodern}
\IfFileExists{upquote.sty}{\usepackage{upquote}}{}
\IfFileExists{microtype.sty}{
  \usepackage[]{microtype}
  \UseMicrotypeSet[protrusion]{basicmath} 
}{}
\makeatletter
\@ifundefined{KOMAClassName}{
  \IfFileExists{parskip.sty}{%
    \usepackage{parskip}
  }{
    \setlength{\parindent}{0pt}
    \setlength{\parskip}{6pt plus 2pt minus 1pt}}
}{
  \KOMAoptions{parskip=half}}
\makeatother
\usepackage{xcolor}
\makeatletter
\ifx\paragraph\undefined\else
  \let\oldparagraph\paragraph
  \renewcommand{\paragraph}{
    \@ifstar
      \xxxParagraphStar
      \xxxParagraphNoStar
  }
  \newcommand{\xxxParagraphStar}[1]{\oldparagraph*{#1}\mbox{}}
  \newcommand{\xxxParagraphNoStar}[1]{\oldparagraph{#1}\mbox{}}
\fi
\ifx\subparagraph\undefined\else
  \let\oldsubparagraph\subparagraph
  \renewcommand{\subparagraph}{
    \@ifstar
      \xxxSubParagraphStar
      \xxxSubParagraphNoStar
  }
  \newcommand{\xxxSubParagraphStar}[1]{\oldsubparagraph*{#1}\mbox{}}
  \newcommand{\xxxSubParagraphNoStar}[1]{\oldsubparagraph{#1}\mbox{}}
\fi
\makeatother

\providecommand{\tightlist}{%
  \setlength{\itemsep}{0pt}\setlength{\parskip}{0pt}}\usepackage{longtable,booktabs,array}
\usepackage{calc} 
\usepackage{etoolbox}
\makeatletter
\patchcmd\longtable{\par}{\if@noskipsec\mbox{}\fi\par}{}{}
\makeatother
\IfFileExists{footnotehyper.sty}{\usepackage{footnotehyper}}{\usepackage{footnote}}
\makesavenoteenv{longtable}
\usepackage{graphicx}
\makeatletter
\def\maxwidth{\ifdim\Gin@nat@width>\linewidth\linewidth\else\Gin@nat@width\fi}
\def\maxheight{\ifdim\Gin@nat@height>\textheight\textheight\else\Gin@nat@height\fi}
\makeatother
\setkeys{Gin}{width=\maxwidth,height=\maxheight,keepaspectratio}
\makeatletter
\def\fps@figure{htbp}
\makeatother
\NewDocumentCommand\citeproctext{}{}

\makeatletter
 \let\@cite@ofmt\@firstofone
 \def\@biblabel#1{}
 \def\@cite#1#2{{#1\if@tempswa , #2\fi}}
\makeatother
\newlength{\cslhangindent}
\newlength{\csllabelwidth}
\newenvironment{CSLReferences}[2] 
 {\begin{list}{}{%
  \setlength{\itemindent}{0pt}
  \setlength{\leftmargin}{0pt}
  \setlength{\parsep}{0pt}
  \ifodd #1
   \setlength{\leftmargin}{\cslhangindent}
   \setlength{\itemindent}{-1\cslhangindent}
  \fi
  \setlength{\itemsep}{#2\baselineskip}}}
 {\end{list}}
\usepackage{calc}

\usepackage{multirow}
\usepackage{multicol}
\usepackage{colortbl}
\usepackage{hhline}
\newlength\Oldarrayrulewidth
\newlength\Oldtabcolsep
\usepackage{longtable}
\usepackage{array}
\usepackage{hyperref}
\usepackage{float}
\usepackage{wrapfig}
\usepackage{fbb}
\usepackage{afterpage}
\usepackage{dblfloatfix}
\usepackage{newunicodechar}
\newunicodechar{λ}{\ensuremath{\lambda}}
\providecommand{\ascline}[3]{\noalign{\global\arrayrulewidth #1}\arrayrulecolor[HTML]{#2}\cmidrule(l){#3}}
\newcommand{\xfnm}[1][]{\ifx!#1!\else\unskip,\space#1\fi}
\KOMAoption{captions}{tableheading}
\makeatletter
\@ifpackageloaded{caption}{}{\usepackage{caption}}
\AtBeginDocument{%
\ifdefined\contentsname
  \renewcommand*\contentsname{Table of contents}
\else
  \newcommand\contentsname{Table of contents}
\fi
\ifdefined\listfigurename
  \renewcommand*\listfigurename{List of Figures}
\else
  \newcommand\listfigurename{List of Figures}
\fi
\ifdefined\listtablename
  \renewcommand*\listtablename{List of Tables}
\else
  \newcommand\listtablename{List of Tables}
\fi
\ifdefined\figurename
  \renewcommand*\figurename{Figure}
\else
  \newcommand\figurename{Figure}
\fi
\ifdefined\tablename
  \renewcommand*\tablename{Table}
\else
  \newcommand\tablename{Table}
\fi
}
\@ifpackageloaded{float}{}{\usepackage{float}}
\floatstyle{ruled}
\@ifundefined{c@chapter}{\newfloat{codelisting}{h}{lop}}{\newfloat{codelisting}{h}{lop}[chapter]}
\floatname{codelisting}{Listing}

\makeatother
\makeatletter
\@ifpackageloaded{caption}{}{\usepackage{caption}}
\@ifpackageloaded{subcaption}{}{\usepackage{subcaption}}
\makeatother

\ifLuaTeX
  \usepackage{selnolig}  
\fi
\usepackage{bookmark}

\IfFileExists{xurl.sty}{\usepackage{xurl}}{} 
\hypersetup{
  pdftitle={No Silver Bullets: Why Understanding Software Cycle Time is Messy, Not Magic},
  pdfauthor={John C. Flournoy; Carol S. Lee; Maggie Wu; Catherine M. Hicks},
  pdfkeywords={cycle time, productivity, software
engineering, collaboration, coding time, task scoping, hierarchical
modeling},
  colorlinks=true,
  linkcolor={blue},
  filecolor={Maroon},
  citecolor={Blue},
  urlcolor={Blue},
  pdfcreator={LaTeX via pandoc}}

\title{No Silver Bullets: Why Understanding Software Cycle Time is
Messy, Not Magic}
\author{John C. Flournoy \and Carol S. Lee \and Maggie Wu \and Catherine
M. Hicks}
\date{2025-10-11}

\begin{document}
\maketitle
\begin{abstract}
Understanding factors that influence software development velocity is
crucial for engineering teams and organizations, yet empirical evidence
at scale remains limited. A more robust understanding of the dynamics of
cycle time may help practitioners avoid pitfalls in relying on velocity
measures while evaluating software work. We analyze cycle time---a
widely-used metric measuring time from ticket creation to
completion---using a dataset of over 55,000 observations across 216
organizations. Through Bayesian hierarchical modeling that appropriately
separates individual and organizational variation, we examine how coding
time, task scoping, and collaboration patterns affect cycle time while
characterizing its substantial variability across contexts. We find
precise but modest associations between cycle time and factors including
coding days per week, number of merged pull requests, and degree of
collaboration. However, these effects are set against considerable
unexplained variation both between and within individuals. Our findings
suggest that while common workplace factors do influence cycle time in
expected directions, any single observation provides limited signal
about typical performance. This work demonstrates methods for analyzing
complex operational metrics at scale while highlighting potential
pitfalls in using such measurements to drive decision-making. We
conclude that improving software delivery velocity likely requires
systems-level thinking rather than individual-focused interventions.
\end{abstract}

\section{Introduction}\label{introduction}

Understanding the factors that affect the delivery of software at an
organizational level offers businesses and engineering teams the
knowledge to deliver value to end-users, maintain competitiveness, and
improve developer experience. Given engineering teams' fundamental role
in software delivery, the velocity of their work---that is, the time it
takes for task completion---has emerged as a focal point of empirical
investigation, particularly through measures like cycle time which
captures the duration between ticket opening and ticket closing.
Moreover, cycle-time is seen by engineers as the most useful metric of
engineering productivity according to a prominent industry report (Carey
2024).

While cycle time is often treated as an indicator of productivity per
se, the concept of productivity remains poorly specified in software
engineering contexts, where outputs fundamentally differ from the more
readily quantifiable measures used in traditional industrial production.
Specific units of work are rarely identical across time for a person,
within a team, or across teams. The interpretation of cycle time as a
proxy for productivity therefore presents particular challenges because
variations could reflect differences in work patterns, task assignment,
task scoping, and organizational contexts rather than differences in
some underlying rate of task completion.

Nevertheless, the intuitive appeal of cycle time and its widespread use
in practice make it a valuable focus for empirical investigation. The
above-mentioned complexities necessitate sophisticated statistical
methods to detect the unique impact of multiple factors, while carefully
characterizing the variability practitioners can expect in day-to-day
and month-to-month observations of cycle time. Through rigorous
statistical modeling of longitudinal data across multiple organizations,
we can both characterize its variability across real-world contexts,
while demonstrating methodological approaches for analyzing such complex
operational metrics. This analysis also allows us to detect systematic
influences from factors commonly believed to affect developer
productivity: task scoping, focused work time, collaboration, and time
of year.

Our investigation leverages a unique dataset comprising over 11,398
contributors at 216 organizations across diverse industries. This work
makes two primary contributions. First, we demonstrate a model for
statistically investigating software activity data at both a larger and
more longitudinal scale than previous empirical research, allowing us to
characterize how cycle time varies across software development contexts
(i.e., individuals, organizations, and variable process factors), using
hierarchical modeling that appropriately separates individual and
organizational variation, combined with the careful disaggregation of
within- and between-person effects. This approach allows us both greater
precision and nuance in describing effects as well as the ability to
highlight potential pitfalls in using such measurements to drive
decision-making. Second, we incorporate these multiple measures of
process factors simultaneously to isolate unique effects, including a
novel measure of collaboration operationalized as degree centrality,
taking initial steps toward reflecting the impact of the interactive
nature of software development in large-scale analyses of activity data.

Our research questions are:

RQ1. How do common workplace and software development process factors
impact cycle time?

RQ2. How much between- and within-person variation is there in cycle
time?

The paper proceeds as follows: We first review the literature on
software productivity measurement, examining cycle time's relationship
to broader discussions of developer performance. We then present our
methodology for analyzing cycle time variation using Bayesian
hierarchical linear models. Our results examine both population-level
effects and the substantial variation observed between individuals and
organizations. We conclude by discussing implications for practice and
future research directions.

\section{Background}\label{background}

\subsection{Productivity}\label{productivity}

The use of cycle time in the academic and industry literature is almost
always as part of a discussion of productivity. This may be in part
because cycle time and related metrics are one of the only so-called
objective quantitative windows we have into the process of software
production (but note that self-reports of perceived productivity are
also potentially valid measures of this process). For this reason, it
behooves us to discuss the literature on productivity, even as we
position the analyses in this report as specifically analyzing what we
consider to be at best a very distal indicator of whatever it is people
mean when they use the word ``productivity.''

Defining software team productivity and performance is a highly
contentious exercise and many different definitions are given by both
practitioners and researchers (Fraser et al. 2007; C. Hicks, Lee, and
Ramsey 2023; C. M. Hicks, Lee, and Ramsey 2024; Murphy-Hill et al. 2021;
Sadowski, Storey, and Feldt 2019). Perceptions of what counts as
successful software work can meaningfully differ across individuals and
roles, as when engineering managers tend to focus on long-term outcomes
and individual developers focus on activity, for example (C. Hicks, Lee,
and Ramsey 2023; Storey, Houck, and Zimmermann 2022b). Across
workplaces, measures of time have been frequently used to assess
productivity even while the shortcomings of these measures are also
widely acknowledged (Griffin 1993). Alternative measures include
self-ratings or peer evaluations (Murphy-Hill et al. 2021; Ramírez and
Nembhard 2004) and in software engineering, operationalizations of code
work such as lines of code (Blackburn, Scudder, and Van Wassenhove 1996;
Maxwell, Van Wassenhove, and Dutta 1996). These have obvious limitations
in that the meaning of a particular unit for any of these metrics may be
different depending on context (Sadowski, Storey, and Feldt 2019). Some
researchers have sought solutions to this problem by asking individuals
to rate their own level of, or satisfaction with, productivity (C.
Hicks, Lee, and Ramsey 2023; Storey et al. 2021). While it is plausible
that perceived productivity could be a good indicator of productivity,
it is still not free of the context effects that are often levied as
critiques of more ``objective'' metrics, and self-report, while perhaps
overcoming some shortcomings of other methods, bring with them another
set of measurement issues.

The difficulty of quantifying productivity arises even prior to the step
of choosing one or several indicators. There is often a lack of clear
distinction between production (quantity of output regardless of
resources provided), productivity (quantity of output given the
resources provided), and performance (flexibility, adaptability,
dependability, sustainability, and quality of output over time) (C.
Hicks, Lee, and Ramsey 2023). As any software developer will be aware,
this conceptual complexity is likely the result of the various ways
their work counts for professional development, for the success of the
product, and for simply meeting deadlines. This piece of research does
not aim to solve the issue of how we conceive of productivity but
instead seeks to take a deep look at a single popular metric in order to
showcase, first, the many factors (themselves, a subset of possible
influences of productivity) that affect cycle time, and second, how
observing this metric over time informs our view of the ways cycle time
varies both within and between people. These views will be helpful both
to illuminate specific properties of cycle time as a measure but also to
demonstrate how one might approach an in-depth analysis of either
``objective'' or self-report metrics of productivity.

\subsection{Evaluating individual developer
performance}\label{evaluating-individual-developer-performance}

Given the difficulty of appropriately defining productivity, the many
metrics that purport to measure it, and the potential cost to an
individual (e.g., career, reputation) of being measured, it is
understandable that software developers have an ambivalent stance about
the measurement of both work activity and productivity, that metrics
adoption can be fraught with failure (Bouwers, van Deursen, and Visser
2013), and that social or socio-technical affordances can be strongly
associated with self-reported productivity and necessary to obtain a
full picture of software team experience beyond project and technical
metrics (C. Hicks, Lee, and Ramsey 2023; Murphy-Hill et al. 2021).

Developers whose teams use metrics generally see those metrics as
helpful, and developers who report agreement which team-level metrics
are measured tend to report higher perceived productivity (C. Hicks,
Lee, and Ramsey 2023; C. M. Hicks, Lee, and Ramsey 2024). However,
paired with this are some indicators of uncertainty in whether and how
metrics are being tracked or used (C. Hicks, Lee, and Ramsey 2023),
there is often backlash against any attempt to define or popularize such
metrics (Bruneaux 2024; Chhuneja 2024; Coté 2023; Finster 2023; Orosz
2024b, 2024a; Riggins 2023; Terhorst-North 2023b, 2023a; Walker 2023b,
2023a), and there is concern about mismeasurement by managers inside of
organizations (C. Hicks, Lee, and Ramsey 2023), which is part of a
broader discussion of surveillance and the discontent it can generate
for workers (Ball 2010; Grisold et al. 2024; Mettler 2024). More
troublingly, recent scholarship on sociocognitive experiences in the
workplace has proposed that severe experiences of employees being
treated by an organization as a ``mere tool'' or a resource may create
organizational dehumanization leading to many negative impacts on both
well-being measures and on work outcomes (Caesens et al. 2017; Lagios et
al. 2022). Moreover, there is evidence that metrics might be used
differently depending on a person's visible identities (e.g, Quadlin
2018).

Likewise, scholarship on employee perceptions of organizational and
procedural justice have long documented that when employees perceive a
context of organizational injustice, this can exacerbate or redefine
experiences of organizational decision-making and performance
evaluations (Brockner et al. 1994, 2007). Given such larger
organizational dynamics, it is likely that whether or not software
metrics adoptions are successful is impacted not only by the choice of
metric but also by larger contextual factors such as teams'
sociocognitive experiences and expectations around measurement, and the
psychological affordances of their environments which may or may not
allow them to address measurement concerns (C. M. Hicks 2024).

We lack holistic evidence about what practitioners in software
development believe about developer performance and ability; some
reports from researchers with samples at large technology companies have
suggested both that definitions of productivity can vary widely between
managers and developers, and that software developers perceive many
potential trade-offs between types of technical goals, e.g.~that quality
and speed may be unattainable together (Storey, Houck, and Zimmermann
2022a).

One ``industry myth'' which is referenced frequently in practitioner
commentary is the idea of a ``10x engineer'': this position alleges that
some small outlier population of software developers consistently
outperform others on key development tasks. Potentially springing from
small case studies examining a handful of developers' time spent solving
small laboratory tasks (Sackman, Erikson, and Grant 1968; discussed in
Nichols 2019), this ``law'' was generalized from only twelve
individuals, uses time spent on the tasks as an estimate of both effort
and cost, has failed to replicate in larger examinations of developer
performance on similar tasks, and failed to acknowledge large
within-individual variation in task performance (Nichols 2019; Shrikanth
et al. 2021).

Nevertheless, the idea that ``10x engineers'' exist and that some
individuals in software engineering outperform others by a ``rule'' of
10x has been cited often and codified in industry commentary, (e.g.,
Brooks 1975). Modern commentary on this idea frequently refers to it as
a myth, but it is also discussed as a potentially real
phenomenon\footnote{For example, see this
  \href{https://news.ycombinator.com/item?id=22349531}{ycombinator
  thread}
  (\href{http://web.archive.org/web/20240917164935/https://stackoverflow.blog/2024/06/19/the-real-10x-developer-makes-their-whole-team-better/}{internet
  archive}), and this
  \href{https://stackoverflow.blog/2024/06/19/the-real-10x-developer-makes-their-whole-team-better}{StackOverflow
  blog post}
  (\href{http://web.archive.org/web/20231209171051/https://news.ycombinator.com/item?id=22349531}{internet
  archive})}. In our previous work, we have noted that some software
practitioners hold field-specific ability beliefs that software
development success and productivity is attributable to a quality of
``innate brilliance'', and that this belief among practitioners may
create a higher likelihood of experiencing threat and anxiety in the
face of rapid role change and technological shifts to developer
workflows (C. M. Hicks, Lee, and Foster-Marks 2024). Broad reviews on
drivers of software development outcomes, particularly frictions in the
form of team ``debt,'' also suggest that social-psychological aspects of
shared work processes may be a significant contributor to these outcomes
separate from individual performance (Ahmad and Gustavsson 2024).

Despite some recognition that the 10x engineer is a problematic concept,
the conflictual measurement of productivity and its use as a tool of
surveillance and punishment contra the interests of individual
contributors (but to the benefit, at least ostensibly, to a company's
profitability) continues with full-throated glee. A recent unpublished
study claims that nearly 10\% of engineers contribute almost no work;
that is to say, it raises the boogeyman of the 0.1x engineer as the 10x
engineer's inverse\footnote{in fact, the original paper perhaps even
  emphasizes the ``low performers'' more than the ``exceptional''
  stating, ``the `horrid' portion of the performance frequency
  distribution is the long tail at the high end, the positively skewed
  part which shows that one poor performer can consume as much time or
  cost as 5, 10, or 20 good ones.'' (Sackman, Erikson, and Grant 1968,
  6)}(Obstbaum and Denisov-Blanch, n.d.). The measure of productivity
used is something half-way between an objective measurement and
self-report: an unspecified machine-learning model trained on expert
ratings of the quality of, and work necessary to complete, 70 commits
(Denisov-Blanch et al. 2024). Unlike prior work, this method lacks both
the transparency of ``objective'' measures and the temperance of
self-report measures.

In taking a deep dive into cycle time, this project does not address
every implementation challenge and organizational affordance that may
define whether organizations can ensure a healthy and sustainable
practice around the measurement of work activity. However, we believe
that a more robust understanding of the dynamics of cycle time may help
practitioners avoid pitfalls in relying on velocity measures while
evaluating software work. We hope to describe the complexity in a way
that at least adds some clarity and aligns with the experience of
software developers in practice.

\subsection{Cycle Time}\label{cycle-time}

Because lower cycle times are thought to indicate faster delivery times
and more efficient software processes, cycle time has long been taken as
a key indicator of team health, developer productivity, and team
efficiency (Clincy 2003; Agrawal and Chari 2007; Carmel 1995; Evers,
Oehler, and Tucker 1998; Gupta and Souder 1998; Nan and Harter 2009;
Ruvimova et al. 2022; Sadowski and Zimmermann 2019; Trendowicz and Münch
2009). This suggests that understanding factors that influence cycle
time may lead to insights into factors that are important to understand
for understanding productivity in general. At minimum, examining cycle
time can provide a description of the complexity of factors that impact
this popular metric.

Cycle time examines one aspect of the speed of software delivery by
measuring the time between task start and task delivery. It has
consistently been described by industry research as one of the best and
most trusted metrics for software productivity (Carey 2024). In this
same report, similar metrics also showed preference, such as lead time,
deploy frequency, and change failure rate. The broader software
engineering community has emphasized similar constructs through the
research program DevOps Research and Assessment (DORA), which identified
four key measures of software delivery performance: lead time,
deployment frequency, change failure rate, and mean time to recovery
(Forsgren, Humble, and Kim 2018). While cycle time is not identical to
these measures, it overlaps conceptually---particularly with lead
time---in capturing aspects of delivery speed. Positioning cycle time
alongside these measures situates it within a family of indicators
concerned with the timeliness and reliability of software delivery, even
though the operational definitions vary across contexts.

The common thread across these metrics is that the unit of work is
defined by the team or company in relation to goals that serve the
strategic interests of the project. While there is a good deal of nuance
with respect to what goes into setting these units up, they are both
discrete (and so ``objective''-feeling) but also defined, often
collaboratively, with respect to the outcomes that matter. This is in
contrast to lines of code, for example, which may or may not be relevant
to the goals of the engineering teams, and which is avoided by 70\% of
respondents in the same industry report. Cycle time may also be
considered an important part of developer experience as a component of
what leads to a fluid-feeling development and release cycle (André N.
Meyer et al. 2021).

In calls to re-examine the complexity of developer productivity,
researchers have argued that velocity measures are highly
task-dependent, and do not represent the quality of work done or other,
longer-term measures of the impact of work (Sadowski, Storey, and Feldt
2019). It is also possible for velocity measures to have multiple
directional relationships with desired outcomes depending on software
developers' larger context. For instance, hypothetically speaking, an
increase in velocity may associate with more success for a software team
when this increase arises because the team engages in process
improvements, creating processes that help them to move more quickly
through development tasks, and thereby meet a critical deadline for a
product launch, leading to business outcomes which then lead to more
resources for the team. However in a different scenario, an increase in
velocity may be associated with more failures for a software team, for
instance, if velocity changes arise because the team begins to eschew
quality control processes, eventually leading to costly critical
business failures.

Nevertheless, time and output-based measures are frequently used as an
outcome measure to make recommendations for software engineering
practices, e.g.~in evaluating the perceived impact of technical debt
(Besker, Martini, and Bosch 2018). These measures have the added benefit
of having a concrete referent that is simple to measure and inexpensive
and convenient for teams trying to track productivity to collect.

The utility of cycle time has subsequently led to numerous industry
experts recommending that engineering managers and leaders track their
teams' cycle times. However, leaders are provided less guidance on how
to analyze and decrease cycle time. As such, leaders are left with the
dilemma of being aware of their cycle times, but not understanding how
to improve their cycle times in an evidence-based way.

In the literature that does directly address this question, four major
areas have been proposed to impact cycle time: (1) organizational
structure and climate, (2) reward system, (3) software development
process and (4) the use of software design and testing tools (Clincy
2003). We focus in this paper on factors from part 3, software
development processes, in part because measurements of these processes
continue to gather significant interest from the technology industry and
are plausibly mobile levers that can be manipulated at the level of an
engineering team. They are also themselves relatively easy to measure
and track at the team level if a software team within a larger
organization were to decide they wanted to try to shift their processes
and take measurements to make sure they were successful. We have argued
elsewhere that organizational structure and climate are also relatively
easy to measure and are powerful levers that should be more often
targeted (C. M. Hicks and Hevesi 2024; C. Hicks, Lee, and Ramsey 2023;
C. M. Hicks, Lee, and Ramsey 2024), though for the present work we focus
on (3) also in part to keep the scope of this analysis manageable.

To reduce cycle times at the level of software development process, the
software industry currently recommends strategies centered around three
major themes:

\begin{enumerate}
\def\labelenumi{\arabic{enumi}.}
\tightlist
\item
  increased coding time
\item
  improved task scoping
\item
  improved collaboration
\end{enumerate}

Industry convention rationalizes that increased coding times increases
the amount of code committed and pull requests merged, thus moving
tickets through their life cycle more quickly. Improved scoping can
similarly yield more efficient teams by breaking work down into more
manageable chunks and reducing the amount of unplanned work from bugs
and defects. Finally, industry reports posit that improved collaboration
can reduce the time it takes for developers to review PRs and increase
review rates (Flow, n.d.; Gralha 2022; Waydev 2021). There has also been
some work looking at this empirically which supports the idea that
collaboration under certain conditions does improve productivity
Gousios, Pinzger, and Deursen (2014). We focus on these three areas as
possible factors that impact cycle time.

\section{Research design and
methodology}\label{research-design-and-methodology}

Code for these analyses is available as \texttt{analyses.qmd}, here:
\url{https://github.com/jflournoy/no-silver-bullets}. Data are
considered proprietary and are not available to be shared. This research
used aggregated, anonymized GitHub activity data routinely collected
through our company's normal operations and permitted by our Terms of
Service. No personal information was gathered specifically for this
study, and strict protocols were followed to prevent re-identification
of individuals or organizations. Because the dataset was pre-existing,
fully anonymized, and did not involve direct interaction with human
subjects, the research is exempt from IRB review under 45 CFR
\S 46.104(d)(4)(ii). All data was stored on secure systems with limited
access, ensuring both data integrity and confidentiality.

\subsection{Data Selection and
Characteristics}\label{data-selection-and-characteristics}

To examine coding time, task scoping, and collaboration as predictors of
cycle time over time, we centered our analysis on a large, real-world
dataset of git and ticketing data. This dataset includes 55,619
observations across 12 months in 2022 from 11,398 users in 216
organizations of varying sizes and industries. We chose to use
longitudinal data across 12 months, as it allowed us to examine
fluctuations within a person's workflow as well as different stable
tendencies between people.This data was available via partnerships
between a software metrics tool\footnote{Formerly Pluralsight Flow, now
  Flow at Appfire. All authors were research scientists or data
  scientists employed at Pluralsight at the time that this data was
  collected and analyzed.} which was incorporated into the workflows of
real working software teams, and the 216 organizations which opted in to
this tool at any point during the 12 month analytic window. Notably,
because this tool was adopted on an organizational level (following
partnership agreements that include organizational opt-in and security
audits), users themselves did not have to be active users of the
software metrics tool itself in order to be included in this dataset,
and git and ticketing data was available retrospectively for dates prior
to the implementation of the tool in the organization. In other words,
the git and ticketing data included in this analysis is not predicated
on being an individual user of the software metrics tool, nor on the
software metrics tool being used at the organization, as our dataset
contains measures both before and after the software metric tool
implementation at the organization, and implementation dates for
organizations vary across the 12 month period.

\begin{figure}[htbp]

\centering{

\includegraphics[width=3in,height=\textheight]{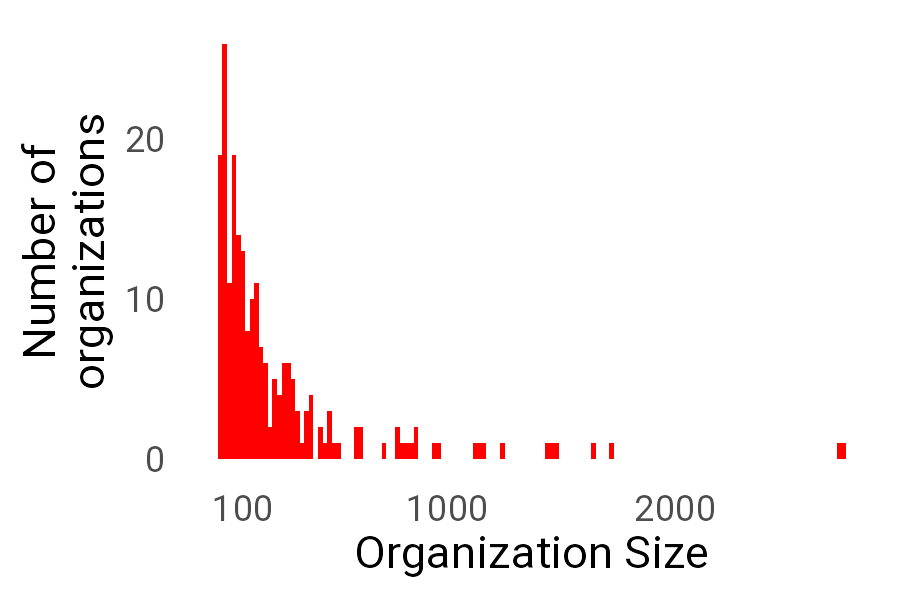}

}

\caption[Histogram of organization sizes in the
dataset.]{\label{fig-org-size}Organization sizes clustered around 130
users, with a long tail of larger organizations. Note that ``users''
generally refers to developers or other individuals creating and closing
tickets.}

\end{figure}

Data were selected for analysis based on whether users actively
contributed code during the time frame of the study. The 216
organizations each had between 1--2,746 individuals in the dataset, with
90\% of organizations being represented by more than 12 users (Median =
130; Figure~\ref{fig-org-size}). In previous pilot surveys used to
inform the design of this project, professional software developer users
from these organizations described their main industries as ranging from
Technology, Finance, Government, Insurance, Retail, and others,
indicating a wide diversity of business use cases and engineering
contexts were present in this sample.

\subsection{Computing study variables}\label{computing-study-variables}

Using the most complete data for each user, we used the mean to
aggregate each variable at the \emph{month} level and the \emph{year}
level (see below for more details specific to each variable). For each
predictor, we then subtracted each person's yearly average from their
monthly data to produce a within-person deviation variable. This allowed
us to disaggregate effects on the outcome due to yearly-level individual
differences and within-person, month-to-month fluctuations (Curran and
Bauer 2011). This also allowed us to avoid averaging between-person and
within-person differences into a single effect estimate. These effects
can be different even in the sign of the effect, for example with a
positive relationship between some time-invariant factor and the outcome
of interest at the between-person level, and a negative relationship
between the same factor measured across time and within-person variation
over time. A common example that is highly relevant to most technical
and knowledge workers is typing speed and errors. Imagine someone trying
to type as fast as they can; it is obvious that they will make more
errors the faster they type, evincing a negative association between
speed and errors. However, if one simply measures the typing speed and
error rate of many people, it should be clear that we would see that
faster typists tend to make fewer errors, perhaps because of differences
in typing experience. In this study, we want to be able to examine
average differences between people's cycle time aggregated at the year
while also examining what is associated with cycle time deviations from
that yearly trend month-to-month. All year-level individual differences
variables were centered at their mean. Exceptions or addenda are
mentioned below. See Table~\ref{tbl-variables} for a brief list of
variables.

\subsubsection{Cycle Time}\label{cycle-time-1}

This is the dependent variable in these analyses. After computing the
cycle time for each closed ticket in seconds, we found the median cycle
time for each month for each user using all tickets \emph{opened} in
that month. For example, a ticket opened on the 9th of April, and closed
on the 3rd of May would contribute 2,246,400 seconds to the calculation
of the median for April. Depending on how organizations actually use
tickets in practice, it is not guaranteed that work has not already
begun prior to ticket opening.

\subsubsection{Unclosed Tickets}\label{unclosed-tickets}

We were not able to observe the closing date for every ticket given our
data collection cutoff of March 7, 2023, and so it is plausible that we
underestimate the median cycle time in a way that depends in part on how
many ticket closing times we do not observe. For this reason, we also
computed the proportion of tickets opened in that month that had not
been closed by the end of our data collection. For example, any ticket
opened in April, 2022 but not closed by March 7, 2023 would count toward
the proportion of unclosed tickets for that month. We transformed
proportions from \([0,1]\) to \((-\infty, \infty)\) using the logistic
quantile function (with minimum and maximum proportions forced to be .01
and .99 respectively). We use this in the regressions below as a control
variable to adjust for this possibility.

\subsubsection{Time (Month, and within-quarter
month)}\label{time-month-and-within-quarter-month}

We examined time in two ways: monthly and quarterly. Months were
represented as numeric values (i.e.~January = 1, February = 2) and
centered at month 7, which allows us to interpret certain quantities
like the intercept as the average cycle time in the middle of the year.
Additionally, because quarters provide meaningful business cadences that
may impact engineering work, for instance in that some organizations set
quarterly goals at the beginning of each quarter and push to meet those
goals at the end of each quarter and that key product deadlines may
occur systematically toward the end of quarters, we accounted for any
effects of quarterly cycles by using an indicator for the within-quarter
month, centered at the middle of the quarter (e.g., -1 for the first
month of the quarter, 0 for the middle month, and 1 for the last month
of the quarter). This approach allowed us to capture a more stable and
realistic trajectory of change over the course of the year.

\subsubsection{Team Size}\label{team-size}

To control for any influence of team size on cycle time, we compute each
individual's team size as the average size of all teams that individual
belongs to as defined by individuals' co-located activity data.
Specifically, in the database used, an individual contributor is given
membership in any team that they have worked in, and this is updated
retroactively. For each individual, we find all teams that person is a
member of, compute the size of that team, and then average across those
team sizes if an individual is a member of multiple teams. As such, this
number is a very rough indicator of the size of teams an individual
tends to be a part of and is static across the year. This is a
limitation of the database. This is then entered as an
individually-varying continuous variable to control for some of the
effect of team size on an individual's cycle time.

\subsubsection{Coding days}\label{coding-days}

Coding days was summarized as the average number of days per week that a
developer made at least one commit. We divided the number of coding days
in a month by the total number of days in that month and multiplied by
seven to aid in interpretation. Based on conversations with software
developers, we understand that making small commits frequently is often
considered best-practice, but the fact that commits can be made
independent of actual coding time means that our proxy measure for
coding days is imperfect.

\subsubsection{Total Merged PRs}\label{total-merged-prs}

One frequently proposed best practice in software work, intended to lead
to outcomes such as improved task scoping, involves breaking work into
smaller and more manageable chunks or pull requests that can be finished
more quickly (Kudrjavets, Nagappan, and Rastogi 2022; Lines 2023; Riosa
2019; Zhang et al. 2022). For a given software development goal, if we
assume the set of commits necessary to accomplish that goals remains the
same, more pull requests suggests that the task was broken down into
smaller discrete goals in a way that groups more closely related
subtasks together than one large pull request. Making smaller, more
frequent pull requests is also itself a way to break up the task for
code reviewers in a way that is thought to improve productivity. As
such, we used the number of total merged pull requests as one measure of
task scoping. To calculate this, we counted the number of merged pull
requests for each user for each month.

\subsubsection{Percent Defect Tickets}\label{percent-defect-tickets}

Another potentially beneficial signal in software activity data is the
reduction of unplanned work on bugs and defect tickets, which is also
proposed as a bottleneck on improving cycle time (Paudel et al. 2024;
Rosser and Norton 2021; Toxboe 2023). As such, we used the percentage of
defect tickets as another measure of task scoping to represent unplanned
work that may interfere with timely completion of planned work. This may
also be a downstream signal of individuals' opportunity for focused work
time and code quality. To account for this possibility, for each user,
for each month, we computed the percent of tickets that were defect
tickets.

\subsubsection{Degree centrality}\label{degree-centrality}

We measured collaboration by calculating degree centrality. To evaluate
degree centrality, a metric derived from network analysis and often used
in the analysis of social networks (Watts 2004), we employed a framework
where developers were treated as nodes within the network, and their
interactions in the form of Pull Requests (PRs) were regarded as
connections. In other words, any contribution of code to the same pull
request constituted a collaboration edge between developers. We
normalized each centrality value by dividing by the total number of
developers constituting the organizational network. The calculations
were executed using the Python package Networkx (Hagberg, Schult, and
Swart 2008). This particular variable serves as an effective proxy for
quantifying the extent of collaboration among developers. We multiply
the normalized degree centrality, which is between 0 and 1, by 100.

\subsubsection{Comments per PR}\label{comments-per-pr}

Another indicator of collaboration is the frequency of comments within
PRs. We undertook a comprehensive examination of all PRs that were
successfully merged in the year 2022 and, for each user, calculated the
average number of comments per PR that they authored each month. This
served as a measure to gauge the depth of collaboration exhibited during
the development and review process. However, see Bacchelli and Bird
(2013) suggesting that more comments may reflect, for example, poorer
code quality.

\begin{longtable}[]{@{}
  >{\raggedright\arraybackslash}p{(\columnwidth - 4\tabcolsep) * \real{0.1393}}
  >{\raggedright\arraybackslash}p{(\columnwidth - 4\tabcolsep) * \real{0.2869}}
  >{\raggedright\arraybackslash}p{(\columnwidth - 4\tabcolsep) * \real{0.5738}}@{}}
\caption{Variable descriptions}\label{tbl-variables}\tabularnewline
\toprule\noalign{}
\begin{minipage}[b]{\linewidth}\raggedright
\end{minipage} & \begin{minipage}[b]{\linewidth}\raggedright
Variable
\end{minipage} & \begin{minipage}[b]{\linewidth}\raggedright
Variable Description
\end{minipage} \\
\midrule\noalign{}
\endfirsthead
\toprule\noalign{}
\begin{minipage}[b]{\linewidth}\raggedright
\end{minipage} & \begin{minipage}[b]{\linewidth}\raggedright
Variable
\end{minipage} & \begin{minipage}[b]{\linewidth}\raggedright
Variable Description
\end{minipage} \\
\midrule\noalign{}
\endhead
\bottomrule\noalign{}
\endlastfoot
Productivity & Cycle Time & Avg time from ticket start to end \\
& Proportion unclosed tickets & Control variable to account for tickets
missing cycle time \\
Time & Month & Continuous time variable coded as month number \\
& Within-quarter month & Index of the month number within each quarter
year \\
Team Context & Team size & Average size across all teams a individual is
on \\
Coding Time & Coding Days per Week & Avg number of coding days a week \\
Task Scoping & Total Merged PRs & Total number of merged PRs per
developer \\
& Percent Defect Tickets & Percent of all tickets that are defect
tickets \\
Collaboration & Degree Centrality & Score based on the number of
reviewers a developer has worked with \\
& Comments per PR & Number of comments per pr a developer is the author
on \\
\end{longtable}

\subsection{Analytic Approach}\label{analytic-approach}

The models described below are fit using \texttt{brms} (v2.21.6, Bürkner
2018, 2017), interface the Stan probabilistic programming language for
Bayesian sampling (v2.35.0, Team 2024), with the \texttt{cmdstanr}
backend (v0.8.0, Gabry et al. 2024), in R (v4.3.2, R Core Team 2023).

We developed a model of monthly average ticket cycle time conditional on
the following predictors: within-quarter month number, team size,
proportion of unclosed tickets, month number, yearly means and
month-level deviations for coding days per week, total merged PRs,
defect ticket percentage, degree centrality, and comments per PR.
Specifically, we modeled cycle time as distributed Weibull with two
parameters, \(\lambda\) (scale), and \emph{k} (shape). The Weibull
distribution is often used to model time-to-event data (Harrell 2015;
Rummel 2017), where \emph{k} determines the change over time in the
probability of an event occurring (often called the ``hazard rate''),
and where \(\lambda\) determines the time-to-event for some proportion
of the cases (or in other words, how spread out the distribution is).
For simplicity, we assume that the shape (hazard rate, \emph{k}) is not
influenced by the factors considered, and focus on how these factors
affect the scale (time-to-event, \(\lambda\)) of ticket closures, though
we did allow the shape, \emph{k}, to vary across organizations. In
short, the Weibull distribution provides flexibility for accurately
describing cycle time data that tend to have a bulk of observations at
the low end, with a very long tail of more extreme observations.

The model for \(\lambda\) is

\begin{equation}
\begin{aligned}
\log(\lambda) &= X\beta + \eta_{\text{org}} + \eta_{\text{org:user}} \\
\eta_{\text{org}} &\sim \mathcal{N}\left(\begin{bmatrix} \mu_1 \\ \mu_2 \end{bmatrix}, \begin{bmatrix} \sigma_{11} & \sigma_{12} \\ \sigma_{21} & \sigma_{22} \end{bmatrix}\right) \\
\eta_{\text{org:user}} &\sim \mathcal{N}\left(\begin{bmatrix} \mu_3 \\ \mu_4 \end{bmatrix}, \begin{bmatrix} \sigma_{33} & \sigma_{34} \\ \sigma_{43} & \sigma_{44} \end{bmatrix}\right) 
\end{aligned}
\end{equation}

where \(X\) is the matrix of predictors, \(\beta\) is the vector of
coefficients, \(\eta_{\text{org}}\) is random intercepts with mean
\(\mu_1\) and linear slopes of month with mean \(\mu_2\) for each
organization, and \(\eta_{\text{org:user}}\) is random intercepts with
mean \(\mu_3\) and linear slopes of month with mean \(\mu_4\) for each
user nested within organization. The specific predictors in \(X\) are
within-quarter month number, team size, proportion of unclosed tickets,
month number, yearly means and month-level deviations for coding days
per week, total merged PRs, defect ticket percentage, degree centrality,
and comments per PR. We also include interactions between month number
and the following: team size, proportion of unclosed tickets, and each
of the yearly mean predictors. This allows us to account as completely
as possible for our control variables (team size and proportion of
unclosed tickets), and allow the effect of month on cycle time to vary
by the individual differences variables (e.g., to account for the
possibility that someone who has higher coding days per week shows a
less steep decrease in cycle time across the year than someone with
lower coding days per week).

The model for \emph{k} is

\begin{equation}
\begin{aligned}
\log(k) &= \zeta_{\text{org}} \\
\zeta_{\text{org}} &\sim \mathcal{N}(\mu_5, \sigma_{5}) 
\end{aligned}
\end{equation}

where \(\zeta_{\text{org}}\) is a random intercept with mean \(\mu_5\)
each organization.

Conceptually, this model allows a unique distribution of cycle times (as
determined by the random intercepts for both \(\lambda\) and \emph{k})
for each organization. It also allows the scale of the distribution of
cycle times to vary for each user due to the random intercept for
\(\lambda\). The effect of time (month number) on the scale of the
distribution of cycle times is also allowed to vary across organizations
as well as users due to the random slopes (with means \(\mu_2\) and
\(\mu_4\)). This strategy allows two advantages: first, we account for
multiple sources of variance that allows our estimates of the effects of
interest to be more precise; and second, we are able to provide
estimates of this variation across organizations and users. This
variation itself is of interest given the various myths mentioned in the
introduction about developer performance.

We model the effect of proportion of unclosed tickets and month number
as smooth functions of the covariate using thin-plate splines for
increased flexibility (Wood 2017). Briefly, thin plate splines
(functions made up of smoothly connected segments) allow for flexible,
non-linear relationships between predictors and the response variable.
These splines are penalized to prevent overfitting, balancing model
flexibility and complexity. The interactions between month number and
our control variables are parameterized as additional smooth functions
of month number multiplied by these variables. While our focal model
parameterizes the interactions between year-level means and month number
as linear coefficients on multiplicative combinations between the two
variables, we also examined a model that uses additional smooth
functions of month number multiplied by these variables to allow for
additional complexity. We provide the model output for this sensitivity
analysis in a supplement.

We set weakly-informative priors centered at zero for all parameters,
except for the intercept for \(\lambda\) and \emph{k} which were
centered on their approximate values in the data (consistent with the
default behavior of \texttt{brms}). We performed prior-predictive checks
to ensure our prior specification generated data that covered and
exceeded the space of our observations. Given the complexity of the
model, we also specified initialization of parameters at small plausible
values (e.g., zero for coefficients, .1 for standard deviations of
random effects). Full prior and initialization specifications are
available in the analysis code.

We sampled from 4 chains with 2,000 total iterations each, discarding
the first 1,000 iterations as warmup. Inferences were made on 4,000
post-warmup draws from the posterior probability distribution from the 4
chains.

\subsection{Inferences}\label{inferences}

We take a Bayesian approach to making claims about the sign of effects
(i.e., whether an association between two variables is positive or
negative), and to describing its magnitude. Instead of the common but
fraught frequentist approach of describing whether an effect size is
unlikely given the assumption of an unrealistic point-null hypothesis,
we try to give the reader a sense of the actual probability that the
sign of an effect is in a particular direction, and what the impact of
the factor is on cycle times in terms that are easy to interpret (Gelman
and Carlin 2014).

In more precise statistical terms, unless otherwise stated we describe
the posterior of parameters and predictions using the median of the
distribution, and characterize its variation using the highest posterior
density interval (HDI) which is defined as the interval that contains a
specified percentage (usually 95\%) of the most probable values of the
parameter (Kruschke 2018). We make general descriptive inferences based
on the probability that a parameter has the sign of the posterior
density's median value. For example, if 80\% of the posterior density of
the slope of the effect of month on cycle time is of the same sign as
the density's median, and that median is negative, we would say
something like, ``given the model and the data, there is an 80\% chance
that there is a decrease in cycle times across the year.''

\subsection{R packages}\label{r-packages}

R packages explicitly loaded in this analysis and manuscript preparation
include brms (v2.22.7, Bürkner 2018, 2021, 2017), cmdstanr (v0.8.0,
Gabry et al. 2024), data.table (v1.15.4, Barrett et al. 2024), ggplot2
(v3.5.0, Wickham 2016), flextable (v0.9.5, Gohel and Skintzos 2024),
knitr (v1.46, Xie 2015, 2014, 2024), marginaleffects (v0.19.0,
Arel-Bundock 2024), mgcv (v1.9.0, Wood 2011, 2017, 2004, 2003; Wood,
Pya, and Saefken 2016), parameters (v0.21.6, Lüdecke et al. 2020),
patchwork (v1.2.0, Pedersen 2024), posterior (v1.5.0, Bürkner et al.
2023; Vehtari et al. 2021), rlang (v1.1.3, Henry and Wickham 2024),
scales (v1.3.0, Wickham, Pedersen, and Seidel 2023), scico (v1.5.0.9000,
Pedersen and Crameri 2025), showtext (v0.9.7, Qiu and details. 2024),
StanHeaders (v2.36.0.9000, Stan Development Team 2020), and tidybayes
(v3.0.6, Kay 2023).

\section{Results}\label{results}

Results from the linear model reported below were highly similar to
those in the more flexible non-linear model sensitivity analysis
described above. Also note that parameters in the table are from a
linear model for the distribution of log(\(\lambda\)) and log(\emph{k}),
while model expectations are on the response scale and can therefore
display curvature even while the model is linear.

The first section of the results concerns the population-level effects
and showcases the expectations of cycle time conditional on the various
co-varying factors we target. The second section explores the
variability in these effects across time, across individuals, and across
organizations.

\textsubscript{Source:
\href{https://jflournoy.github.io/no-silver-bullets/index.qmd.html}{Article
Notebook}}

\subsection{Population-level effects}\label{population-level-effects}

\global\setlength{\Oldarrayrulewidth}{\arrayrulewidth}

\global\setlength{\Oldtabcolsep}{\tabcolsep}

\setlength{\tabcolsep}{2pt}

\renewcommand*{\arraystretch}{1.5}

\providecommand{\ascline}[3]{\noalign{\global\arrayrulewidth #1}\arrayrulecolor[HTML]{#2}\cline{#3}}

\begin{longtable}[c]{|p{1.25in}|p{0.75in}|p{0.75in}|p{0.75in}|p{0.75in}}

\caption{\label{tbl-results-fe}Population-level effect estimates}

\tabularnewline

\ascline{1.5pt}{666666}{1-5}

\multicolumn{1}{>{\raggedright}m{\dimexpr 1.25in+0\tabcolsep}}{\textcolor[HTML]{000000}{\fontsize{8}{8}\selectfont{\textbf{Parameter}}}} & \multicolumn{1}{>{\raggedleft}m{\dimexpr 0.75in+0\tabcolsep}}{\textcolor[HTML]{000000}{\fontsize{8}{8}\selectfont{\textbf{Posterior\ Median}}}\textcolor[HTML]{000000}{\fontsize{8}{8}\selectfont{\textbf{\textsuperscript{1}}}}} & \multicolumn{1}{>{\raggedleft}m{\dimexpr 0.75in+0\tabcolsep}}{\textcolor[HTML]{000000}{\fontsize{8}{8}\selectfont{\textbf{Lower\ 95\%\ HDI}}}\textcolor[HTML]{000000}{\fontsize{8}{8}\selectfont{\textbf{\textsuperscript{2}}}}} & \multicolumn{1}{>{\raggedleft}m{\dimexpr 0.75in+0\tabcolsep}}{\textcolor[HTML]{000000}{\fontsize{8}{8}\selectfont{\textbf{Upper\ 95\%\ HDI}}}\textcolor[HTML]{000000}{\fontsize{8}{8}\selectfont{\textbf{\textsuperscript{2}}}}} & \multicolumn{1}{>{\raggedleft}m{\dimexpr 0.75in+0\tabcolsep}}{\textcolor[HTML]{000000}{\fontsize{8}{8}\selectfont{\textbf{Sign\ Probability}}}\textcolor[HTML]{000000}{\fontsize{8}{8}\selectfont{\textbf{\textsuperscript{3}}}}} \\

\ascline{1.5pt}{666666}{1-5}\endfirsthead 

\ascline{1.5pt}{666666}{1-5}

\multicolumn{1}{>{\raggedright}m{\dimexpr 1.25in+0\tabcolsep}}{\textcolor[HTML]{000000}{\fontsize{8}{8}\selectfont{\textbf{Parameter}}}} & \multicolumn{1}{>{\raggedleft}m{\dimexpr 0.75in+0\tabcolsep}}{\textcolor[HTML]{000000}{\fontsize{8}{8}\selectfont{\textbf{Posterior\ Median}}}\textcolor[HTML]{000000}{\fontsize{8}{8}\selectfont{\textbf{\textsuperscript{1}}}}} & \multicolumn{1}{>{\raggedleft}m{\dimexpr 0.75in+0\tabcolsep}}{\textcolor[HTML]{000000}{\fontsize{8}{8}\selectfont{\textbf{Lower\ 95\%\ HDI}}}\textcolor[HTML]{000000}{\fontsize{8}{8}\selectfont{\textbf{\textsuperscript{2}}}}} & \multicolumn{1}{>{\raggedleft}m{\dimexpr 0.75in+0\tabcolsep}}{\textcolor[HTML]{000000}{\fontsize{8}{8}\selectfont{\textbf{Upper\ 95\%\ HDI}}}\textcolor[HTML]{000000}{\fontsize{8}{8}\selectfont{\textbf{\textsuperscript{2}}}}} & \multicolumn{1}{>{\raggedleft}m{\dimexpr 0.75in+0\tabcolsep}}{\textcolor[HTML]{000000}{\fontsize{8}{8}\selectfont{\textbf{Sign\ Probability}}}\textcolor[HTML]{000000}{\fontsize{8}{8}\selectfont{\textbf{\textsuperscript{3}}}}} \\

\ascline{1.5pt}{666666}{1-5}\endhead

\multicolumn{1}{>{\raggedright}m{\dimexpr 1.25in+0\tabcolsep}}{\textcolor[HTML]{000000}{\fontsize{8}{8}\selectfont{Intercept\ log(λ)}}} & \multicolumn{1}{>{\raggedleft}m{\dimexpr 0.75in+0\tabcolsep}}{\textcolor[HTML]{000000}{\fontsize{8}{8}\selectfont{14.3484}}} & \multicolumn{1}{>{\raggedleft}m{\dimexpr 0.75in+0\tabcolsep}}{\textcolor[HTML]{000000}{\fontsize{8}{8}\selectfont{14.2727}}} & \multicolumn{1}{>{\raggedleft}m{\dimexpr 0.75in+0\tabcolsep}}{\textcolor[HTML]{000000}{\fontsize{8}{8}\selectfont{14.4282}}} & \multicolumn{1}{>{\raggedleft}m{\dimexpr 0.75in+0\tabcolsep}}{\textcolor[HTML]{000000}{\fontsize{8}{8}\selectfont{\ 100\%}}} \\

\multicolumn{1}{>{\raggedright}m{\dimexpr 1.25in+0\tabcolsep}}{\textcolor[HTML]{000000}{\fontsize{8}{8}\selectfont{Intercept\ log(k)}}} & \multicolumn{1}{>{\raggedleft}m{\dimexpr 0.75in+0\tabcolsep}}{\textcolor[HTML]{000000}{\fontsize{8}{8}\selectfont{0.1214}}} & \multicolumn{1}{>{\raggedleft}m{\dimexpr 0.75in+0\tabcolsep}}{\textcolor[HTML]{000000}{\fontsize{8}{8}\selectfont{0.0807}}} & \multicolumn{1}{>{\raggedleft}m{\dimexpr 0.75in+0\tabcolsep}}{\textcolor[HTML]{000000}{\fontsize{8}{8}\selectfont{0.1585}}} & \multicolumn{1}{>{\raggedleft}m{\dimexpr 0.75in+0\tabcolsep}}{\textcolor[HTML]{000000}{\fontsize{8}{8}\selectfont{\ 100\%}}} \\

\multicolumn{1}{>{\raggedright}m{\dimexpr 1.25in+0\tabcolsep}}{\textcolor[HTML]{000000}{\fontsize{8}{8}\selectfont{Within-quarter\ month}}} & \multicolumn{1}{>{\raggedleft}m{\dimexpr 0.75in+0\tabcolsep}}{\textcolor[HTML]{000000}{\fontsize{8}{8}\selectfont{-0.0085}}} & \multicolumn{1}{>{\raggedleft}m{\dimexpr 0.75in+0\tabcolsep}}{\textcolor[HTML]{000000}{\fontsize{8}{8}\selectfont{-0.0188}}} & \multicolumn{1}{>{\raggedleft}m{\dimexpr 0.75in+0\tabcolsep}}{\textcolor[HTML]{000000}{\fontsize{8}{8}\selectfont{0.0013}}} & \multicolumn{1}{>{\raggedleft}m{\dimexpr 0.75in+0\tabcolsep}}{\textcolor[HTML]{000000}{\fontsize{8}{8}\selectfont{\ 95\%}}} \\

\multicolumn{1}{>{\raggedright}m{\dimexpr 1.25in+0\tabcolsep}}{\textcolor[HTML]{000000}{\fontsize{8}{8}\selectfont{Team\ size}}} & \multicolumn{1}{>{\raggedleft}m{\dimexpr 0.75in+0\tabcolsep}}{\textcolor[HTML]{000000}{\fontsize{8}{8}\selectfont{0.0001}}} & \multicolumn{1}{>{\raggedleft}m{\dimexpr 0.75in+0\tabcolsep}}{\textcolor[HTML]{000000}{\fontsize{8}{8}\selectfont{-0.0644}}} & \multicolumn{1}{>{\raggedleft}m{\dimexpr 0.75in+0\tabcolsep}}{\textcolor[HTML]{000000}{\fontsize{8}{8}\selectfont{0.0560}}} & \multicolumn{1}{>{\raggedleft}m{\dimexpr 0.75in+0\tabcolsep}}{\textcolor[HTML]{000000}{\fontsize{8}{8}\selectfont{\ 50\%}}} \\

\multicolumn{1}{>{\raggedright}m{\dimexpr 1.25in+0\tabcolsep}}{\textcolor[HTML]{000000}{\fontsize{8}{8}\selectfont{Avg.\ coding\ days/week\ (within-person)}}} & \multicolumn{1}{>{\raggedleft}m{\dimexpr 0.75in+0\tabcolsep}}{\textcolor[HTML]{000000}{\fontsize{8}{8}\selectfont{-0.0794}}} & \multicolumn{1}{>{\raggedleft}m{\dimexpr 0.75in+0\tabcolsep}}{\textcolor[HTML]{000000}{\fontsize{8}{8}\selectfont{-0.0911}}} & \multicolumn{1}{>{\raggedleft}m{\dimexpr 0.75in+0\tabcolsep}}{\textcolor[HTML]{000000}{\fontsize{8}{8}\selectfont{-0.0677}}} & \multicolumn{1}{>{\raggedleft}m{\dimexpr 0.75in+0\tabcolsep}}{\textcolor[HTML]{000000}{\fontsize{8}{8}\selectfont{\ 100\%}}} \\

\multicolumn{1}{>{\raggedright}m{\dimexpr 1.25in+0\tabcolsep}}{\textcolor[HTML]{000000}{\fontsize{8}{8}\selectfont{Avg.\ coding\ days/week}}} & \multicolumn{1}{>{\raggedleft}m{\dimexpr 0.75in+0\tabcolsep}}{\textcolor[HTML]{000000}{\fontsize{8}{8}\selectfont{-0.0839}}} & \multicolumn{1}{>{\raggedleft}m{\dimexpr 0.75in+0\tabcolsep}}{\textcolor[HTML]{000000}{\fontsize{8}{8}\selectfont{-0.1100}}} & \multicolumn{1}{>{\raggedleft}m{\dimexpr 0.75in+0\tabcolsep}}{\textcolor[HTML]{000000}{\fontsize{8}{8}\selectfont{-0.0587}}} & \multicolumn{1}{>{\raggedleft}m{\dimexpr 0.75in+0\tabcolsep}}{\textcolor[HTML]{000000}{\fontsize{8}{8}\selectfont{\ 100\%}}} \\

\multicolumn{1}{>{\raggedright}m{\dimexpr 1.25in+0\tabcolsep}}{\textcolor[HTML]{000000}{\fontsize{8}{8}\selectfont{Total\ merged\ PRs\ (within-person)}}} & \multicolumn{1}{>{\raggedleft}m{\dimexpr 0.75in+0\tabcolsep}}{\textcolor[HTML]{000000}{\fontsize{8}{8}\selectfont{-0.0127}}} & \multicolumn{1}{>{\raggedleft}m{\dimexpr 0.75in+0\tabcolsep}}{\textcolor[HTML]{000000}{\fontsize{8}{8}\selectfont{-0.0155}}} & \multicolumn{1}{>{\raggedleft}m{\dimexpr 0.75in+0\tabcolsep}}{\textcolor[HTML]{000000}{\fontsize{8}{8}\selectfont{-0.0097}}} & \multicolumn{1}{>{\raggedleft}m{\dimexpr 0.75in+0\tabcolsep}}{\textcolor[HTML]{000000}{\fontsize{8}{8}\selectfont{\ 100\%}}} \\

\multicolumn{1}{>{\raggedright}m{\dimexpr 1.25in+0\tabcolsep}}{\textcolor[HTML]{000000}{\fontsize{8}{8}\selectfont{Total\ merged\ PRs}}} & \multicolumn{1}{>{\raggedleft}m{\dimexpr 0.75in+0\tabcolsep}}{\textcolor[HTML]{000000}{\fontsize{8}{8}\selectfont{-0.0083}}} & \multicolumn{1}{>{\raggedleft}m{\dimexpr 0.75in+0\tabcolsep}}{\textcolor[HTML]{000000}{\fontsize{8}{8}\selectfont{-0.0139}}} & \multicolumn{1}{>{\raggedleft}m{\dimexpr 0.75in+0\tabcolsep}}{\textcolor[HTML]{000000}{\fontsize{8}{8}\selectfont{-0.0027}}} & \multicolumn{1}{>{\raggedleft}m{\dimexpr 0.75in+0\tabcolsep}}{\textcolor[HTML]{000000}{\fontsize{8}{8}\selectfont{\ 100\%}}} \\

\multicolumn{1}{>{\raggedright}m{\dimexpr 1.25in+0\tabcolsep}}{\textcolor[HTML]{000000}{\fontsize{8}{8}\selectfont{Defect\ tickets\ \%\ (within-person)}}} & \multicolumn{1}{>{\raggedleft}m{\dimexpr 0.75in+0\tabcolsep}}{\textcolor[HTML]{000000}{\fontsize{8}{8}\selectfont{-0.0019}}} & \multicolumn{1}{>{\raggedleft}m{\dimexpr 0.75in+0\tabcolsep}}{\textcolor[HTML]{000000}{\fontsize{8}{8}\selectfont{-0.0023}}} & \multicolumn{1}{>{\raggedleft}m{\dimexpr 0.75in+0\tabcolsep}}{\textcolor[HTML]{000000}{\fontsize{8}{8}\selectfont{-0.0014}}} & \multicolumn{1}{>{\raggedleft}m{\dimexpr 0.75in+0\tabcolsep}}{\textcolor[HTML]{000000}{\fontsize{8}{8}\selectfont{\ 100\%}}} \\

\multicolumn{1}{>{\raggedright}m{\dimexpr 1.25in+0\tabcolsep}}{\textcolor[HTML]{000000}{\fontsize{8}{8}\selectfont{Defect\ tickets\ \%}}} & \multicolumn{1}{>{\raggedleft}m{\dimexpr 0.75in+0\tabcolsep}}{\textcolor[HTML]{000000}{\fontsize{8}{8}\selectfont{0.0060}}} & \multicolumn{1}{>{\raggedleft}m{\dimexpr 0.75in+0\tabcolsep}}{\textcolor[HTML]{000000}{\fontsize{8}{8}\selectfont{0.0049}}} & \multicolumn{1}{>{\raggedleft}m{\dimexpr 0.75in+0\tabcolsep}}{\textcolor[HTML]{000000}{\fontsize{8}{8}\selectfont{0.0070}}} & \multicolumn{1}{>{\raggedleft}m{\dimexpr 0.75in+0\tabcolsep}}{\textcolor[HTML]{000000}{\fontsize{8}{8}\selectfont{\ 100\%}}} \\

\multicolumn{1}{>{\raggedright}m{\dimexpr 1.25in+0\tabcolsep}}{\textcolor[HTML]{000000}{\fontsize{8}{8}\selectfont{Degree\ centrality\ (within-person)}}} & \multicolumn{1}{>{\raggedleft}m{\dimexpr 0.75in+0\tabcolsep}}{\textcolor[HTML]{000000}{\fontsize{8}{8}\selectfont{-0.0023}}} & \multicolumn{1}{>{\raggedleft}m{\dimexpr 0.75in+0\tabcolsep}}{\textcolor[HTML]{000000}{\fontsize{8}{8}\selectfont{-0.0040}}} & \multicolumn{1}{>{\raggedleft}m{\dimexpr 0.75in+0\tabcolsep}}{\textcolor[HTML]{000000}{\fontsize{8}{8}\selectfont{-0.0006}}} & \multicolumn{1}{>{\raggedleft}m{\dimexpr 0.75in+0\tabcolsep}}{\textcolor[HTML]{000000}{\fontsize{8}{8}\selectfont{\ 100\%}}} \\

\multicolumn{1}{>{\raggedright}m{\dimexpr 1.25in+0\tabcolsep}}{\textcolor[HTML]{000000}{\fontsize{8}{8}\selectfont{Degree\ centrality}}} & \multicolumn{1}{>{\raggedleft}m{\dimexpr 0.75in+0\tabcolsep}}{\textcolor[HTML]{000000}{\fontsize{8}{8}\selectfont{-0.0040}}} & \multicolumn{1}{>{\raggedleft}m{\dimexpr 0.75in+0\tabcolsep}}{\textcolor[HTML]{000000}{\fontsize{8}{8}\selectfont{-0.0063}}} & \multicolumn{1}{>{\raggedleft}m{\dimexpr 0.75in+0\tabcolsep}}{\textcolor[HTML]{000000}{\fontsize{8}{8}\selectfont{-0.0015}}} & \multicolumn{1}{>{\raggedleft}m{\dimexpr 0.75in+0\tabcolsep}}{\textcolor[HTML]{000000}{\fontsize{8}{8}\selectfont{\ 100\%}}} \\

\multicolumn{1}{>{\raggedright}m{\dimexpr 1.25in+0\tabcolsep}}{\textcolor[HTML]{000000}{\fontsize{8}{8}\selectfont{Comments\ per\ PR\ (within-person)}}} & \multicolumn{1}{>{\raggedleft}m{\dimexpr 0.75in+0\tabcolsep}}{\textcolor[HTML]{000000}{\fontsize{8}{8}\selectfont{0.0046}}} & \multicolumn{1}{>{\raggedleft}m{\dimexpr 0.75in+0\tabcolsep}}{\textcolor[HTML]{000000}{\fontsize{8}{8}\selectfont{0.0037}}} & \multicolumn{1}{>{\raggedleft}m{\dimexpr 0.75in+0\tabcolsep}}{\textcolor[HTML]{000000}{\fontsize{8}{8}\selectfont{0.0054}}} & \multicolumn{1}{>{\raggedleft}m{\dimexpr 0.75in+0\tabcolsep}}{\textcolor[HTML]{000000}{\fontsize{8}{8}\selectfont{\ 100\%}}} \\

\multicolumn{1}{>{\raggedright}m{\dimexpr 1.25in+0\tabcolsep}}{\textcolor[HTML]{000000}{\fontsize{8}{8}\selectfont{Comments\ per\ PR}}} & \multicolumn{1}{>{\raggedleft}m{\dimexpr 0.75in+0\tabcolsep}}{\textcolor[HTML]{000000}{\fontsize{8}{8}\selectfont{0.0098}}} & \multicolumn{1}{>{\raggedleft}m{\dimexpr 0.75in+0\tabcolsep}}{\textcolor[HTML]{000000}{\fontsize{8}{8}\selectfont{0.0075}}} & \multicolumn{1}{>{\raggedleft}m{\dimexpr 0.75in+0\tabcolsep}}{\textcolor[HTML]{000000}{\fontsize{8}{8}\selectfont{0.0120}}} & \multicolumn{1}{>{\raggedleft}m{\dimexpr 0.75in+0\tabcolsep}}{\textcolor[HTML]{000000}{\fontsize{8}{8}\selectfont{\ 100\%}}} \\

\multicolumn{1}{>{\raggedright}m{\dimexpr 1.25in+0\tabcolsep}}{\textcolor[HTML]{000000}{\fontsize{8}{8}\selectfont{Avg.\ coding\ days/week\ ×\ Month}}} & \multicolumn{1}{>{\raggedleft}m{\dimexpr 0.75in+0\tabcolsep}}{\textcolor[HTML]{000000}{\fontsize{8}{8}\selectfont{-0.0047}}} & \multicolumn{1}{>{\raggedleft}m{\dimexpr 0.75in+0\tabcolsep}}{\textcolor[HTML]{000000}{\fontsize{8}{8}\selectfont{-0.0098}}} & \multicolumn{1}{>{\raggedleft}m{\dimexpr 0.75in+0\tabcolsep}}{\textcolor[HTML]{000000}{\fontsize{8}{8}\selectfont{-0.0001}}} & \multicolumn{1}{>{\raggedleft}m{\dimexpr 0.75in+0\tabcolsep}}{\textcolor[HTML]{000000}{\fontsize{8}{8}\selectfont{\ 97\%}}} \\

\multicolumn{1}{>{\raggedright}m{\dimexpr 1.25in+0\tabcolsep}}{\textcolor[HTML]{000000}{\fontsize{8}{8}\selectfont{Total\ merged\ PRs\ ×\ Month}}} & \multicolumn{1}{>{\raggedleft}m{\dimexpr 0.75in+0\tabcolsep}}{\textcolor[HTML]{000000}{\fontsize{8}{8}\selectfont{-0.0007}}} & \multicolumn{1}{>{\raggedleft}m{\dimexpr 0.75in+0\tabcolsep}}{\textcolor[HTML]{000000}{\fontsize{8}{8}\selectfont{-0.0017}}} & \multicolumn{1}{>{\raggedleft}m{\dimexpr 0.75in+0\tabcolsep}}{\textcolor[HTML]{000000}{\fontsize{8}{8}\selectfont{0.0003}}} & \multicolumn{1}{>{\raggedleft}m{\dimexpr 0.75in+0\tabcolsep}}{\textcolor[HTML]{000000}{\fontsize{8}{8}\selectfont{\ 90\%}}} \\

\multicolumn{1}{>{\raggedright}m{\dimexpr 1.25in+0\tabcolsep}}{\textcolor[HTML]{000000}{\fontsize{8}{8}\selectfont{Defect\ tickets\ \%\ ×\ Month}}} & \multicolumn{1}{>{\raggedleft}m{\dimexpr 0.75in+0\tabcolsep}}{\textcolor[HTML]{000000}{\fontsize{8}{8}\selectfont{-0.0001}}} & \multicolumn{1}{>{\raggedleft}m{\dimexpr 0.75in+0\tabcolsep}}{\textcolor[HTML]{000000}{\fontsize{8}{8}\selectfont{-0.0003}}} & \multicolumn{1}{>{\raggedleft}m{\dimexpr 0.75in+0\tabcolsep}}{\textcolor[HTML]{000000}{\fontsize{8}{8}\selectfont{0.0001}}} & \multicolumn{1}{>{\raggedleft}m{\dimexpr 0.75in+0\tabcolsep}}{\textcolor[HTML]{000000}{\fontsize{8}{8}\selectfont{\ 78\%}}} \\

\multicolumn{1}{>{\raggedright}m{\dimexpr 1.25in+0\tabcolsep}}{\textcolor[HTML]{000000}{\fontsize{8}{8}\selectfont{Degree\ centrality\ ×\ Month}}} & \multicolumn{1}{>{\raggedleft}m{\dimexpr 0.75in+0\tabcolsep}}{\textcolor[HTML]{000000}{\fontsize{8}{8}\selectfont{-0.0001}}} & \multicolumn{1}{>{\raggedleft}m{\dimexpr 0.75in+0\tabcolsep}}{\textcolor[HTML]{000000}{\fontsize{8}{8}\selectfont{-0.0005}}} & \multicolumn{1}{>{\raggedleft}m{\dimexpr 0.75in+0\tabcolsep}}{\textcolor[HTML]{000000}{\fontsize{8}{8}\selectfont{0.0002}}} & \multicolumn{1}{>{\raggedleft}m{\dimexpr 0.75in+0\tabcolsep}}{\textcolor[HTML]{000000}{\fontsize{8}{8}\selectfont{\ 71\%}}} \\

\multicolumn{1}{>{\raggedright}m{\dimexpr 1.25in+0\tabcolsep}}{\textcolor[HTML]{000000}{\fontsize{8}{8}\selectfont{Comments\ per\ PR\ ×\ Month}}} & \multicolumn{1}{>{\raggedleft}m{\dimexpr 0.75in+0\tabcolsep}}{\textcolor[HTML]{000000}{\fontsize{8}{8}\selectfont{0.0001}}} & \multicolumn{1}{>{\raggedleft}m{\dimexpr 0.75in+0\tabcolsep}}{\textcolor[HTML]{000000}{\fontsize{8}{8}\selectfont{-0.0004}}} & \multicolumn{1}{>{\raggedleft}m{\dimexpr 0.75in+0\tabcolsep}}{\textcolor[HTML]{000000}{\fontsize{8}{8}\selectfont{0.0005}}} & \multicolumn{1}{>{\raggedleft}m{\dimexpr 0.75in+0\tabcolsep}}{\textcolor[HTML]{000000}{\fontsize{8}{8}\selectfont{\ 59\%}}} \\

\ascline{1.5pt}{666666}{1-5}

\multicolumn{5}{>{\raggedright}m{\dimexpr 4.25in+8\tabcolsep}}{\textcolor[HTML]{000000}{\fontsize{8}{8}\selectfont{}}\textcolor[HTML]{000000}{\fontsize{8}{8}\selectfont{\textsuperscript{1}}}\textcolor[HTML]{000000}{\fontsize{8}{8}\selectfont{Median\ of\ the\ posterior\ distribution,\ used\ as\ point\ estimate}}\textcolor[HTML]{000000}{\fontsize{8}{8}\selectfont{.\ }}\textcolor[HTML]{000000}{\fontsize{8}{8}\selectfont{\textsuperscript{2}}}\textcolor[HTML]{000000}{\fontsize{8}{8}\selectfont{95\%\ Highest\ Density\ Interval,\ containing\ the\ most\ probable\ parameter\ values\ with\ 95\%\ posterior\ probability\ mass}}\textcolor[HTML]{000000}{\fontsize{8}{8}\selectfont{.\ }}\textcolor[HTML]{000000}{\fontsize{8}{8}\selectfont{\textsuperscript{3}}}\textcolor[HTML]{000000}{\fontsize{8}{8}\selectfont{Probability\ that\ the\ effect\ is\ in\ the\ reported\ direction,\ calculated\ as\ the\ proportion\ of\ posterior\ samples\ with\ the\ same\ sign\ as\ the\ point\ estimate}}\textcolor[HTML]{000000}{\fontsize{8}{8}\selectfont{.\ }}} \\

\end{longtable}

\arrayrulecolor[HTML]{000000}

\global\setlength{\arrayrulewidth}{\Oldarrayrulewidth}

\global\setlength{\tabcolsep}{\Oldtabcolsep}

\renewcommand*{\arraystretch}{1}

\textsubscript{Source:
\href{https://jflournoy.github.io/no-silver-bullets/index.qmd.html}{Article
Notebook}}

\begin{figure}[htbp]

\begin{minipage}{0.49\linewidth}

\centering{

\includegraphics{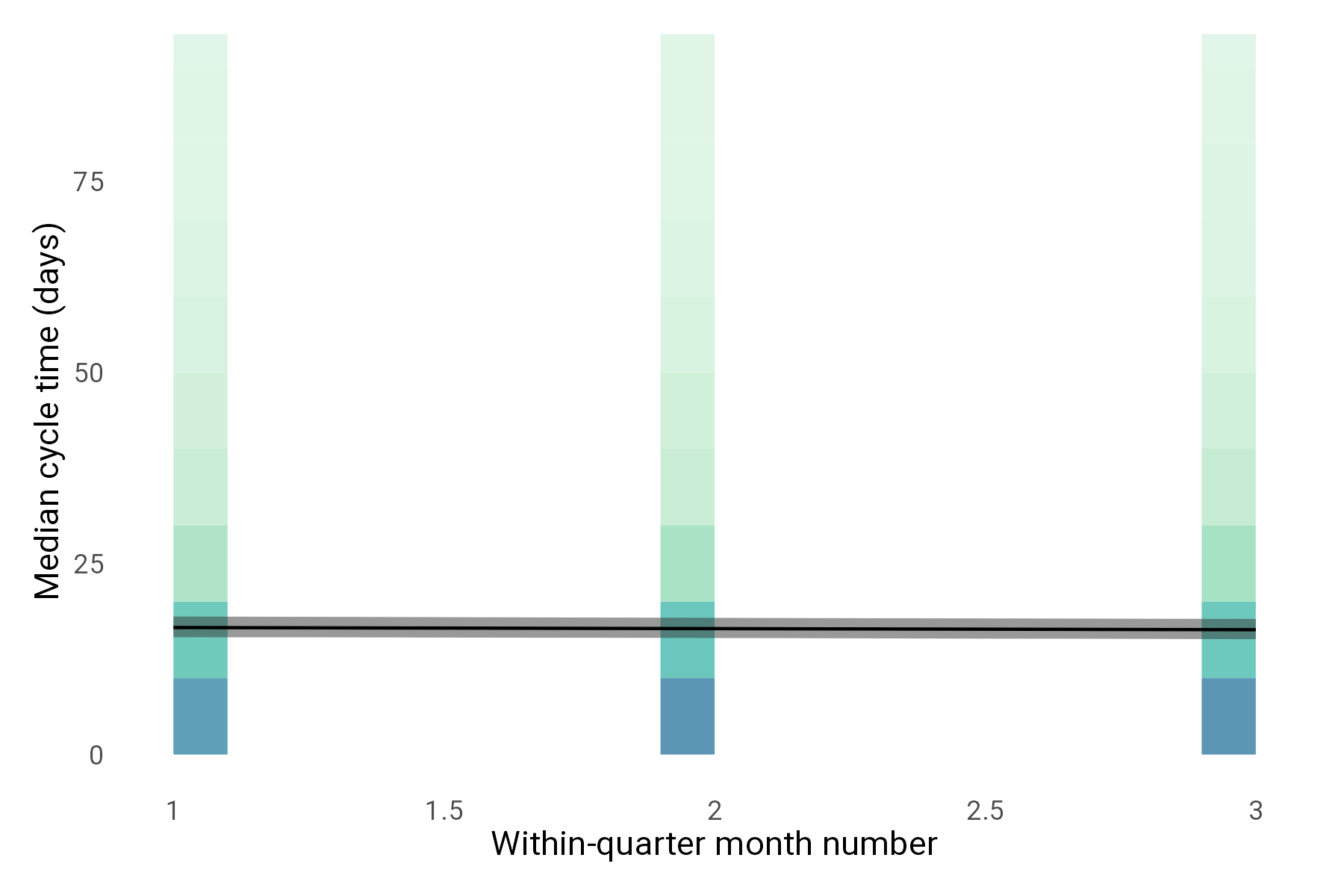}

}

\subcaption{\label{fig-quarter}Within-quarter month doesn't affect cycle
time. Background pixels represent density of data, with darker colors
indicating greater density. Lines are median posterior expectations,
with 95\% credible interval ribbons.}

\end{minipage}%
\begin{minipage}{0.02\linewidth}
~\end{minipage}%
\begin{minipage}{0.49\linewidth}

\centering{

\includegraphics{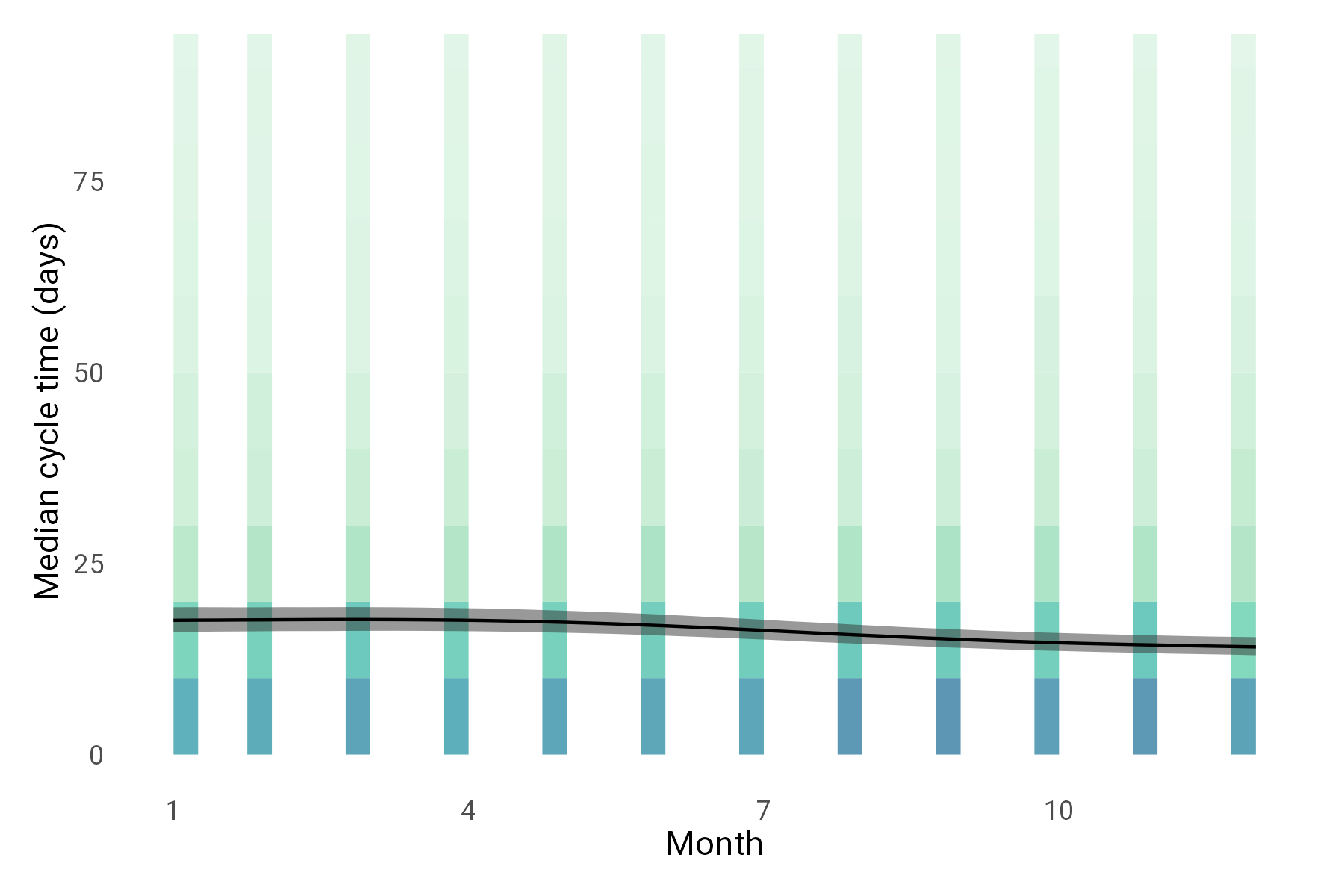}

}

\subcaption{\label{fig-month}Slight reduction of cycle time across the
year. Background hexagons represent density of data, with darker colors
indicating greater density. Lines are median posterior expectations,
with 95\% credible interval ribbons.}

\end{minipage}%

\end{figure}

We find that all measured factors, both individual-difference and
within-person deviations, have a non-zero association with cycle time,
with 100\% of the posterior distribution for these parameters sharing
the same sign (see Table~\ref{tbl-results-fe} for point estimates and
uncertainty intervals). Within-quarter month showed a very small
(relative to other effects shown below) negative association with cycle
time, indicating that time to ticket completion is shorter at the end of
quarters (Figure~\ref{fig-quarter}). Team size had almost no effect on
cycle time (Table~\ref{tbl-results-fe}). Cycle times tended to decrease
slightly over the year (Figure~\ref{fig-month}). The proportion of
unclosed tickets, an important control variable, on average had close to
no effect on our measure of average monthly cycle time but interacted
with month.

\begin{figure}[htbp]

\centering{

\includegraphics[width=4in,height=\textheight]{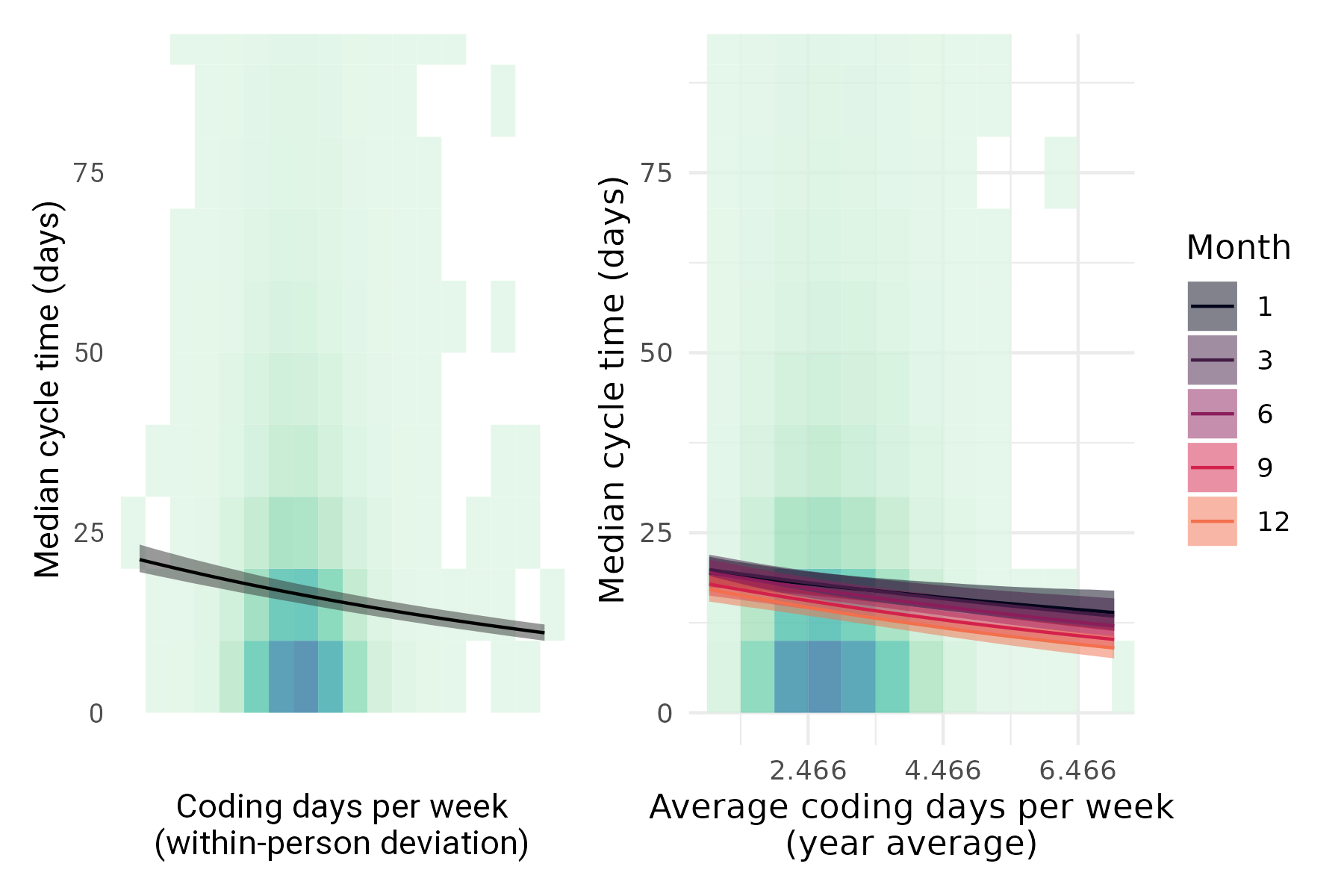}

}

\caption[More coding days is associated with shorter cycle
times]{\label{fig-codingdays}More coding days is associated with shorter
cycle times. Background hexagons represent density of data, with darker
colors indicating greater density. Lines are median posterior
expectations, with 95\% credible interval ribbons.}

\end{figure}

Specifically, when individuals increased average coding days per week
month-to-month they also tended to have lower cycle times, and
individuals with more average coding days per week across the year
tended to have lower cycle times (Figure~\ref{fig-codingdays}). The
association between coding days and cycle time also tended to increase
in strength across months, with 97\% of the posterior in this direction.

\begin{figure}[htbp]

\centering{

\includegraphics[width=4in,height=\textheight]{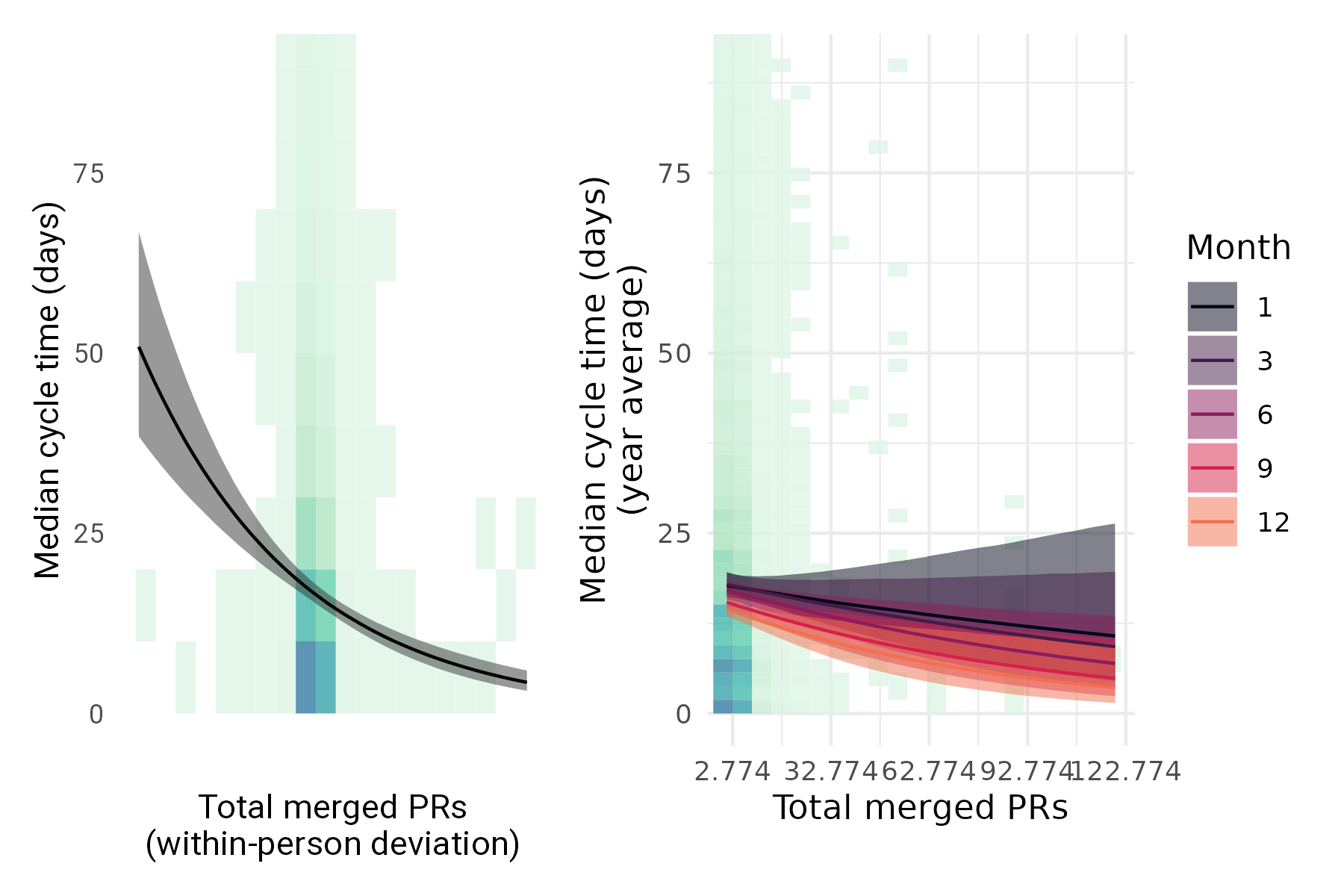}

}

\caption[More merged PRs is associated with shorter cycle
times]{\label{fig-mergedprs}More merged PRs is associated with shorter
cycle times. Background hexagons represent density of data, with darker
colors indicating greater density. Lines are median posterior
expectations, with 95\% credible interval ribbons.}

\end{figure}

More merged PRs was associated with lower cycle time for both individual
average differences and within-person differences. This effect also may
get stronger across the year with 90\% of the posterior in this
direction (Figure~\ref{fig-mergedprs}).

\begin{figure}[htbp]

\centering{

\includegraphics[width=4in,height=\textheight]{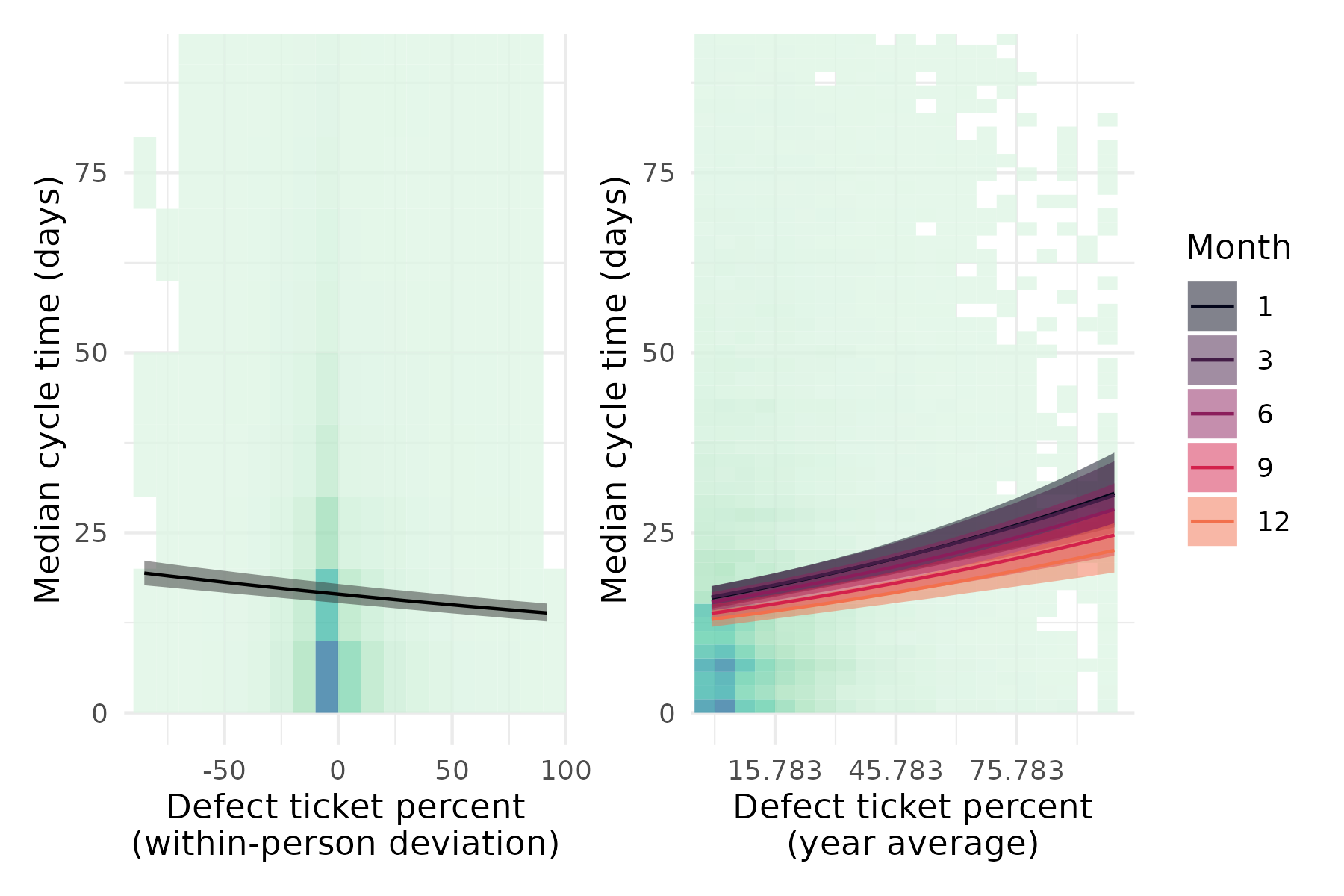}

}

\caption[Higher-than-average proportion of defect tickets in a month is
associated with shorter cycle times, while individuals with more defect
tickets on average show longer cycle
times]{\label{fig-defecttickets}Higher-than-average proportion of defect
tickets in a month is associated with shorter cycle times, while
individuals with more defect tickets on average show longer cycle times.
Background hexagons represent density of data, with darker colors
indicating greater density. Lines are median posterior expectations,
with 95\% credible interval ribbons.}

\end{figure}

The percent of defect tickets showed a negative association with cycle
time for within-person deviations and a positive association for
individual differences. In other words, individuals who tended to have
more defect tickets as a proportion of their work across the course of
the year also tended to have longer cycle times. However, for any given
person, an increase in the proportion of defect tickets in a month was
associated with lower cycle times (Figure~\ref{fig-defecttickets}). The
interaction with month number for this effect was centered close to
zero, with only 78\% of the posterior in the negative direction with a
fairly narrow distribution around zero (95\% HDI = {[}-0.0003,
0.0001{]}).

\begin{figure}[htbp]

\centering{

\includegraphics[width=4in,height=\textheight]{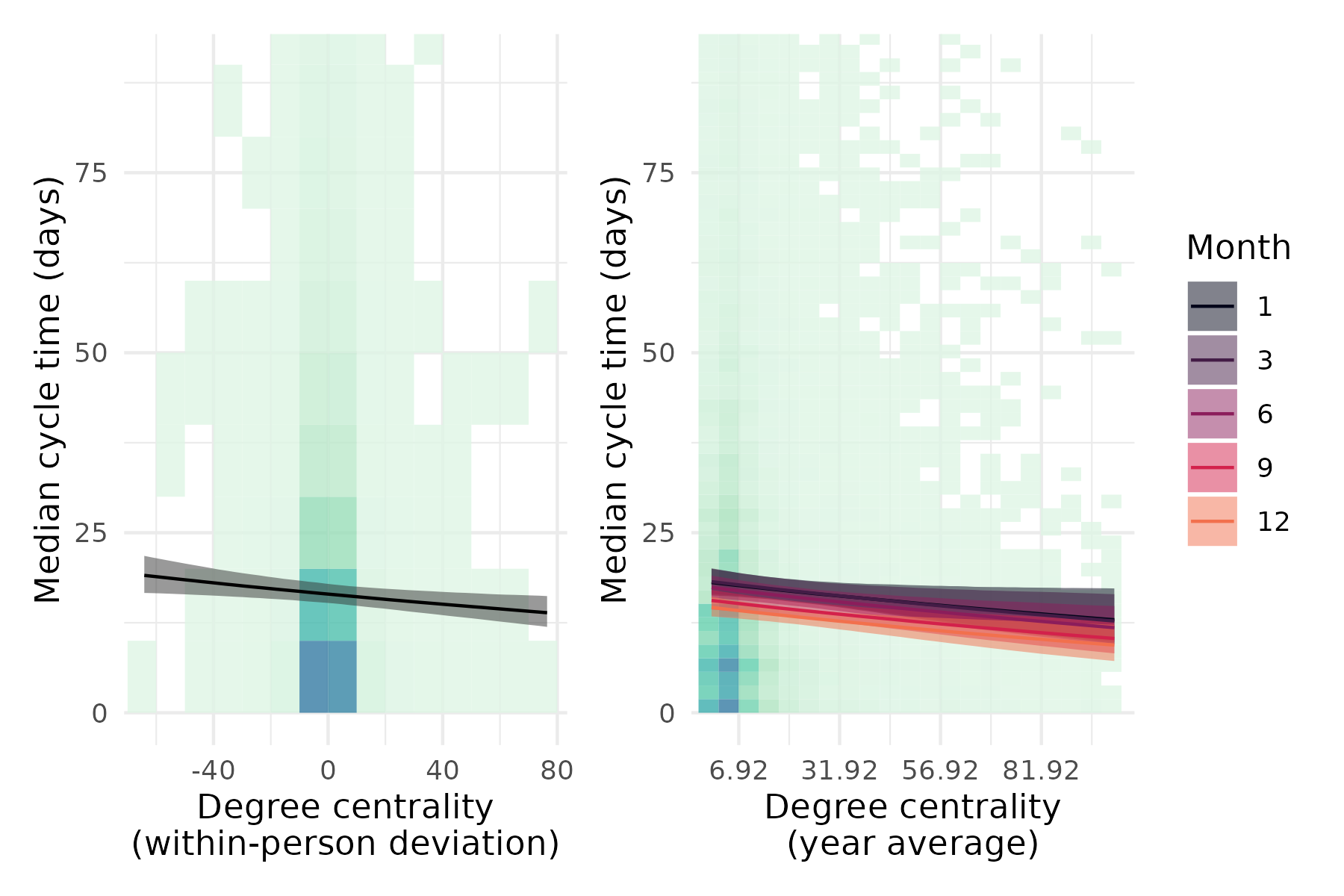}

}

\caption[Higher degree centrality is associated with shorter cycle
times]{\label{fig-degree}Higher degree centrality is associated with
shorter cycle times. Background hexagons represent density of data, with
darker colors indicating greater density. Lines are median posterior
expectations, with 95\% credible interval ribbons.}

\end{figure}

Degree centrality, as measured both by year-averaged individual
differences and within-person deviations, showed a negative association
with cycle time (Figure~\ref{fig-degree}). In other words, individuals
who on average contribute code to PRs that have a lot of other
contributors tend to have lower cycle times for tickets they own.
Similarly, when individuals' collaboration on PRs increases in a given
month, their cycle time tends to go down. This effect does not
unambiguously strengthen or weaken across the year with 71\% of the
posterior for the interaction effect having negative sign with a fairly
narrow distribution around zero (95\% HDI = {[}-0.0005, 0.0002{]}).

\begin{figure}[htbp]

\centering{

\includegraphics[width=4in,height=\textheight]{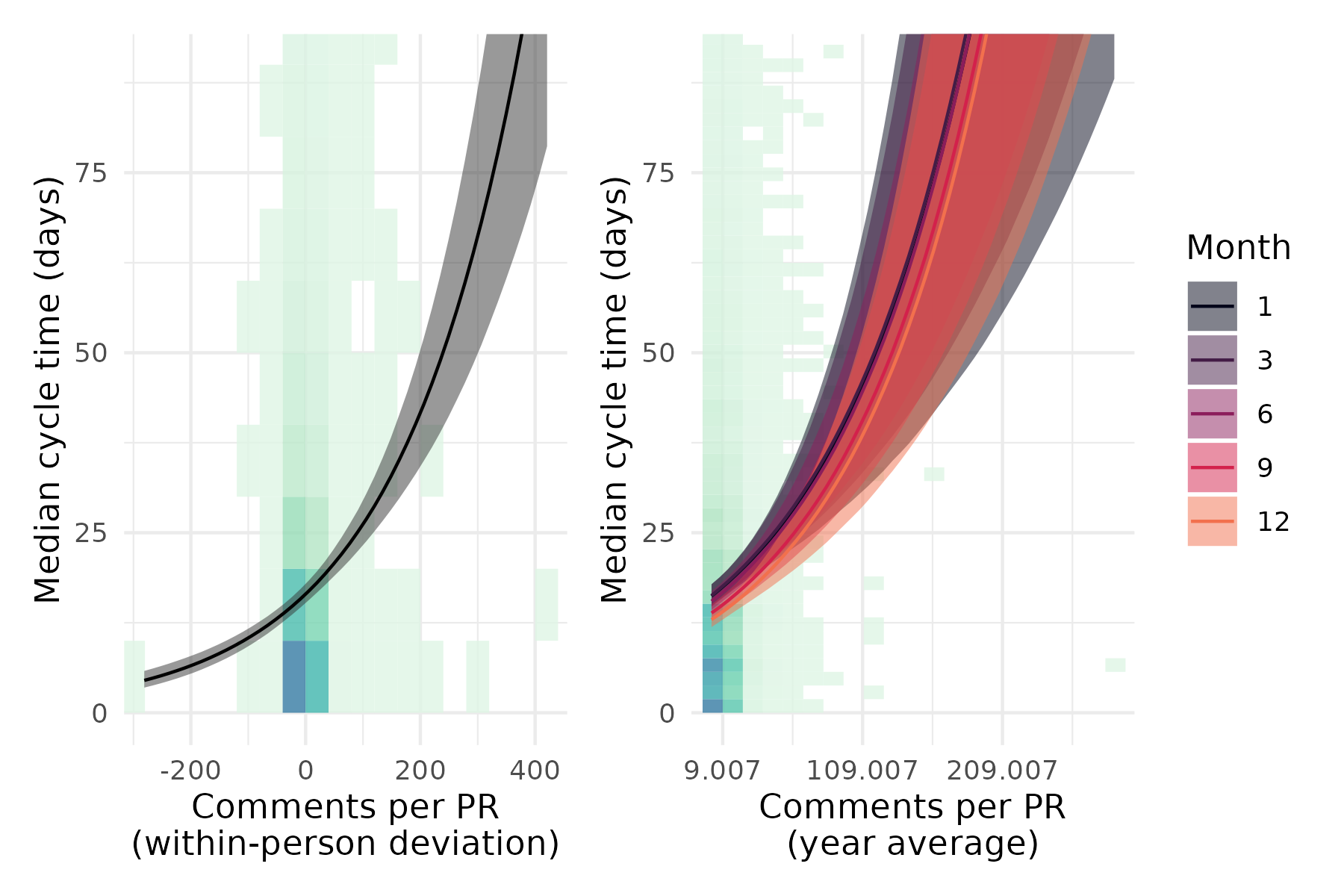}

}

\caption[More comments per PR is associated with longer cycle
times]{\label{fig-comments}More comments per PR is associated with
longer cycle times. Background hexagons represent density of data, with
darker colors indicating greater density. Lines are median posterior
expectations, with 95\% credible interval ribbons.}

\end{figure}

Finally, the number of comments per PR showed a positive association
with cycle time. Individuals who tended to garner more comments on their
PRs also tended to have higher cycle times, and within a given month, a
higher number of comments per PR relative to a person's average was also
associated with higher cycle times (Figure~\ref{fig-comments}). This
effect also does not unambiguously strengthen or weaken across the year
with 59\% of the posterior for the interaction effect having negative
sign with a fairly narrow distribution around zero (95\% HDI =
{[}-0.0004, 0.0005{]}).

\subsection{Effect sizes and
heterogeneity}\label{effect-sizes-and-heterogeneity}

Given the inherent non-linearity of the Weibull distribution, the
effects of the predictors on cycle time are not constant across the
range of the data. For example, the expected difference in cycle-time
for a unit difference for within-person coding days per week will be
different at different times of the year simply as a by-product that we
are modeling the log of the scale parameter. Notice that this is true
even in the absence of interactive effects, which further complicate the
interpretation of the effect sizes for the year-average variables.
Indeed, the random effects which allow intercept and month-effect
variance both organizations and individuals also adds to the complexity
of interpreting the effect sizes.

\begin{figure}[htbp]

\begin{minipage}{0.33\linewidth}

\centering{

\includegraphics{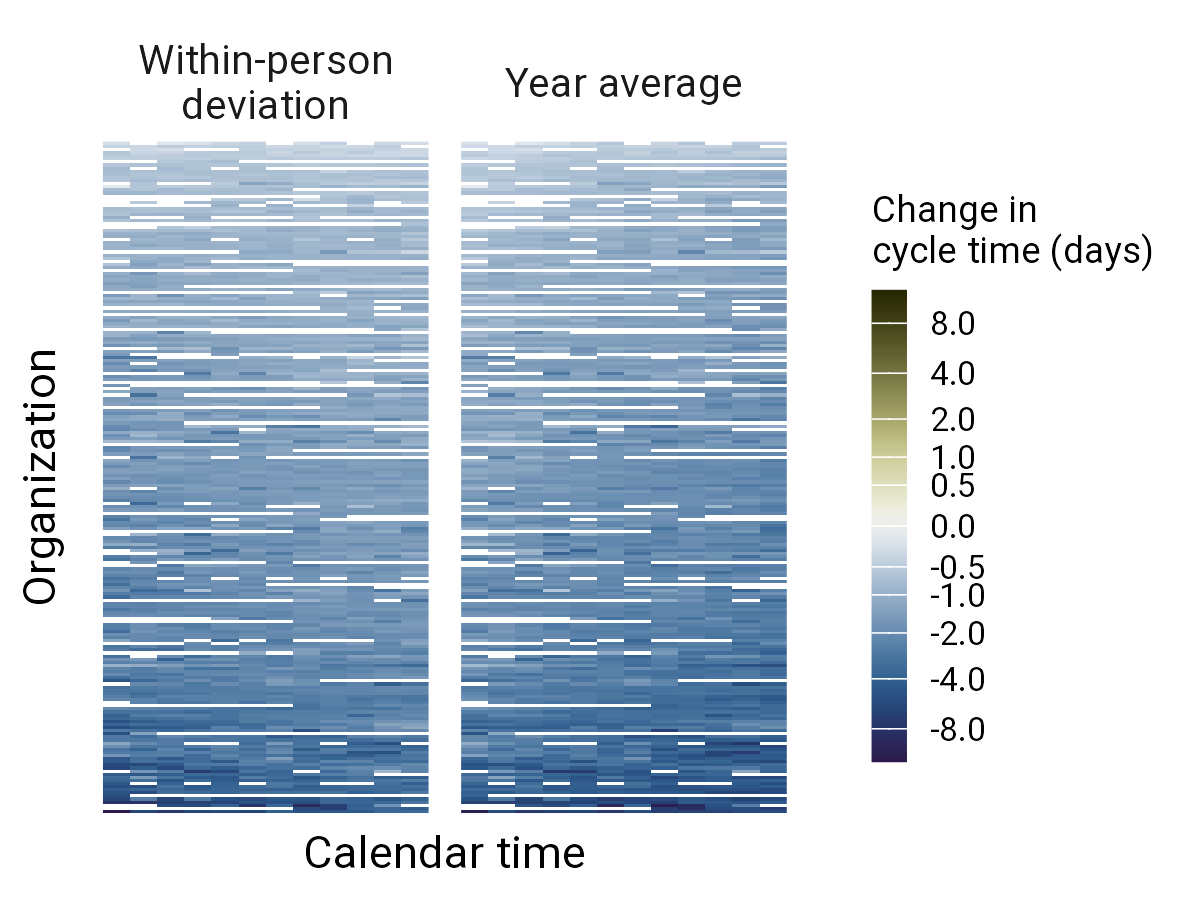}

}

\subcaption{\label{fig-heatmaps-coding}Avg. coding days per Week}

\end{minipage}%
\begin{minipage}{0.33\linewidth}

\centering{

\includegraphics{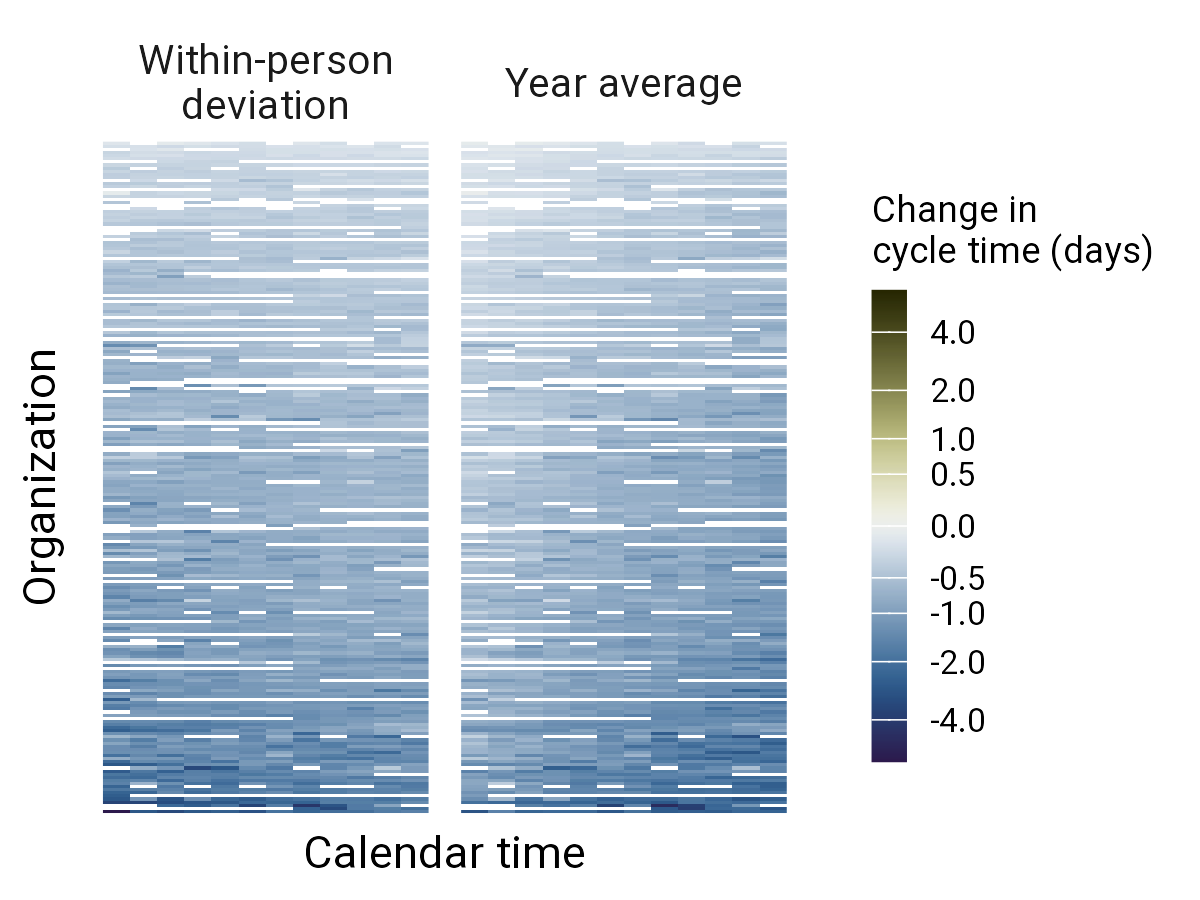}

}

\subcaption{\label{fig-heatmaps-prs}Total merged PRs}

\end{minipage}%
\begin{minipage}{0.33\linewidth}

\centering{

\includegraphics{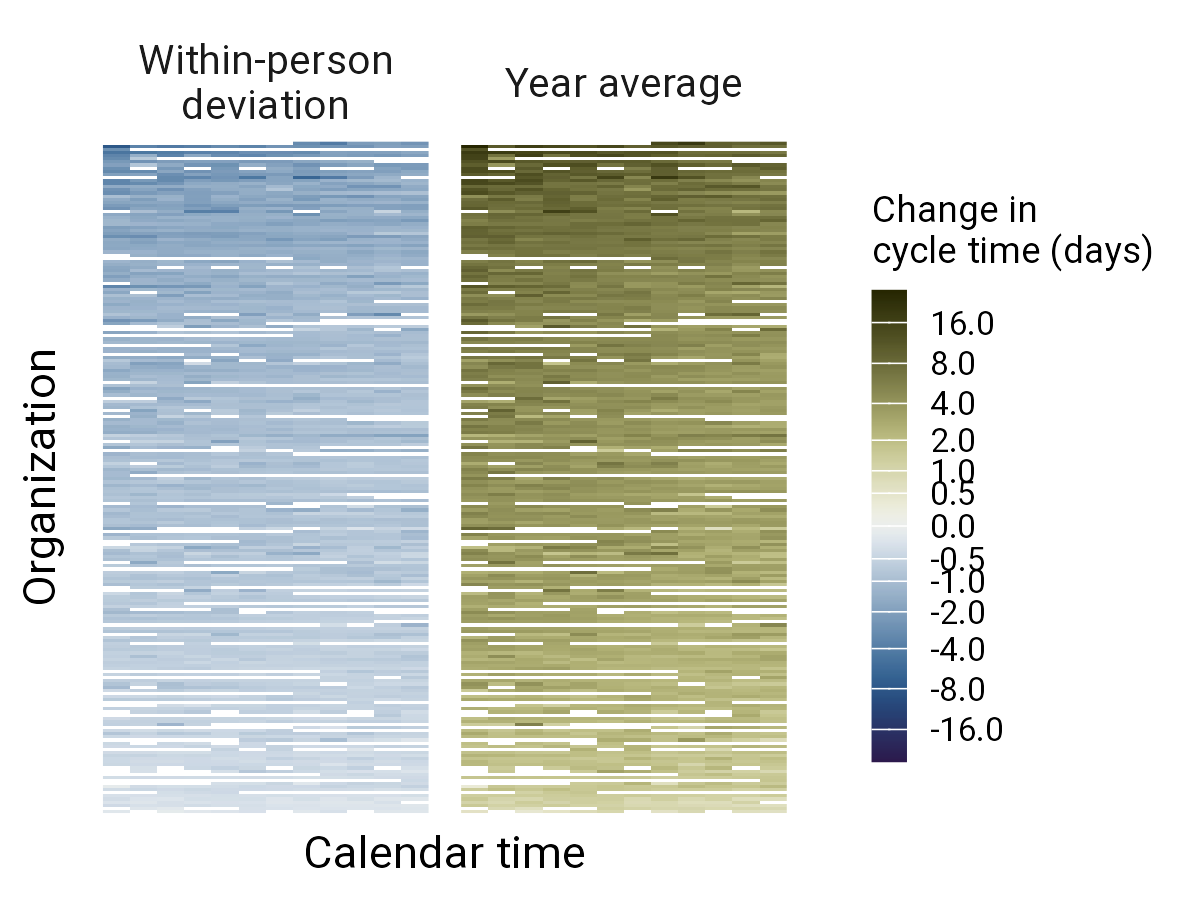}

}

\subcaption{\label{fig-heatmaps-defects}Defect tickets percentage}

\end{minipage}%
\newline
\begin{minipage}{0.17\linewidth}
~\end{minipage}%
\begin{minipage}{0.33\linewidth}

\centering{

\includegraphics{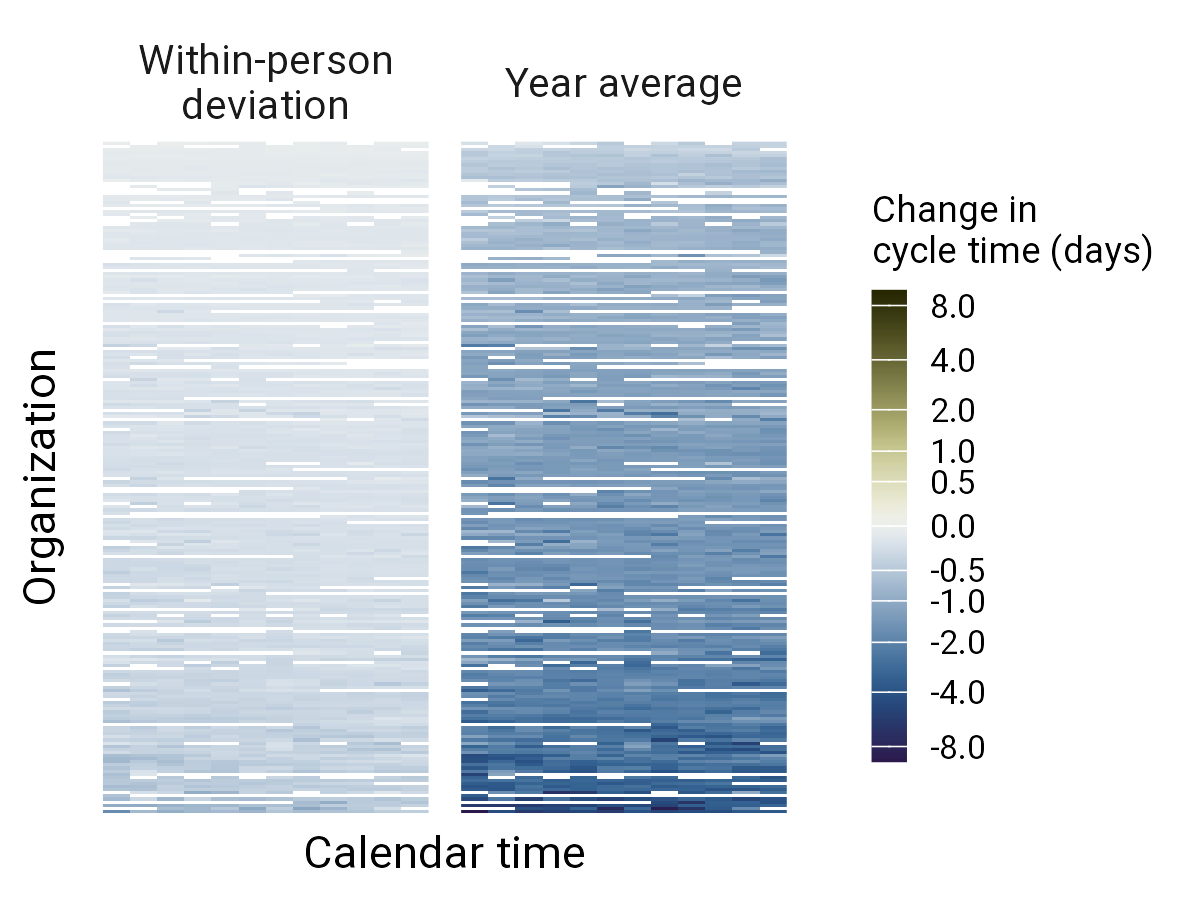}

}

\subcaption{\label{fig-heatmaps-centrality}Degree centrality}

\end{minipage}%
\begin{minipage}{0.33\linewidth}

\centering{

\includegraphics{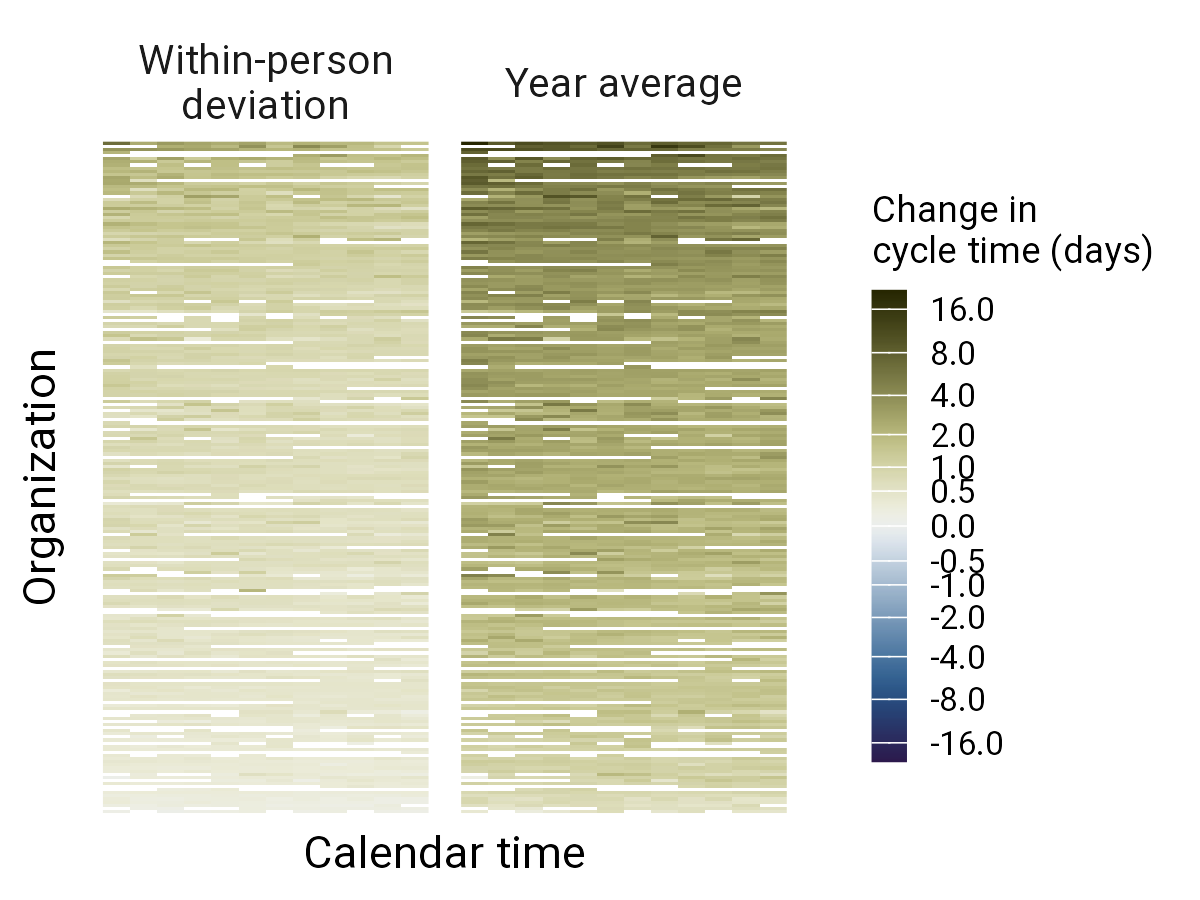}

}

\subcaption{\label{fig-heatmaps-comments}Comments per PR}

\end{minipage}%
\begin{minipage}{0.17\linewidth}
~\end{minipage}%

\caption{\label{fig-heatmaps}Effect sizes for each variable are
heterogeneous across time, organizations, and values of other
predictors. These plots show the expected change in cycle time from 50th
to 90th percentile of each variable, by organization and month. The
color represents the expected change in cycle time, with warm indicating
an increase in cycle time and cool indicating a decrease. The scale is
the same across all plots.}

\end{figure}

To give the reader a sense for how these associations play out across
organizations, we plot a range of expected changes in cycle time given a
counterfactual change from the 50th percentile to the 90th percentile on
the variable of interest, all else held equal
(Figure~\ref{fig-heatmaps}). We do this for each organization, for each
month, and then plot these as a heatmap where the color represents the
expected change in cycle time. This allows us to see how the effect of a
variable on cycle time changes across organizations and across time.

\subsection{Variability in effects}\label{variability-in-effects}

\begin{figure}

\centering{

\includegraphics[width=4in,height=\textheight]{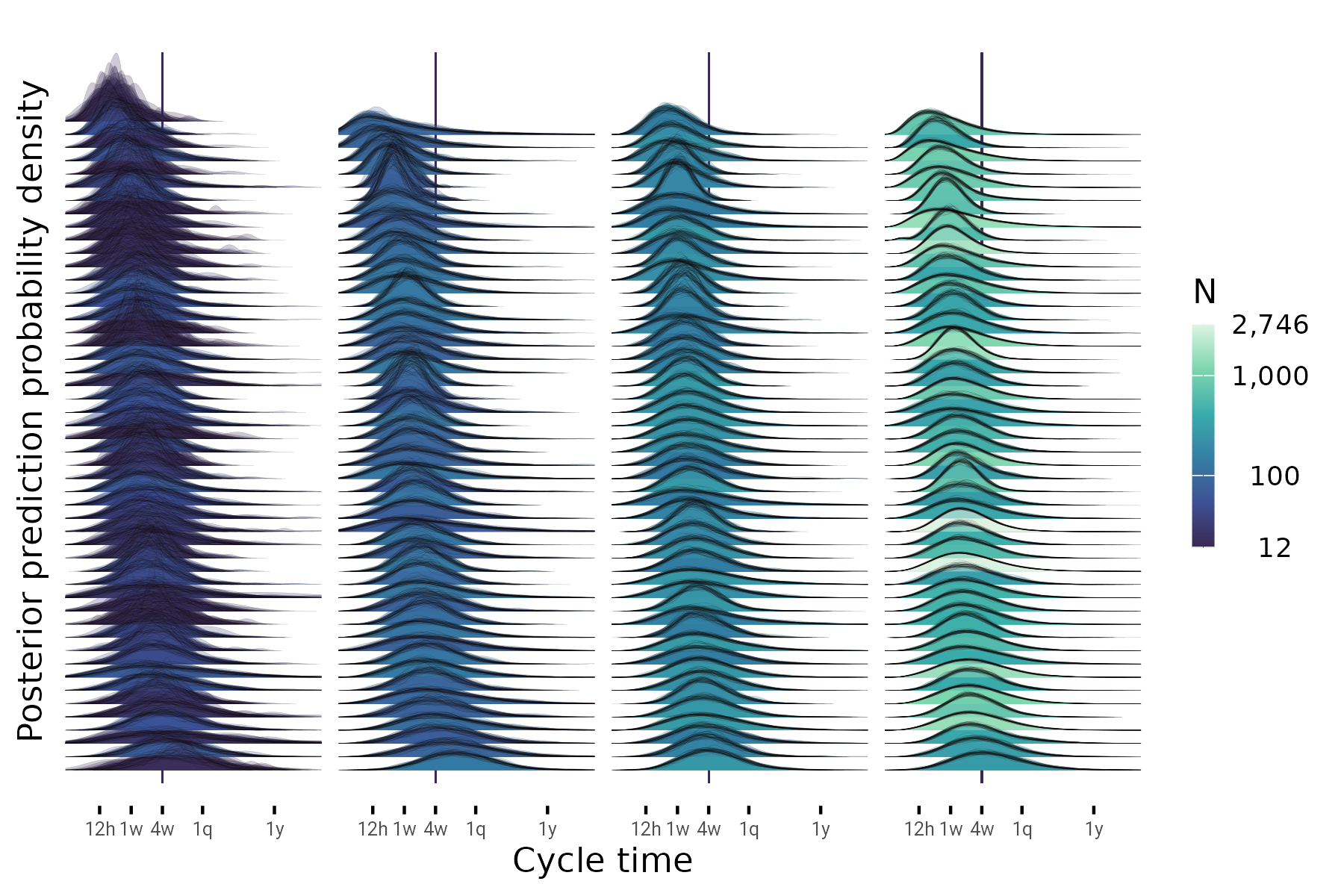}

}

\caption[Distributions of cycle time vary widely across organizations
both in their central tendency and
spread]{\label{fig-pp-check-org}Distributions of cycle time vary widely
across organizations both in their central tendency and spread.
Posterior prediction densities for model-expected distributions of cycle
time are shown across all organizations with at least 10 observations.
Each density represents one organization, showing model-predicted cycle
times from 50 posterior draws. The scale has been transformed slightly
to better show the spread of data. Columns have been ordered by sample
size, and rows are ordered by median predicted cycle time. Density fill
colors reflect sample size. Note that distributions from larger orgs
have less variability in their posterior predictions. Line at 4 weeks is
set arbitrarily to aid in comparisons.}

\end{figure}

\begin{figure}

\centering{

\includegraphics[width=4in,height=\textheight]{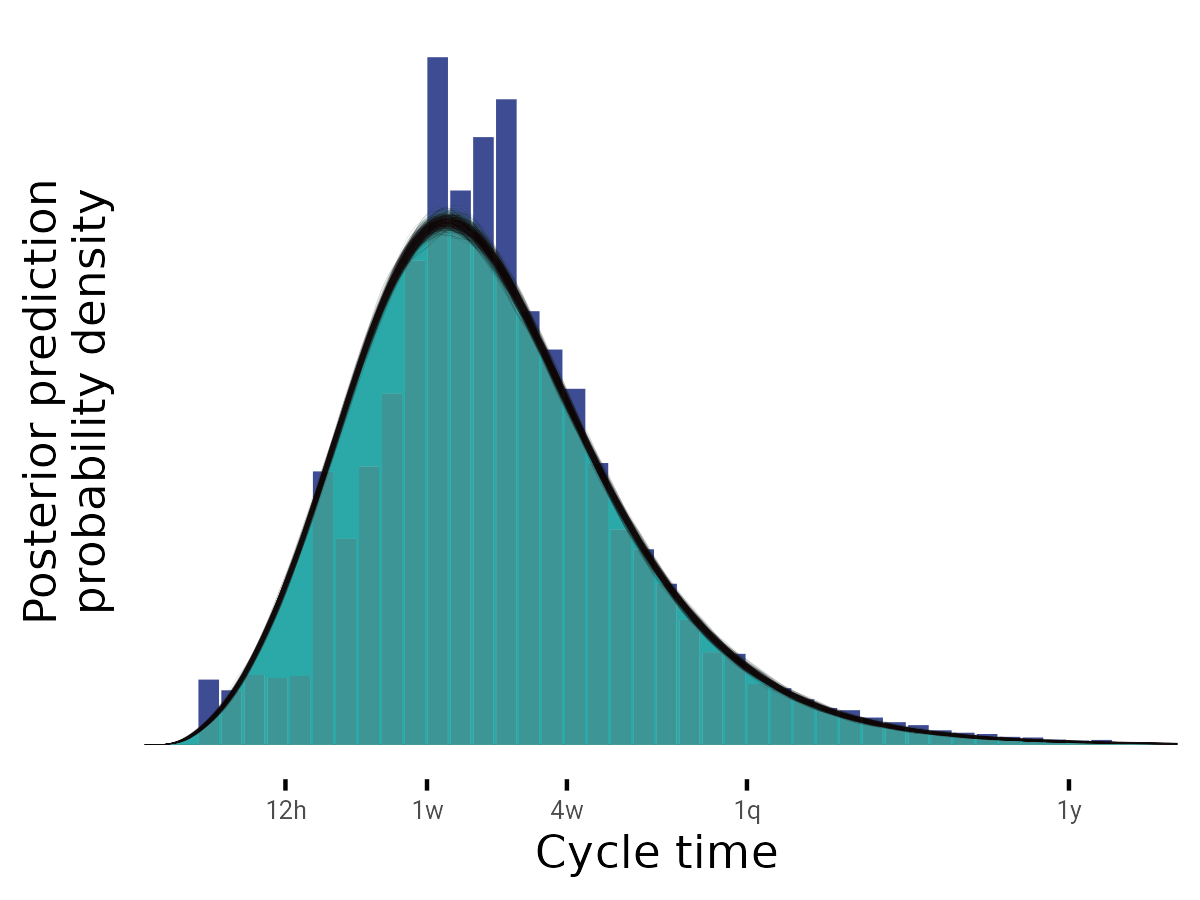}

}

\caption[Model-predicted posterior distributions of cycle time captures
the data distribution well]{\label{fig-pp-check-sum}Model-predicted
posterior distributions of cycle time captures the data distribution
well. The posterior prediction density for cycle time is shown across
all data points. The scale has been transformed slightly to better show
the spread of data.}

\end{figure}

There is heterogeneity in the distribution of cycle time across
organizations in both the scale and shape parameters, which suggests
that comparisons of cycle time trends between teams across different
organizations may be difficult and should include careful approaches
that take into account this variation (Table~\ref{tbl-results-re}). This
can be seen visually as differences in the posterior predictive
distribution of cycle times across organizations especially when viewed
on a log scale to emphasize differences at the low end of cycle times
where the distribution tends to be most dense
(Figure~\ref{fig-pp-check-org}). Different organizations likely have
different guidelines and cultures around using tickets and this may show
up as this kind of heterogeneity in cycle time. Incidentally we can also
see, looking across the full data-set, that the posterior distribution
of our model captures well our data distribution
(Figure~\ref{fig-pp-check-sum}).

\providecommand{\ascline}[3]{\noalign{\global\arrayrulewidth #1}\arrayrulecolor[HTML]{#2}\cline{#3}}

\begin{longtable}[c]{|p{0.40in}|p{0.40in}|p{0.50in}|p{0.50in}|p{0.50in}|p{0.75in}|p{0.50in}}

\caption{\label{tbl-results-re}Variance and covariance of organization
and individual-level effects}

\tabularnewline

\ascline{1.5pt}{666666}{1-7}

\multicolumn{1}{>{\raggedright}m{\dimexpr 0.4in+0\tabcolsep}}{\textcolor[HTML]{000000}{\fontsize{8}{8}\selectfont{\textbf{Par}}}\textcolor[HTML]{000000}{\fontsize{8}{8}\selectfont{\textbf{\textsuperscript{1}}}}} & \multicolumn{1}{>{\raggedright}m{\dimexpr 0.4in+0\tabcolsep}}{\textcolor[HTML]{000000}{\fontsize{8}{8}\selectfont{\textbf{Grp}}}\textcolor[HTML]{000000}{\fontsize{8}{8}\selectfont{\textbf{\textsuperscript{2}}}}} & \multicolumn{1}{>{\raggedright}m{\dimexpr 0.5in+0\tabcolsep}}{\textcolor[HTML]{000000}{\fontsize{8}{8}\selectfont{\textbf{Stat}}}\textcolor[HTML]{000000}{\fontsize{8}{8}\selectfont{\textbf{\textsuperscript{3}}}}} & \multicolumn{1}{>{\raggedright}m{\dimexpr 0.5in+0\tabcolsep}}{\textcolor[HTML]{000000}{\fontsize{8}{8}\selectfont{\textbf{Pred}}}\textcolor[HTML]{000000}{\fontsize{8}{8}\selectfont{\textbf{\textsuperscript{4}}}}} & \multicolumn{1}{>{\raggedleft}m{\dimexpr 0.5in+0\tabcolsep}}{\textcolor[HTML]{000000}{\fontsize{8}{8}\selectfont{\textbf{Post\ Med}}}\textcolor[HTML]{000000}{\fontsize{8}{8}\selectfont{\textbf{\textsuperscript{5}}}}} & \multicolumn{1}{>{\raggedleft}m{\dimexpr 0.75in+0\tabcolsep}}{\textcolor[HTML]{000000}{\fontsize{8}{8}\selectfont{\textbf{95\%\ HDI}}}\textcolor[HTML]{000000}{\fontsize{8}{8}\selectfont{\textbf{\textsuperscript{6}}}}} & \multicolumn{1}{>{\raggedleft}m{\dimexpr 0.5in+0\tabcolsep}}{\textcolor[HTML]{000000}{\fontsize{8}{8}\selectfont{\textbf{Sign\ Prob}}}\textcolor[HTML]{000000}{\fontsize{8}{8}\selectfont{\textbf{\textsuperscript{7}}}}} \\

\ascline{1.5pt}{666666}{1-7}\endfirsthead 

\ascline{1.5pt}{666666}{1-7}

\multicolumn{1}{>{\raggedright}m{\dimexpr 0.4in+0\tabcolsep}}{\textcolor[HTML]{000000}{\fontsize{8}{8}\selectfont{\textbf{Par}}}\textcolor[HTML]{000000}{\fontsize{8}{8}\selectfont{\textbf{\textsuperscript{1}}}}} & \multicolumn{1}{>{\raggedright}m{\dimexpr 0.4in+0\tabcolsep}}{\textcolor[HTML]{000000}{\fontsize{8}{8}\selectfont{\textbf{Grp}}}\textcolor[HTML]{000000}{\fontsize{8}{8}\selectfont{\textbf{\textsuperscript{2}}}}} & \multicolumn{1}{>{\raggedright}m{\dimexpr 0.5in+0\tabcolsep}}{\textcolor[HTML]{000000}{\fontsize{8}{8}\selectfont{\textbf{Stat}}}\textcolor[HTML]{000000}{\fontsize{8}{8}\selectfont{\textbf{\textsuperscript{3}}}}} & \multicolumn{1}{>{\raggedright}m{\dimexpr 0.5in+0\tabcolsep}}{\textcolor[HTML]{000000}{\fontsize{8}{8}\selectfont{\textbf{Pred}}}\textcolor[HTML]{000000}{\fontsize{8}{8}\selectfont{\textbf{\textsuperscript{4}}}}} & \multicolumn{1}{>{\raggedleft}m{\dimexpr 0.5in+0\tabcolsep}}{\textcolor[HTML]{000000}{\fontsize{8}{8}\selectfont{\textbf{Post\ Med}}}\textcolor[HTML]{000000}{\fontsize{8}{8}\selectfont{\textbf{\textsuperscript{5}}}}} & \multicolumn{1}{>{\raggedleft}m{\dimexpr 0.75in+0\tabcolsep}}{\textcolor[HTML]{000000}{\fontsize{8}{8}\selectfont{\textbf{95\%\ HDI}}}\textcolor[HTML]{000000}{\fontsize{8}{8}\selectfont{\textbf{\textsuperscript{6}}}}} & \multicolumn{1}{>{\raggedleft}m{\dimexpr 0.5in+0\tabcolsep}}{\textcolor[HTML]{000000}{\fontsize{8}{8}\selectfont{\textbf{Sign\ Prob}}}\textcolor[HTML]{000000}{\fontsize{8}{8}\selectfont{\textbf{\textsuperscript{7}}}}} \\

\ascline{1.5pt}{666666}{1-7}\endhead

\multicolumn{1}{>{\raggedright}p{\dimexpr 0.4in+0\tabcolsep}}{} & \multicolumn{1}{>{\raggedright}p{\dimexpr 0.4in+0\tabcolsep}}{} & \multicolumn{1}{>{\raggedright}p{\dimexpr 0.5in+0\tabcolsep}}{} & \multicolumn{1}{>{\raggedright}m{\dimexpr 0.5in+0\tabcolsep}}{\textcolor[HTML]{000000}{\fontsize{8}{8}\selectfont{Intercept}}} & \multicolumn{1}{>{\raggedleft}m{\dimexpr 0.5in+0\tabcolsep}}{\textcolor[HTML]{000000}{\fontsize{8}{8}\selectfont{0.474}}} & \multicolumn{1}{>{\raggedleft}m{\dimexpr 0.75in+0\tabcolsep}}{\textcolor[HTML]{000000}{\fontsize{8}{8}\selectfont{[0.424,\ 0.529]}}} & \multicolumn{1}{>{\raggedleft}m{\dimexpr 0.5in+0\tabcolsep}}{\textcolor[HTML]{000000}{\fontsize{8}{8}\selectfont{100\%}}} \\

\multicolumn{1}{>{\raggedright}p{\dimexpr 0.4in+0\tabcolsep}}{} & \multicolumn{1}{>{\raggedright}p{\dimexpr 0.4in+0\tabcolsep}}{} & \multicolumn{1}{>{\raggedright}p{\dimexpr 0.5in+0\tabcolsep}}{\multirow[t]{-2}{*}{\parbox{0.5in}{\raggedright \textcolor[HTML]{000000}{\fontsize{8}{8}\selectfont{SD}}}}} & \multicolumn{1}{>{\raggedright}m{\dimexpr 0.5in+0\tabcolsep}}{\textcolor[HTML]{000000}{\fontsize{8}{8}\selectfont{Month}}} & \multicolumn{1}{>{\raggedleft}m{\dimexpr 0.5in+0\tabcolsep}}{\textcolor[HTML]{000000}{\fontsize{8}{8}\selectfont{0.034}}} & \multicolumn{1}{>{\raggedleft}m{\dimexpr 0.75in+0\tabcolsep}}{\textcolor[HTML]{000000}{\fontsize{8}{8}\selectfont{[0.027,\ 0.042]}}} & \multicolumn{1}{>{\raggedleft}m{\dimexpr 0.5in+0\tabcolsep}}{\textcolor[HTML]{000000}{\fontsize{8}{8}\selectfont{100\%}}} \\

\multicolumn{1}{>{\raggedright}p{\dimexpr 0.4in+0\tabcolsep}}{} & \multicolumn{1}{>{\raggedright}p{\dimexpr 0.4in+0\tabcolsep}}{\multirow[t]{-3}{*}{\parbox{0.4in}{\raggedright \textcolor[HTML]{000000}{\fontsize{8}{8}\selectfont{org}}}}} & \multicolumn{1}{>{\raggedright}p{\dimexpr 0.5in+0\tabcolsep}}{\textcolor[HTML]{000000}{\fontsize{8}{8}\selectfont{Correlation}}} & \multicolumn{1}{>{\raggedright}m{\dimexpr 0.5in+0\tabcolsep}}{\textcolor[HTML]{000000}{\fontsize{8}{8}\selectfont{Intercept-Month}}} & \multicolumn{1}{>{\raggedleft}m{\dimexpr 0.5in+0\tabcolsep}}{\textcolor[HTML]{000000}{\fontsize{8}{8}\selectfont{-0.123}}} & \multicolumn{1}{>{\raggedleft}m{\dimexpr 0.75in+0\tabcolsep}}{\textcolor[HTML]{000000}{\fontsize{8}{8}\selectfont{[-0.335,\ 0.110]}}} & \multicolumn{1}{>{\raggedleft}m{\dimexpr 0.5in+0\tabcolsep}}{\textcolor[HTML]{000000}{\fontsize{8}{8}\selectfont{86.15\%}}} \\

\multicolumn{1}{>{\raggedright}p{\dimexpr 0.4in+0\tabcolsep}}{} & \multicolumn{1}{>{\raggedright}p{\dimexpr 0.4in+0\tabcolsep}}{} & \multicolumn{1}{>{\raggedright}p{\dimexpr 0.5in+0\tabcolsep}}{} & \multicolumn{1}{>{\raggedright}m{\dimexpr 0.5in+0\tabcolsep}}{\textcolor[HTML]{000000}{\fontsize{8}{8}\selectfont{Intercept}}} & \multicolumn{1}{>{\raggedleft}m{\dimexpr 0.5in+0\tabcolsep}}{\textcolor[HTML]{000000}{\fontsize{8}{8}\selectfont{0.645}}} & \multicolumn{1}{>{\raggedleft}m{\dimexpr 0.75in+0\tabcolsep}}{\textcolor[HTML]{000000}{\fontsize{8}{8}\selectfont{[0.631,\ 0.661]}}} & \multicolumn{1}{>{\raggedleft}m{\dimexpr 0.5in+0\tabcolsep}}{\textcolor[HTML]{000000}{\fontsize{8}{8}\selectfont{100\%}}} \\

\multicolumn{1}{>{\raggedright}p{\dimexpr 0.4in+0\tabcolsep}}{} & \multicolumn{1}{>{\raggedright}p{\dimexpr 0.4in+0\tabcolsep}}{} & \multicolumn{1}{>{\raggedright}p{\dimexpr 0.5in+0\tabcolsep}}{\multirow[t]{-2}{*}{\parbox{0.5in}{\raggedright \textcolor[HTML]{000000}{\fontsize{8}{8}\selectfont{SD}}}}} & \multicolumn{1}{>{\raggedright}m{\dimexpr 0.5in+0\tabcolsep}}{\textcolor[HTML]{000000}{\fontsize{8}{8}\selectfont{Month}}} & \multicolumn{1}{>{\raggedleft}m{\dimexpr 0.5in+0\tabcolsep}}{\textcolor[HTML]{000000}{\fontsize{8}{8}\selectfont{0.067}}} & \multicolumn{1}{>{\raggedleft}m{\dimexpr 0.75in+0\tabcolsep}}{\textcolor[HTML]{000000}{\fontsize{8}{8}\selectfont{[0.062,\ 0.071]}}} & \multicolumn{1}{>{\raggedleft}m{\dimexpr 0.5in+0\tabcolsep}}{\textcolor[HTML]{000000}{\fontsize{8}{8}\selectfont{100\%}}} \\

\multicolumn{1}{>{\raggedright}p{\dimexpr 0.4in+0\tabcolsep}}{\multirow[t]{-6}{*}{\parbox{0.4in}{\raggedright \textcolor[HTML]{000000}{\fontsize{8}{8}\selectfont{log(λ)}}}}} & \multicolumn{1}{>{\raggedright}p{\dimexpr 0.4in+0\tabcolsep}}{\multirow[t]{-3}{*}{\parbox{0.4in}{\raggedright \textcolor[HTML]{000000}{\fontsize{8}{8}\selectfont{org:user}}}}} & \multicolumn{1}{>{\raggedright}p{\dimexpr 0.5in+0\tabcolsep}}{\textcolor[HTML]{000000}{\fontsize{8}{8}\selectfont{Correlation}}} & \multicolumn{1}{>{\raggedright}m{\dimexpr 0.5in+0\tabcolsep}}{\textcolor[HTML]{000000}{\fontsize{8}{8}\selectfont{Intercept-Month}}} & \multicolumn{1}{>{\raggedleft}m{\dimexpr 0.5in+0\tabcolsep}}{\textcolor[HTML]{000000}{\fontsize{8}{8}\selectfont{-0.131}}} & \multicolumn{1}{>{\raggedleft}m{\dimexpr 0.75in+0\tabcolsep}}{\textcolor[HTML]{000000}{\fontsize{8}{8}\selectfont{[-0.192,\ -0.077]}}} & \multicolumn{1}{>{\raggedleft}m{\dimexpr 0.5in+0\tabcolsep}}{\textcolor[HTML]{000000}{\fontsize{8}{8}\selectfont{100\%}}} \\

\multicolumn{1}{>{\raggedright}p{\dimexpr 0.4in+0\tabcolsep}}{\textcolor[HTML]{000000}{\fontsize{8}{8}\selectfont{log(k)}}} & \multicolumn{1}{>{\raggedright}p{\dimexpr 0.4in+0\tabcolsep}}{\textcolor[HTML]{000000}{\fontsize{8}{8}\selectfont{org}}} & \multicolumn{1}{>{\raggedright}p{\dimexpr 0.5in+0\tabcolsep}}{\textcolor[HTML]{000000}{\fontsize{8}{8}\selectfont{SD}}} & \multicolumn{1}{>{\raggedright}m{\dimexpr 0.5in+0\tabcolsep}}{\textcolor[HTML]{000000}{\fontsize{8}{8}\selectfont{Intercept}}} & \multicolumn{1}{>{\raggedleft}m{\dimexpr 0.5in+0\tabcolsep}}{\textcolor[HTML]{000000}{\fontsize{8}{8}\selectfont{0.277}}} & \multicolumn{1}{>{\raggedleft}m{\dimexpr 0.75in+0\tabcolsep}}{\textcolor[HTML]{000000}{\fontsize{8}{8}\selectfont{[0.247,\ 0.310]}}} & \multicolumn{1}{>{\raggedleft}m{\dimexpr 0.5in+0\tabcolsep}}{\textcolor[HTML]{000000}{\fontsize{8}{8}\selectfont{100\%}}} \\

\ascline{1.5pt}{666666}{1-7}

\multicolumn{7}{>{\raggedright}m{\dimexpr 3.55in+12\tabcolsep}}{\textcolor[HTML]{000000}{\fontsize{8}{8}\selectfont{}}\textcolor[HTML]{000000}{\fontsize{8}{8}\selectfont{\textsuperscript{1}}}\textcolor[HTML]{000000}{\fontsize{8}{8}\selectfont{Response\ distribution\ parameter\ (λ:\ rate\ parameter,\ k:\ shape\ parameter)}}\textcolor[HTML]{000000}{\fontsize{8}{8}\selectfont{.\ }}\textcolor[HTML]{000000}{\fontsize{8}{8}\selectfont{\textsuperscript{2}}}\textcolor[HTML]{000000}{\fontsize{8}{8}\selectfont{Random\ effects\ grouping\ structure}}\textcolor[HTML]{000000}{\fontsize{8}{8}\selectfont{.\ }}\textcolor[HTML]{000000}{\fontsize{8}{8}\selectfont{\textsuperscript{3}}}\textcolor[HTML]{000000}{\fontsize{8}{8}\selectfont{Statistic\ type\ (SD:\ standard\ deviation\ of\ random\ effect,\ Correlation:\ correlation\ between\ random\ effects)}}\textcolor[HTML]{000000}{\fontsize{8}{8}\selectfont{.\ }}\textcolor[HTML]{000000}{\fontsize{8}{8}\selectfont{\textsuperscript{4}}}\textcolor[HTML]{000000}{\fontsize{8}{8}\selectfont{Fixed\ or\ random\ effect\ term}}\textcolor[HTML]{000000}{\fontsize{8}{8}\selectfont{.\ }}\textcolor[HTML]{000000}{\fontsize{8}{8}\selectfont{\textsuperscript{5}}}\textcolor[HTML]{000000}{\fontsize{8}{8}\selectfont{Median\ of\ the\ posterior\ distribution,\ used\ as\ point\ estimate}}\textcolor[HTML]{000000}{\fontsize{8}{8}\selectfont{.\ }}\textcolor[HTML]{000000}{\fontsize{8}{8}\selectfont{\textsuperscript{6}}}\textcolor[HTML]{000000}{\fontsize{8}{8}\selectfont{95\%\ Highest\ Density\ Interval,\ containing\ the\ most\ probable\ parameter\ values\ with\ 95\%\ posterior\ probability\ mass}}\textcolor[HTML]{000000}{\fontsize{8}{8}\selectfont{.\ }}\textcolor[HTML]{000000}{\fontsize{8}{8}\selectfont{\textsuperscript{7}}}\textcolor[HTML]{000000}{\fontsize{8}{8}\selectfont{Probability\ that\ the\ effect\ is\ in\ the\ reported\ direction,\ calculated\ as\ the\ proportion\ of\ posterior\ samples\ with\ the\ same\ sign\ as\ the\ point\ estimate}}\textcolor[HTML]{000000}{\fontsize{8}{8}\selectfont{.\ }}} \\

\end{longtable}

\textsubscript{Source:
\href{https://jflournoy.github.io/no-silver-bullets/index.qmd.html}{Article
Notebook}}

Variation across individuals' scale parameter after accounting for
organization heterogeneity is also substantial, and greater in magnitude
than organizational heterogeneity (sd = 0.65 {[}0.63, 0.66{]} versus sd
= 0.47 {[}0.42, 0.53{]}; Table~\ref{tbl-results-re}). There is also
notable heterogeneity in the effect of month both at the organization
level and the level of individuals. There is a small negative
correlation between individuals' (and less credibly, organizations')
scale intercepts and the effect of month meaning that a person who has a
higher cycle time at month 7 also tends to decrease more steeply in
their cycle time across the year (Table~\ref{tbl-results-re}).

\begin{figure}[htbp]

\centering{

\includegraphics[width=4in,height=\textheight]{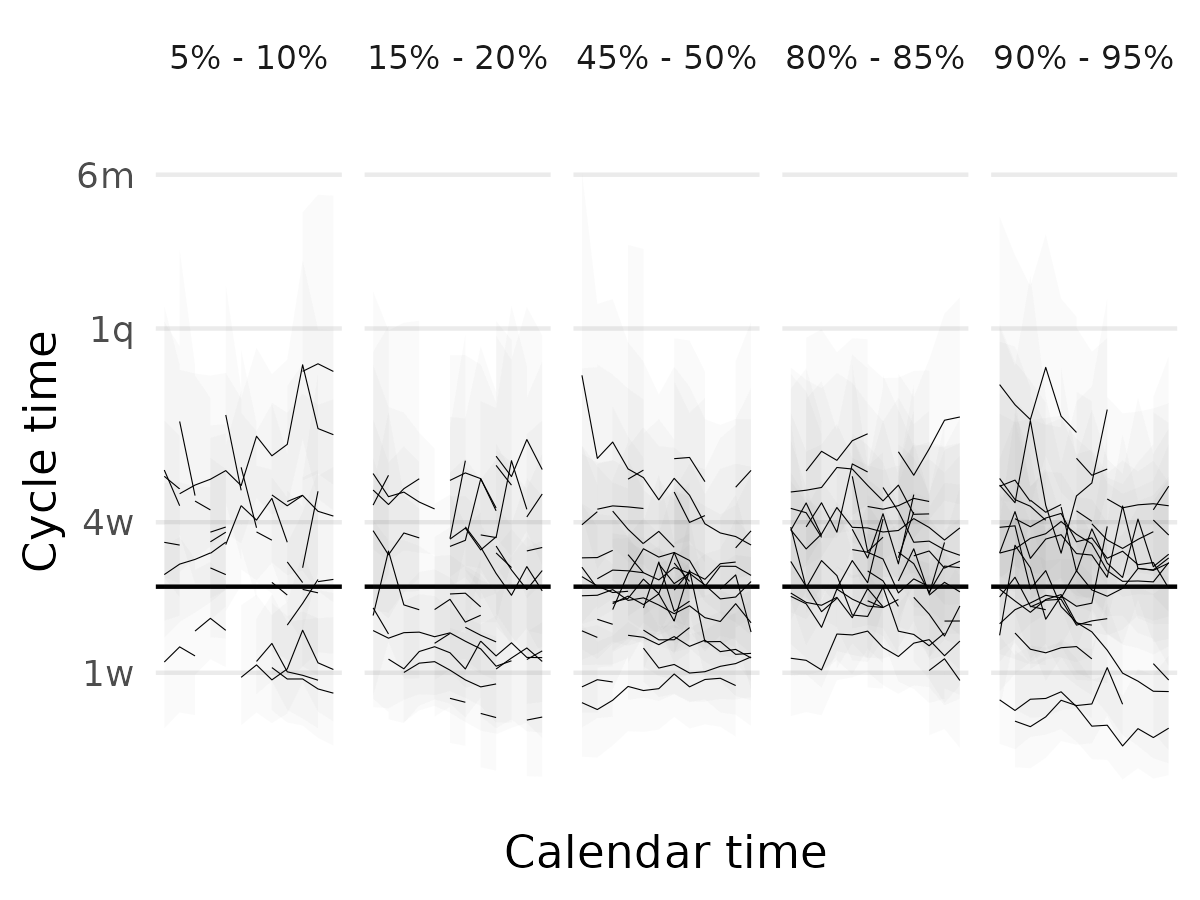}

}

\caption[Individual observations of cycle time are highly variable
across the year]{\label{fig-bw-ct-traj}Individual observations of cycle
time are highly variable across the year. Each line tracks cycle time
month-to-month for one randomly sampled individual. Each facet shows the
trajectories for individuals who have year-average
average-coding-days-per-week in the quantile specified at the top of the
facet. Shading represents prediction intervals from the model for
plausible cycle time values for these individuals.}

\end{figure}

We also examine the variability around our population effects.
Figure~\ref{fig-bw-ct-traj} shows model-expected cycle-time trajectories
across the year for randomly-sampled individuals. One of the stronger
effects at the population-level is the effect of the average number of
coding-days-per-week (both averaged over the year, and month by month).
To begin, we examine the effect of year-average coding-days-per-week. We
split the sampled individuals into 5\%-wide quantiles based on their
yearly average coding days per week. The population-level effect
discussed above reveals that, on average, individuals who have more
coding-days-per-week also tend to have lower cycle times. This shows up
subtly in Figure 9 as a decline in cycle-time from left-to-right across
these quantiles, especially when examining the amount to which the
trajectories occur below the thick black median-cycle-time line.

What is important about this illustration when interpreting these
results in the context of a real software development team is that it
shows a single measurement for a person on any given day or even
averaged across a month may not be representative of that individual's
long-term trend. While some of this uncertainty reflects measurement
error, much of it is irreducible given the factors we've considered in
this analysis. It may be possible to reduce it by adding further
information to the model. However, at present the conclusion must be
that one must take great care in comparing cycle-time between
individuals even in the same organization, or even against themselves.

\begin{figure}[htbp]

\centering{

\includegraphics[width=4in,height=\textheight]{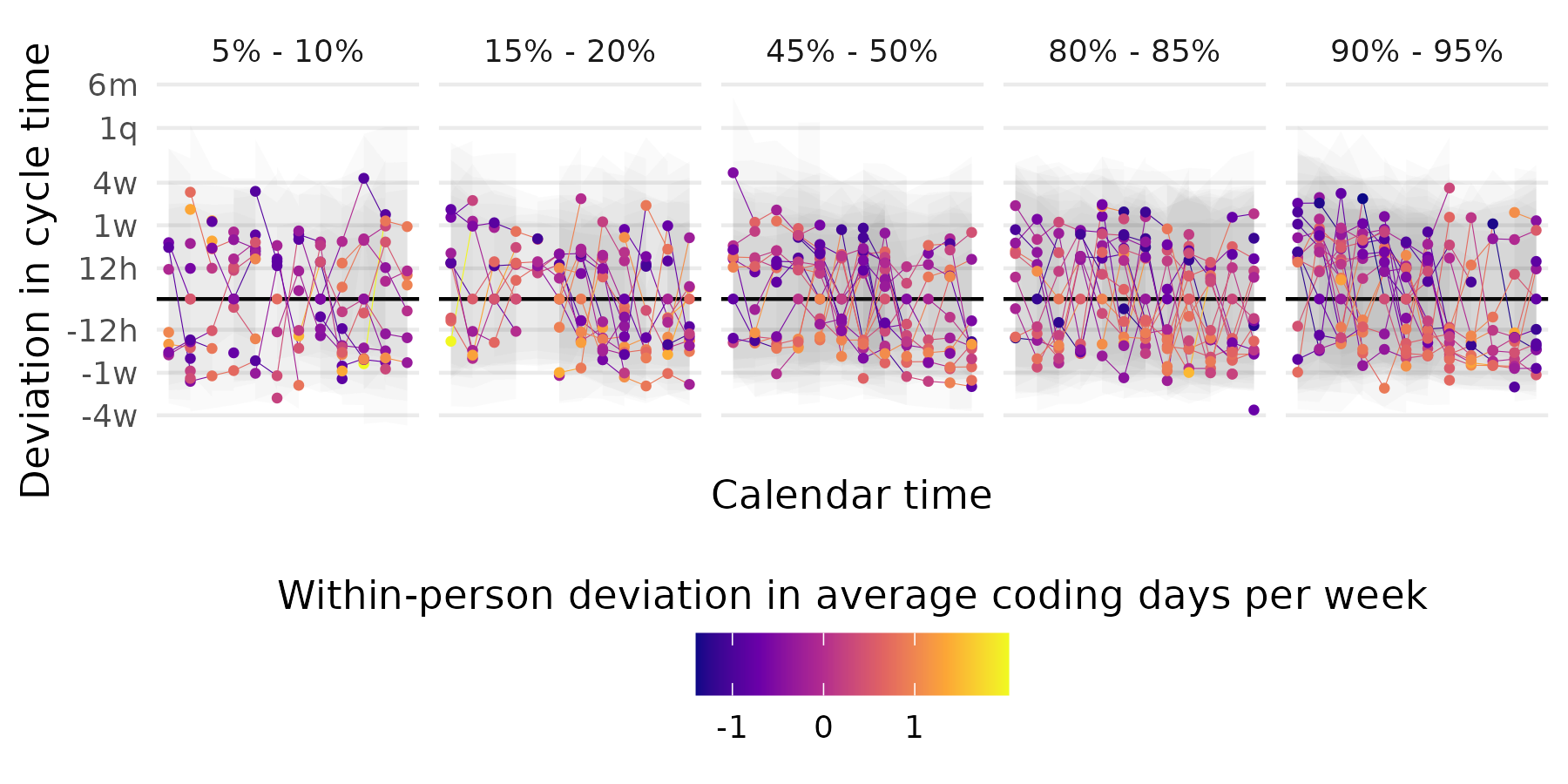}

}

\caption[Individual month-to-month deviations from year-average cycle
time are highly variable across the
year]{\label{fig-wi-ct-traj}Individual month-to-month deviations from
year-average cycle time are highly variable across the year. Each line
tracks cycle time month-to-month for one randomly sampled individual.
Each facet shows the trajectories for individuals who have year-average
average-coding-days-per-week in the quantile specified at the top of the
facet. Shading represents prediction intervals from the model for
plausible cycle time values for these individuals.}

\end{figure}

We can also examine the effect of month-by-month deviations in
coding-days-per-week at the level of these example individuals.
Figure~\ref{fig-wi-ct-traj} shows data for the same quantiles as above,
but now cycle time is normalized around each person's median to better
visualize within-person deviations in cycle time. Again, across the year
we see that variation within-person is substantial. While we can see the
tendency for within-person increases and decreases in
average-coding-days-per-week to affect cycle time, there is considerable
variation still, with many yellow-colored points above the 0 line and
darker points below the zero line.

\begin{figure}[htbp]

\centering{

\includegraphics[width=5in,height=\textheight]{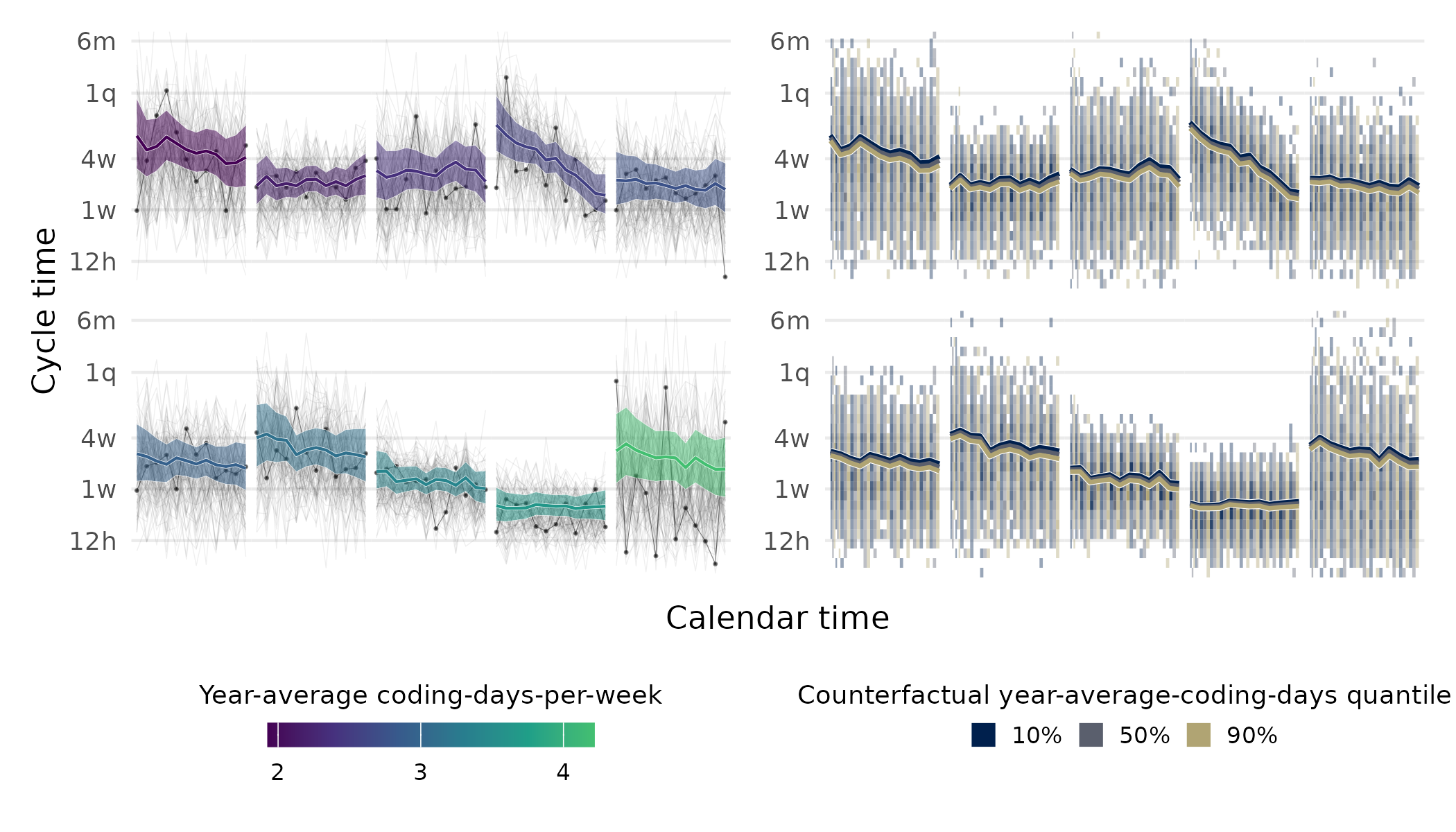}

}

\caption[Variability in cycle time within-person across time overwhelms
subtle differences in averages]{\label{fig-ct-pp-traj}Variability in
cycle time within-person across time overwhelms subtle differences in
averages. These plots show posterior predictions of reasonable
monthly-median cycle time and counterfactual predictions. The
\textbf{left} plot shows model expectations (our best guess at the
central tendency) and 95\% credible intervals (colored lines and
ribbons) of cycle time over the year for 10 randomly selected
participants. Lines and ribbons are colored by the participants observed
year-average average-coding-days-per-week. Black points and the stronger
black line indicates observed data for these participants. Light gray
lines represent model-derived posterior predictions of cycle times we
might expect to see from these participants. The \textbf{right} plot
shows these same participants under three counterfactual conditions:
with their year-average-coding-days set to the 10th, 50th, and 90th
quantile values. Each line represents the model expectation under these
conditions, all else being equal. Posterior predictions of reasonable
values for observed median-monthly cycle times are binned by these
quantiles and by month in the pixels behind the expectations.}

\end{figure}

We can unpack this variability further with the help of our model. We
will again examine a random subset of 10 participants with complete data
across the year. Again, each one is sampled from a different bin of
values of one of our strongest predictors, year-average
average-coding-days-per-week. Figure~\ref{fig-ct-pp-traj} (left) shows
the model-expected central tendency (gloss as a sort of moving-average)
of each of these participants, along with their observed data. We can
also ask the model to generate plausible values for cycle time that we
might observe; these are the gray lines behind everything. What is clear
here is that there is some systematic effect on the central tendency for
different levels of year-average coding days. Again, we see a lot of
overlap across individuals. And when we examine both the observed data,
and the model predictions, we see a whole lot more overlap.

The next thing we can do with the model is ask it to give us these
expectations and predictions under the counterfactual condition of each
of these participants having a different number of year-average coding
days. This is just a simulation, not an experiment, so we should heed
this caveat. In Figure~\ref{fig-ct-pp-traj} (right), we see model
expectations and posterior predictions were each of these participants
induced to take on the yearly-average number of coding days from the
10th, 50th, and 90th quantile of our data. This is a huge spread, but we
see only small, incremental changes in the model expectations for each
of them. More importantly, the spread of the posterior predictions,
binned for ease of visualization in the pixels behind the expectation
lines, are nearly indistinguishable. Taken together, these results once
again suggest caution in the use of this highly variable software
metric: taking a relatively unsophisticated monthly average of an
individual developer's cycle time data is not likely to be able to tell
you what their cycle time will be like in the future. This further
suggests that future work investigating the impact of interventions on
factors that impact cycle time may expect to see relatively small or
even invisible changes at the individual level, even when inducing
changes that do scale to desired impact at the organizational level.

\section{Limitations}\label{limitations}

Throughout the results, we have included commentary on the larger
context of software work and associated cautions regarding the
interpretation of our findings. In the following discussion section, we
further elaborate on future research directions which may build on the
evidence in the current paper. As noted in the introduction, there are
many limitations to the usage of an output-based velocity metric,
including that such a metric does not capture business outcomes, the
quality of task performance, and perceived value of software development
work. Nevertheless, monitoring cycle time is frequently recommended as a
measurement practice for software management. Below, we elaborate on
four key limitations of the current findings in greater detail, using
the framework of internal, external, construct, and statistical
conclusion validity (Vazire, Schiavone, and Bottesini 2022). In general,
these limitations point to the need for more and better empirical
evidence for software engineering (Devanbu, Zimmermann, and Bird 2016;
Kitchenham, Dyba, and Jorgensen 2004; Sjoberg, Dyba, and Jorgensen
2007).

\subsection{Construct validity}\label{construct-validity}

First, measurement validity challenges in ticket data and sample
representativeness constrain our analysis. A primary contribution of
this paper is to present findings from the analysis of a large (more
than 55k observations across 216 organizations) and longitudinal (a
calendar year) dataset of software work activity data. Metrics which are
git-based can enhance the measurement validity of our project by
providing this large-scale observation, unfiltered by individual
perception, in the ecologically valid context of real working teams.
Nevertheless, measurement validity may also be threatened by the way
that data are defined and constructed in the process leading to the
creation of variables around tickets: for instance, we rely on a general
assumption that teams mark tickets in a fairly accurate and timely
manner. Our dataset lacked contextual information that could serve to
validate team practices around ticket entry, and while we sought to
align our assumptions with known practices around these ticketing tools
and software workflows, it is important to note that differences in the
timeliness and accuracy of ticket entries may be an important factor
complicating our ability to understand cycle time.

\subsection{External validity}\label{external-validity}

This context limitation also applies to our sample of organizations
which is non-random and reflects organizations that invested in the
software metrics tool that provided that data, and does threaten the
generalizability of these results (Baltes and Ralph 2021). Moreover,
systematic differences in ticket assignment patterns may complicate
cycle time interpretations. How a ticket and its associated work tasks
are assigned inside of an organization is likely an important factor in
how tickets are completed. Assignment of tickets, especially defect
tickets, is complex over time and both the initial and subsequent
assignments of tickets can be impacted by many factors. For example,
after an initial assignment to a developer, bugs may frequently be
reassigned due to diverse factors such as determining ownership, time
constraints, and identifying developers who may have a particular view
into the root cause of the bug (Guo et al. 2011). Some individuals may
get assigned defect tickets systematically more often than others, and
changes in the proportion of work that consists of defect tickets may
change how fast someone is able to work in general, either because
defect tickets tend to be larger or smaller in scope than other tickets,
or because of disruptions.

\subsection{Internal validity}\label{internal-validity}

Our simplified measurement of team structures fails to capture the
complex collaborative reality of software development. In this analysis,
given limitations in our dataset about contexts such as internal team
structures within organizations, we have operationalized team size with
a generous attribution of team membership across shared activity, which
almost certainly represents a very rough estimate of team size. We have
also calculated team size as averaged in a static measure for the entire
year, rather than captured over time. Given that team size likely
impacts how quickly work is completed, we may miss out on nuanced
effects here; for instance, it is reasonable to imagine that further
variables around how resources are distributed and allocated to teams
may provide further predictive value to our understanding of team-level
factors that change cycle time. Team ``assignment'' itself is also
highly complex in technical work, as cross-team collaboration is
typical, and collaborations across software projects may frequently not
match explicit organizational hierarchies. In our previous research
using self-reports from software teams, we have found that over 60\% of
individual contributors on software teams report working closely with
other individual collaborators who do not share their same manager (C.
Hicks, Lee, and Ramsey 2023). In short, software development is a
collaborative exercise that is embedded in an ecosystem with history and
many actors (C. M. Hicks and Hevesi 2024).

Another major issue that threatens internal validity is invisible
collaborative activities and contributor role ambiguity limit
interpretations of cycle time variations. Developers communicate and
collaborate in many ways that may not be captured in a ticket,
e.g.~impromptu or formal mentorship, paired or mob programming, and
planning meetings. Where activity data are collected for some of these
activities, it is likely that such data may be perceived in our data
collection as ``belonging'' to a single individual, but work may often
reflect joint work (e.g., in the case of mentorship, paired, and
planning activities where one developer ``logs'' the work of many).
These organizational communication activities and social norms may
provide informative context for interpreting cycle time activities. For
instance, organizational planning data may provide a useful next step in
understanding how planning meetings do or do not cause downstream cycle
time to progress more efficiently. Also many people that work on
software may not be represented in this ticketing data. We also do not
know for certain the job titles of each contributor to this dataset and
use the term software developer broadly.

\subsection{Statistical conclusion
validity}\label{statistical-conclusion-validity}

Even using an outcome distribution appropriate to the context (Weibull,
in this case), hierarchical structure to account for nested
non-independence, and an emphasis on effect sizes over dichotomous
significance, our statistical conclusions remain conditional on a
particular modeling representation. First, the model encodes causal and
functional-form assumptions (e.g., additivity, smoothness, and the
exogeneity of our target predictors) that may be only approximately
true; good fit does not preclude biased inferences if these assumptions
are wrong. Second, unmodeled confounders may bias what we infer to be
causal associations. Third, cycle time is a time-evolving, heavy-tailed
process; departures from assumed variance and tail behavior, lagged
dependencies, or feedback (e.g., long tickets begetting process changes
that alter future tickets) can distort estimated associations. Fourth,
inferences depend on operationalization and scaling (e.g., static versus
time-varying team size; alternative definitions of ticket start and
close); different defensible specifications can yield materially
different effect sizes (Giudice and Gangestad 2021). Finally, our
results test a narrow class of models. Together, these considerations
suggest treating our estimates as conditional on the specific model
chosen.

\section{Discussion}\label{discussion}

\subsection{Take-aways for practitioners and
anti-patterns}\label{take-aways-for-practitioners-and-anti-patterns}

In this section, we summarize parts of our more thorough discussion
(below) in plain language that we hope will be useful for software
practitioners, highlighting not only the evidence at hand, but also
cautionary anti-patterns we think may be likely when using cycle time
data at scale.

\begin{itemize}
\tightlist
\item
  We observe many robust influences on cycle time but with small
  effects. This means implementing a change targeting any one of these
  factors may help a little bit, but don't expect a magic bullet. We
  recommend that practitioners consider the trade-offs of implementing
  any particular change by incorporating more context about developers'
  work tasks that goes beyond activity data, as well as designing an
  iterative, experimental approach to evaluating changes that includes
  measures that go beyond cycle time, such as incorporating developers'
  self-report on the utility of measurement practices.
\item
  Any single observation of cycle time is a very noisy indication of
  what is typical. Again, remember that each ticket's time-to-close is
  influenced by myriad factors, most of which are beyond the individual
  contributor's control. Ensuring a standard and repeatable practice
  around how tickets are created and managed across teams is likely to
  be a necessary precondition to relying on ticket data to make valid
  comparisons.
\item
  Practitioners can use this research to feel justified in seeking more
  shared and environmental explanations for the speed of work while
  doing complex software development, rather than locating their
  explanations in individual blame, or praise.
\item
  We can start to build up a sense of cycle times for teams and
  individuals if we are willing to observe patiently over long periods
  of time. Since we rarely observe identical task repetition, we need
  more observations to establish ``typical'' performance and identify
  factors that influence cycle time. Similar to public health research,
  understanding software work requires studying diverse contexts across
  industries and companies. This broad scope also creates opportunities
  for natural experiments. While organizations may hesitate to share
  data, these ``secrets'' typically diffuse naturally, and ``free
  revealing\footnote{''When we suggest that an innovator---be it an
    individual or a firm---`freely reveals' proprietary information, we
    mean that all intellectual property rights to that information are
    voluntarily given up by that innovator and all parties are given
    equal access to it---the information becomes a public good (Harhoff
    et al., 2003).'' (Von Hippel and Von Krogh 2006, 295)}'' may
  ultimately prove more profitable, as described by Von Hippel and Von
  Krogh (2006).
\item
  We should understand that cycle time is still quite distant from the
  ultimate objective: efficient delivery of value in a context that
  supports those who produce that value.
\item
  Although we did not have self-report data from individual contributors
  in the context of this analysis, it is possible that their perceptions
  and experience of doing software work gives them an accurate sense of
  what can help or hinder their problem solving. The measurement and
  attempt to change any particular objective indicator of software work,
  such as cycle time, should happen in conversation with these experts.
\item
  If you collect your own data on this, be aware that sometimes data can
  be too noisy to draw any credible conclusions. That in itself is a
  signal you can use to improve how and what you measure. At the end of
  the day, software development problem-solving is a social, human
  activity, and these are notoriously complex.
\end{itemize}

\subsection{General Discussion}\label{general-discussion}

Our analyses revealed precise directional associations between the
factors of interest and cycle time, though the magnitude of these
effects warrants careful consideration. These effects were generally in
an intuitive direction for coding time, task scoping, and collaboration.
More coding days, both on average across the year (between-person), and
month-to-month variations from that average (within-person) was
associated with lower cycle times. Moreover, the effect of average
coding days on cycle-time had a larger effect later in the year. Having
more merged PRs both on average, and month-to-month was also associated
with lower cycle-time. Greater collaboration at both the between- and
within-person level, and as measured by shared work on the same pull
request (and operationalized as degree centrality), were associated with
lower cycle-times as well. Two main results will require more nuance in
their interpretation. The effect of task scoping as measured by defect
tickets (where a lower percentage of defect tickets, representing
unplanned work, was taken as better task scoping) was associated with
faster cycle times at the within-person level and greater cycle times at
the between-person level. Finally, collaboration as measured by
comments-per-PR was associated with longer cycle times. These main
effects, while quite precise, were also somewhat small, and are set
against a backdrop of substantial unexplained between- and within-person
variation. This suggests that while there are a number of factors that
do push around cycle-time, each the life cycle of each ticket is a
complex function that is not necessarily captured by the metrics we have
available, some of which are quite commonly recommended as important
inputs to cycle time.

It is perhaps not surprising that cycle time benefits when individuals
have more time to code, either as a deviation from their usual, or than
other people doing similar work. When task completion requires writing
code, more time to code will obviously decrease time until completion.
Similarly, being able to break work into more discrete chunks and
completing more of these chunks should also benefit cycle time. Keep in
mind that these results both control for the other. That is, this is the
effect of more PRs for the same amount of coding time, and vice versa
(we discuss this benefit of our modeling approach more below).

Perhaps not as obviously, contributing code to a PR that is being
contributed to by others (what is captured by our degree centrality
measure) also benefits cycle time. Contrary to the possibility that
there are too many cooks in the kitchen, we show that on average cycle
time is lower when people work together on a shared problem. Again, this
is all-else-equal controlling for coding time, total PRs, and the other
variables in our model. However, we must also consider that more
comments per PR is associated with higher cycle time. This is another
signal of coordination and collaboration, but has an opposite effect.
Although speculative, we think that this reflects cases where a
particular PR is attempting to solve a difficult problem, and where
discussion is needed. Of course, this may also be a signal that
communication can sometimes become problematic. Future work would have
to examine the content of the PR and communication to disambiguate these
possibilities as well as others.

To finish our preliminary discussion of the main effects, we consider
the effect of defect tickets. Individuals with a higher proportion of
defect tickets on average over the course of the year tend to also have
longer cycle times, while a person who suddenly has more defect tickets
in a given month tends to have lower cycle times that month. We
speculate that the between-person effect may be a result of different
roles: in other words, people whose general workload is bug squashing
may simply be given more difficult bugs to squash; alternatively, if
less experienced programmers are generally given more bugs as a share of
work, they may be slower at completing their tasks. Organizational
factors may also be a culprit: if one's work is, on average over the
course of a year, swamped by defect tickets, one doesn't have time to
complete other tasks.

Considering the association of lower cycle-times during months when a
person tends to have more defect tickets, we speculate that this may be
a result of a shift in work from larger pieces of generative work to
fixing a lot of small errors. Again, as is the case with the
between-person effect here, and with the effect of comments per PR
above, these results suggest that a holistic and diagnostic approach to
understanding cycle time changes and their relationship to tickets would
need to include measures of the content of tickets (and context around
ticket assignment) in a more granular way than was possible within the
scope of this project.

The substantial individual differences observed in our analysis, coupled
with considerable residual variance after accounting for all measured
factors, presents both methodological and practical implications for the
assessment of software work. This heterogeneity in software metrics
suggests that detecting the impact of specific interventions on cycle
time may prove challenging unless the effect size substantially exceeds
the natural variation in individual and team performance. The magnitude
of unexplained variance serves as a crucial caveat for practitioners and
researchers attempting to implement and evaluate software work
interventions in software development contexts.

The observed variability itself constitutes a meaningful signal that
resonates with the lived experience of software development
practitioners. Our findings suggest that there exists no universal
formula for optimizing cycle time or enhancing software work across all
contexts and individuals. Rather, the path toward improved development
efficiency likely requires a nuanced approach that acknowledges this
inherent variability while simultaneously pursuing refined measurement
methodologies and targeted interventions. Success in this domain may
depend on our ability to identify and mitigate confounding factors while
developing increasingly sophisticated metrics that capture the
complexity of software development processes.

Indeed, the amount of variability apparent in these ticket data is
likely the result of the complexity of each unit of work in software
development. The process of setting a particular goal, and then of
planning how to reach that goal by breaking the work into discrete
tickets and tasks, is itself both conceptually and socially complex, and
may impact cycle-time. Moreover, the preconditions set by previous work
on a codebase determines what a software developer is able to do to
reach a particular goal and close a ticket. The discussion of the
process of ticket assignment in the limitations, above, is also relevant
here. Though crucially important as inputs to software work, these
sources of variability are not often measured and may be considered by
some to be immeasurable.

This work shows some clear signals about what might get in the way of
closing tickets. Developers may lack enough time for coding, may not
have enough collaborators, or may be bogged down with defect tickets.
These are not necessarily factors individual contributors (ICs) have any
control over. Again, the agency that any one person has to meaningfully
alter cycle-time is likely limited. This points toward the need for
systems-level thinking at the level of teams, units, organizations, and
interactions between these elements over time, rather than models which
only measure interventions on particular ICs in snapshots at a single
point in time.

What about the individual? Many practitioners reading this will wonder
what this means for their own productivity, or that of their direct
reports. Is cycle-time a good way to measure productivity? Can I just
give my team an extra head-down coding day and boost productivity? We
think that some of these signals are straightforward and will likely
help. As is often the case, more research is necessary. Testing actual
changes (i.e., interventions) with good measurement would help us figure
this out. However, for any one person, the effect on cycle-time will
likely be small and hard to see unless one is taking careful data over a
long period of time. That is perhaps the strongest message this research
sends to the practitioner: cycle-time is a very poor-resolution tool for
taking a snapshot of software work velocity because there are so many
inputs that go into the time it takes to close a particular ticket.
Resolution increases with more measurements across time, and across
people. As we refine our measures of software work velocity it may be
possible to make inferences about individuals in shorter time-scales but
we do not know of a measure with such properties and the present work
very strongly shows that cycle-time, as useful as it is, and as much as
it can tell us in the aggregate, is not such a measure. One practical
take-away for a practitioner may in fact be to feel justified in seeking
more shared and environmental explanations for the speed of work while
doing complex software development, rather than locating their
explanations in individual blame, or praise.

The present work demonstrates numerous methodological strengths that
also reveal ways in which inferences can be biased when data
complexities are not accounted for. This is important to understand both
for research like the current report but also for teams and
organizations modeling their data internally. Primarily, the analyses
presented here appropriately model cycle-time as Weibull-distributed.
This is a probability distribution that has a lower bound at 0, and no
upper bound. The distribution's shape parameter characterizes the
temporal dynamics of ticket closure probability (whether tickets become
more or less likely to close as time passes), while the scale parameter
(our focus in these analyses) determines the typical time window in
which tickets tend to close. The shape of observed cycle-times, with the
majority clustering around the low end but with a long right tail, is
well described by this distribution. The more common Gaussian (normal)
distribution has support on all real numbers (i.e., assumes negative
cycle-times are plausible!), and the bulk of the distribution is
centered symmetrically about the mean. Attempting to represent the
influence of various factors on cycle-time data using an inappropriate
distribution will have unintended consequences on the observed
relationships, which may be biased, and can lead to mode predictions
outside the range of possible values (Collett 1994; Lawless 2003;
McElreath 2020; Nelson 1982).

Another strength that is relevant to both researchers and practitioners
is the inclusion, simultaneously, of multiple factors in the same
models. Splitting the data into groups using some factor like coding
days and taking the average cycle time for each group, as a basis for
making inferences about differences between groups, is not sufficient to
account for the possibility that other factors systematically differ
between those groups. Nevertheless, this is an approach which we have
frequently seen in the industry. While not necessarily uncommon in this
literature, it is worth mentioning that modeling these factors
simultaneously as continuous variables allows us to get closer to the
idea of ``X has an effect on Y, all else being held equal.'' This is
necessary for starting to figure out the effect of a particular factor
in isolation. Of course, if one has access to even more sophisticated
observational-causal-inference methods (e.g., natural experiments, or
matching) or even interventional approaches, these would be even better.
If this sounds a little bit difficult, it's because, like many things in
software development, it is. There's no free lunch; and there are no
solutions, only tradeoffs. If one wants easy quantitative answers, one
must be willing to trade-off accuracy (and we almost never can say how
much accuracy is lost until we do it the hard way too).

We have emphasized in this work the small size of these observed
associations, and the vast variability in the outcome of interest. This
may belie the utility of what can be gained by further investigation
into these constructs. What we cannot know from this work is what a
change in one of our workplace factors would really mean for cycle time,
and what that might mean for an organization. We also have not yet
emphasized how small effects can accumulate if sustained. For example,
the 50th and 90th percentile in year-average
average-coding-days-per-week is about 1 day. If we look at the
model-expected association with an increase of 1 coding day per week, we
reduce cycle time by roughly 2 days (compared to a raw median of 13
days; Figure~\ref{fig-heatmaps}). Across an entire organization, this
average may mean a lot. It is also worth noting that in the software
literature, developers have reported substantial social-psychological
benefits from increases to focus time, deep work, and days spent coding
that may provide real and meaningful impacts on the quality of software
work whether or not they result in cycle time change (Andre N. Meyer et
al. 2021).

\subsection{Future Research
Directions}\label{future-research-directions}

Looking forward, our findings suggest two primary directions for future
research. In the domain of observational studies, increased attention to
a ticket's lifecycle, and process analysis appears as a tractable area
for new longitudinal investigations. There are a number of measurable
properties of tickets and code contributions that would be useful for
distinguishing between different kinds of work and different
interactions. For example, determining the expected scope or difficulty
of tickets would allow for important statistical control, ensuring we
compare like with like. Different kinds of work may also show different
pitfalls and benefits, and could plausibly be determined from the
content of a ticket or by assignment by those creating the tickets (for
example, not all tickets will have a programming solution).
Understanding and quantifying the type of content in PR comments would
be useful in determining the causes and consequences of this avenue for
collaboration. Ticket scope may also be contrasted with the ultimate
complexity of the work that closes that ticket. Because of the creative
and complex nature of software development, understanding what gets it
stuck and what helps it flow will almost certainly have to wrestle with
this complexity.

In the domain of intervention studies, we have argued that while
software metrics may play a role in evaluating the impact of changes
made to engineering organizations (or broader product and other
cross-functional partnerships that include engineering organizations),
determining this impact is not as simple as expecting a coherent,
consistent and average increase in a metric such as cycle time to result
from a change. Changes which may be meaningful at the scale of an
organization may be relatively small or invisible on the individual
level, and changes which are meaningful to individuals may not reflect
an intervention that scales to an organization. Nevertheless,
engineering (and other) orgs are currently seeking to become data-driven
and use their own activity data as a point of reflection on both sides
of this question. Increased attention to developing robust efficacy
measures for behavioral interventions at scale, and detailing the
nuances of how these changes show up in software activity metrics, will
be needed to answer these questions. Detailing the potential
methodological and statistical pitfalls of these data may also play a
critical role in steering organizations away from misleading and
inaccurate summaries, and toward appropriate methodologies.
Understanding how to implement a software work intervention across an
organization in a standardized way presents its own set of research
questions for the future. For this, we believe that increased attention
to developers' own within-person growth, wellbeing, and work will also
be necessary to provide a full understanding of software work.

As the findings of the current work have demonstrated, moving toward a
greater understanding of how to improve software development will likely
require a plurality of methods, measurements, and thoughtful
experimental practices within software engineering organizations, rather
than silver bullet, isolated metrics.

\section{Acknowledgements}\label{acknowledgements}

We would like to thank Kristen Foster-Marks for advice on the project
and helpful feedback on a draft; also thanks to Bennet Cook for his help
navigating many databases.

\section{Declarations}\label{declarations}

\subsection{Ethical approval}\label{ethical-approval}

This research used aggregated, anonymized GitHub activity data routinely
collected through our company's normal operations and permitted by our
Terms of Service. No personal information was gathered specifically for
this study, and strict protocols were followed to prevent
re-identification of individuals or organizations. Because the dataset
was pre-existing, fully anonymized, and did not involve direct
interaction with human subjects, the research is exempt from IRB review
under 45 CFR \S 46.104(d)(4)(ii). All data was stored on secure systems
with limited access, ensuring both data integrity and confidentiality.

\subsection{Author Contributions}\label{author-contributions}

Contribution roles listed are defined in Group (2022).

John C. Flournoy: Conceptualization, Data curation, Formal analysis,
Investigation, Methodology, Visualization, Writing -- original draft,
Writing -- review \& editing. Carol S. Lee: Conceptualization, Data
curation, Formal analysis, Investigation, Methodology, Writing --
original draft, Writing -- review \& editing. Maggie Wu:
Conceptualization, Data curation, Formal analysis, Investigation,
Methodology. Catherine M. Hicks: Conceptualization, Funding acquisition,
Investigation, Methodology, Project administration, Resources,
Supervision, Writing -- review \& editing.

\subsection{Data Availability
Statement}\label{data-availability-statement}

Data are considered proprietary and are not available to be shared. Code
for these analyses is available as \texttt{analyses.qmd}, here:
\url{https://github.com/jflournoy/no-silver-bullets}.

\subsection{Conflict of Interest}\label{conflict-of-interest}

The authors are employed by Pluralsight, which is the source of the data
used in this study. However, the authors' roles do not involve product
sales or marketing, and this research does not evaluate product
performance. The company approved the use of anonymized data for this
research but did not influence the analysis methods, findings, or
conclusions presented in this paper.

\subsection{Other declarations}\label{other-declarations}

\begin{itemize}
\tightlist
\item
  Funding: Not applicable.
\item
  Informed consent: Not applicable.
\item
  Clinical trial number: not applicable.
\end{itemize}

\section*{References}\label{references}
\addcontentsline{toc}{section}{References}

\phantomsection\label{refs}


\begin{CSLReferences}{1}{0}
\bibitem[\citeproctext]{ref-agrawalSoftwareEffortQuality2007}
Agrawal, Manish, and Kaushal Chari. 2007. {``Software {Effort},
{Quality}, and {Cycle Time}: {A Study} of {CMM Level} 5 {Projects}.''}
\emph{IEEE Transactions on Software Engineering} 33 (3): 145--56.
\url{https://doi.org/10.1109/TSE.2007.29}.

\bibitem[\citeproctext]{ref-ahmadPandorasBoxSocial2024}
Ahmad, Muhammad Ovais, and Tomas Gustavsson. 2024. {``The {Pandora}'s
Box of Social, Process, and People Debts in Software Engineering.''}
\emph{Journal of Software: Evolution and Process} 36 (2): e2516.
\url{https://doi.org/10.1002/smr.2516}.

\bibitem[\citeproctext]{ref-arel-bundockMarginaleffectsPredictionsComparisons2024}
Arel-Bundock, Vincent. 2024. \emph{Marginaleffects: {Predictions},
Comparisons, Slopes, Marginal Means, and Hypothesis Tests}. Manual.
\url{https://marginaleffects.com/}.

\bibitem[\citeproctext]{ref-bacchelliExpectationsOutcomesChallenges2013}
Bacchelli, Alberto, and Christian Bird. 2013. {``Expectations, Outcomes,
and Challenges of Modern Code Review.''} In \emph{2013 35th
{International Conference} on {Software Engineering} ({ICSE})}, 712--21.
\url{https://doi.org/10.1109/ICSE.2013.6606617}.

\bibitem[\citeproctext]{ref-ballWorkplaceSurveillanceOverview2010}
Ball, Kirstie. 2010. {``Workplace Surveillance: An Overview.''}
\emph{Labor History} 51 (1): 87--106.
\url{https://doi.org/10.1080/00236561003654776}.

\bibitem[\citeproctext]{ref-baltesSamplingSoftwareEngineering2021a}
Baltes, Sebastian, and Paul Ralph. 2021. {``Sampling in {Software
Engineering Research}: {A Critical Review} and {Guidelines}.''} October
20, 2021. \url{https://doi.org/10.48550/arXiv.2002.07764}.

\bibitem[\citeproctext]{ref-barrettDatatableExtensionDataframe2024}
Barrett, Tyson, Matt Dowle, Arun Srinivasan, Jan Gorecki, Michael
Chirico, and Toby Hocking. 2024. \emph{Data.table: {Extension} of
`Data.frame`}. Manual. \url{https://r-datatable.com}.

\bibitem[\citeproctext]{ref-beskerTechnicalDebtCripples2018}
Besker, Terese, Antonio Martini, and Jan Bosch. 2018. {``Technical Debt
Cripples Software Developer Productivity: A Longitudinal Study on
Developers' Daily Software Development Work.''} In \emph{Proceedings of
the 2018 {International Conference} on {Technical Debt}}, 105--14.
Gothenburg Sweden: ACM. \url{https://doi.org/10.1145/3194164.3194178}.

\bibitem[\citeproctext]{ref-blackburnImprovingSpeedProductivity1996}
Blackburn, J. D., G. D. Scudder, and L. N. Van Wassenhove. 1996.
{``Improving Speed and Productivity of Software Development: A Global
Survey of Software Developers.''} \emph{IEEE Transactions on Software
Engineering} 22 (12): 875--85. \url{https://doi.org/10.1109/32.553636}.

\bibitem[\citeproctext]{ref-bouwersSoftwareMetricsPitfalls2013}
Bouwers, Eric, Arie van Deursen, and Joost Visser. 2013. {``Software
Metrics: {Pitfalls} and Best Practices.''} In \emph{2013 35th
{International Conference} on {Software Engineering} ({ICSE})},
1491--92. \url{https://doi.org/10.1109/ICSE.2013.6606755}.

\bibitem[\citeproctext]{ref-brocknerProceduralFairnessOutcome2007}
Brockner, Joel, Ariel Y. Fishman, Jochen Reb, Barry Goldman, Scott
Spiegel, and Charlee Garden. 2007. {``Procedural Fairness, Outcome
Favorability, and Judgments of an Authority's Responsibility.''}
\emph{Journal of Applied Psychology} 92 (6): 1657--71.
\url{https://doi.org/10.1037/0021-9010.92.6.1657}.

\bibitem[\citeproctext]{ref-brocknerInteractiveEffectsProcedural1994}
Brockner, Joel, Mary Konovsky, Rochelle Cooper-Schneider, Robert Folger,
Christopher Martin, and Robert J. Bies. 1994. {``Interactive {Effects}
of {Procedural Justice} and {Outcome Negativity} on {Victims} and
{Survivors} of {Job Loss}.''} \emph{Academy of Management Journal} 37
(2): 397--409. \url{https://doi.org/10.5465/256835}.

\bibitem[\citeproctext]{ref-brooksMythicalManmonthEssays1975}
Brooks, Frederick P. 1975. \emph{The Mythical Man-Month: Essays on
Software Engineering}. Reading, Mass.: Addison-Wesley Pub. Co.

\bibitem[\citeproctext]{ref-bruneauxWhatMcKinseyHas2024}
Bruneaux, Taylor. 2024. {``What {McKinsey} Has to Say about Developer
Productivity.''} May 9, 2024.
\url{https://getdx.com/blog/mckinsey-developer-productivity/}.

\bibitem[\citeproctext]{ref-burknerBrmsPackageBayesian2017}
Bürkner, Paul-Christian. 2017. {``Brms: {An R Package} for {Bayesian
Multilevel Models Using Stan}.''} \emph{Journal of Statistical Software}
80 (1, 1): 1--28. \url{https://doi.org/10.18637/jss.v080.i01}.

\bibitem[\citeproctext]{ref-burknerAdvancedBayesianMultilevel2018}
---------. 2018. {``Advanced {Bayesian} Multilevel Modeling with the {R}
Package {brms}.''} \emph{The R Journal} 10 (1): 395--411.
\url{https://doi.org/10.32614/RJ-2018-017}.

\bibitem[\citeproctext]{ref-burknerBayesianItemResponse2021}
---------. 2021. {``Bayesian {Item Response Modeling} in {R} with Brms
and {Stan}.''} \emph{Journal of Statistical Software} 100 (November):
1--54. \url{https://doi.org/10.18637/jss.v100.i05}.

\bibitem[\citeproctext]{ref-burknerPosteriorToolsWorking2023}
Bürkner, Paul-Christian, Jonah Gabry, Matthew Kay, and Aki Vehtari.
2023. {``Posterior: {Tools} for Working with Posterior Distributions.''}
\url{https://mc-stan.org/posterior/}.

\bibitem[\citeproctext]{ref-caesensPerceivedOrganizationalSupport2017}
Caesens, Gaëtane, Florence Stinglhamber, Stéphanie Demoulin, and
Matthias De Wilde. 2017. {``Perceived Organizational Support and
Employees' Well-Being: The Mediating Role of Organizational
Dehumanization.''} \emph{European Journal of Work and Organizational
Psychology} 26 (4): 527--40.
\url{https://doi.org/10.1080/1359432X.2017.1319817}.

\bibitem[\citeproctext]{ref-careyWhy70Engineers2024}
Carey, Scott. 2024. {``Why 70\% of Engineers Avoid Measuring Lines of
Code.''} LeadDev. December 5, 2024.
\url{https://leaddev.com/reporting/why-70-of-engineers-avoid-measuring-lines-of-code}.

\bibitem[\citeproctext]{ref-carmelCycleTimePackaged1995}
Carmel, Erran. 1995. {``Cycle {Time} in {Packaged Software Firms}.''}
\emph{Journal of Product Innovation Management} 12 (2): 110--23.
\url{https://doi.org/10.1111/j.1540-5885.1995.jpim122_0110.xml.x}.

\bibitem[\citeproctext]{ref-cataldoSociotechnicalCongruenceFramework2008}
Cataldo, Marcelo, James D. Herbsleb, and Kathleen M. Carley. 2008.
{``Socio-Technical Congruence: A Framework for Assessing the Impact of
Technical and Work Dependencies on Software Development Productivity.''}
In \emph{Proceedings of the {Second ACM-IEEE} International Symposium on
{Empirical} Software Engineering and Measurement}, 2--11. {ESEM} '08.
New York, NY, USA: Association for Computing Machinery.
\url{https://doi.org/10.1145/1414004.1414008}.

\bibitem[\citeproctext]{ref-chhunejaWhy50Developers2024}
Chhuneja, Shivam. 2024. {``Why 50\% {Developers Hate DORA Metrics}?''}
April 19, 2024.
\url{https://middlewarehq.com/blog/why-50-developers-hate-dora-metrics}.

\bibitem[\citeproctext]{ref-clincySoftwareDevelopmentProductivity2003}
Clincy, Victor. 2003. {``Software {Development Productivity} and {Cycle
Time Reduction}.''} \emph{Journal of Computing Sciences in Colleges} 19
(2): 278--87. \url{https://digitalcommons.kennesaw.edu/facpubs/1581}.

\bibitem[\citeproctext]{ref-collettModellingSurvivalData1994}
Collett, D. 1994. \emph{Modelling Survival Data in Medical Research}.
1st ed. Texts in Statistical Science. London ; New York: Chapman \&
Hall.

\bibitem[\citeproctext]{ref-coteOnlyPeopleWho2023}
Coté. 2023. {``The Only People Who Don't Like Metrics Are the People
Being Measured, or, Developer Productivity Metrics Quicksand.''}
February 2, 2023.
\url{https://newsletter.cote.io/p/the-only-people-who-dont-like-metrics}.

\bibitem[\citeproctext]{ref-curranDisaggregationWithinPersonBetweenPerson2011}
Curran, Patrick J., and Daniel J. Bauer. 2011. {``The {Disaggregation}
of {Within-Person} and {Between-Person Effects} in {Longitudinal Models}
of {Change}.''} \emph{Annual Review of Psychology} 62 (1): 583--619.
\url{https://doi.org/10.1146/annurev.psych.093008.100356}.

\bibitem[\citeproctext]{ref-denisov-blanchPredictingExpertEvaluations2024}
Denisov-Blanch, Yegor, Igor Ciobanu, Simon Obstbaum, and Michal
Kosinski. 2024. {``Predicting {Expert Evaluations} in {Software Code
Reviews}.''} September 23, 2024.
\url{https://doi.org/10.48550/arXiv.2409.15152}.

\bibitem[\citeproctext]{ref-devanbuBeliefEvidenceEmpirical2016}
Devanbu, Prem, Thomas Zimmermann, and Christian Bird. 2016. {``Belief \&
Evidence in Empirical Software Engineering.''} In \emph{Proceedings of
the 38th {International Conference} on {Software Engineering}}, 108--19.
{ICSE} '16. New York, NY, USA: Association for Computing Machinery.
\url{https://doi.org/10.1145/2884781.2884812}.

\bibitem[\citeproctext]{ref-eversImprovingEngineeringProductivity1998}
Evers, J. H., G. M. Oehler, and M. G. Tucker. 1998. {``Improving
Engineering Productivity: A Time Study of an Engineer's Typical Work
Day.''} In \emph{{IEMC} '98 {Proceedings}. {International Conference} on
{Engineering} and {Technology Management}. {Pioneering New
Technologies}: {Management Issues} and {Challenges} in the {Third
Millennium} ({Cat}. {No}.{98CH36266})}, 377--83.
\url{https://doi.org/10.1109/IEMC.1998.727789}.

\bibitem[\citeproctext]{ref-finster5MinuteDevOps2023}
Finster, Bryan. 2023. {``5 {Minute DevOps}: {McKinsey Gets Developer
Productivity Wrong}.''} Medium. September 8, 2023.
\url{https://bdfinst.medium.com/5-minute-devops-mckinsey-gets-developer-productivity-wrong-573b57cd6f6a}.

\bibitem[\citeproctext]{ref-flowHowIncreaseSoftware}
Flow. n.d. {``How to Increase Software Delivery Speeds by Reducing Cycle
Time.''}
\url{https://www.pluralsight.com/product/flow/flow-academy/how-to-improve-cycle-time}.

\bibitem[\citeproctext]{ref-forsgrenAccelerateScienceDevOps2018}
Forsgren, Nicole, Jez Humble, and Gene Kim. 2018. \emph{Accelerate: The
Science Behind {DevOps}: Building and Scaling High Performing Technology
Organizations}. First edition. Portland, Oregon: IT Revolution.

\bibitem[\citeproctext]{ref-fraserNoSilverBullet2007}
Fraser, Steven D., Frederick P. Brooks, Martin Fowler, Ricardo Lopez,
Aki Namioka, Linda Northrop, David Lorge Parnas, and David Thomas. 2007.
{``"{No} Silver Bullet" Reloaded: Retrospective on "Essence and
Accidents of Software Engineering".''} In \emph{Companion to the 22nd
{ACM SIGPLAN} Conference on {Object-oriented} Programming Systems and
Applications Companion}, 1026--30. {OOPSLA} '07. New York, NY, USA:
Association for Computing Machinery.
\url{https://doi.org/10.1145/1297846.1297973}.

\bibitem[\citeproctext]{ref-gabryCmdstanrInterfaceCmdStan2024}
Gabry, Jonah, Rok Češnovar, Andrew Johnson, and Steve Bronder. 2024.
\emph{Cmdstanr: {R Interface} to '{CmdStan}'}. Manual.
\url{https://mc-stan.org/cmdstanr/}.

\bibitem[\citeproctext]{ref-GelmanPowerCalculationsAssessing2014}
Gelman, Andrew, and John Carlin. 2014. {``Beyond {Power Calculations
Assessing Type S} ({Sign}) and {Type M} ({Magnitude}) {Errors}.''}
\emph{Perspectives on Psychological Science} 9 (6): 641--51.
\url{https://doi.org/10.1177/1745691614551642}.

\bibitem[\citeproctext]{ref-giudiceTravelersGuideMultiverse2021}
Giudice, Marco Del, and Steven W. Gangestad. 2021. {``A {Traveler}'s
{Guide} to the {Multiverse}: {Promises}, {Pitfalls}, and a {Framework}
for the {Evaluation} of {Analytic Decisions}.''} \emph{Advances in
Methods and Practices in Psychological Science} 4 (1): 251524592095492.
\url{https://doi.org/10.1177/2515245920954925}.

\bibitem[\citeproctext]{ref-gohelFlextableFunctionsTabular2024}
Gohel, David, and Panagiotis Skintzos. 2024. \emph{Flextable:
{Functions} for Tabular Reporting}. Manual.
\url{https://ardata-fr.github.io/flextable-book/}.

\bibitem[\citeproctext]{ref-gousiosExploratoryStudyPullbased2014}
Gousios, Georgios, Martin Pinzger, and Arie Van Deursen. 2014. {``An
Exploratory Study of the Pull-Based Software Development Model.''} In
\emph{Proceedings of the 36th {International Conference} on {Software
Engineering}}, 345--55. Hyderabad India: ACM.
\url{https://doi.org/10.1145/2568225.2568260}.

\bibitem[\citeproctext]{ref-gralhaReduceCycleTime2022}
Gralha, Catarina. 2022. {``Reduce Cycle Time - Best Practices.''} August
18, 2022. \url{https://blog.codacy.com/reducing-cycle-time}.

\bibitem[\citeproctext]{ref-griffinMetricsMeasuringProduct1993}
Griffin, Abbie. 1993. {``Metrics for {Measuring Product Development
Cycle Time}.''} \emph{Journal of Product Innovation Management} 10 (2):
112--25. \url{https://doi.org/10.1111/1540-5885.1020112}.

\bibitem[\citeproctext]{ref-grisoldDigitalSurveillanceOrganizations2024}
Grisold, Thomas, Stefan Seidel, Markus Heck, and Nicholas Berente. 2024.
{``Digital {Surveillance} in {Organizations}.''} \emph{Business \&
Information Systems Engineering} 66 (3): 401--10.
\url{https://doi.org/10.1007/s12599-024-00866-7}.

\bibitem[\citeproctext]{ref-nisocreditworkinggroupANSINISOZ3910420222022}
Group, NISO CRediT Working. 2022. \emph{{ANSI}/{NISO Z39}.104-2022,
{CRediT}, {Contributor Roles Taxonomy}}. 3600 Clipper Mill Road Suite
302 Baltimore, MD 21211: NISO.
\url{https://doi.org/10.3789/ansi.niso.z39.104-2022}.

\bibitem[\citeproctext]{ref-guoNotMyBug2011}
Guo, Philip J., Thomas Zimmermann, Nachiappan Nagappan, and Brendan
Murphy. 2011. {``"{Not} My Bug!" and Other Reasons for Software Bug
Report Reassignments.''} In \emph{Proceedings of the {ACM} 2011
Conference on {Computer} Supported Cooperative Work}, 395--404. Hangzhou
China: ACM. \url{https://doi.org/10.1145/1958824.1958887}.

\bibitem[\citeproctext]{ref-guptaKeyDriversReduced1998}
Gupta, Ashok K., and William E. Souder. 1998. {``Key {Drivers} of
{Reduced Cycle Time}.''} \emph{Research-Technology Management} 41 (4):
38--43. \url{https://doi.org/10.1080/08956308.1998.11671221}.

\bibitem[\citeproctext]{ref-SciPyProceedings_11}
Hagberg, Aric A., Daniel A. Schult, and Pieter J. Swart. 2008.
{``Exploring Network Structure, Dynamics, and Function Using
{NetworkX}.''} In \emph{Proceedings of the 7th Python in Science
Conference}, edited by Gaël Varoquaux, Travis Vaught, and Jarrod
Millman, 11--15. Pasadena, CA USA.

\bibitem[\citeproctext]{ref-harrellRegressionModelingStrategies2015}
Harrell, Frank E. 2015. \emph{Regression {Modeling Strategies}: {With
Applications} to {Linear Models}, {Logistic} and {Ordinal Regression},
and {Survival Analysis}}. Springer {Series} in {Statistics}. Cham:
Springer International Publishing.
\url{https://doi.org/10.1007/978-3-319-19425-7}.

\bibitem[\citeproctext]{ref-henryRlangFunctionsBase2024}
Henry, Lionel, and Hadley Wickham. 2024. \emph{Rlang: {Functions} for
Base Types and Core {R} and 'Tidyverse' Features}. Manual.
\url{https://rlang.r-lib.org}.

\bibitem[\citeproctext]{ref-hicksPsychologicalAffordancesCan2024}
Hicks, Catherine M. 2024. {``Psychological {Affordances Can Provide} a
{Missing Explanatory Layer} for {Why Interventions} to {Improve
Developer Experience Take Hold} or {Fail}.''} January 25, 2024.
\url{https://doi.org/10.31234/osf.io/qz43x}.

\bibitem[\citeproctext]{ref-hicksCumulativeCultureTheory2024}
Hicks, Catherine M., and Ana Hevesi. 2024. {``A {Cumulative Culture
Theory} for {Developer Problem-Solving}.''} November 21, 2024.
\url{https://doi.org/10.31234/osf.io/tfjyw}.

\bibitem[\citeproctext]{ref-hicksDeveloperThrivingFour2024}
Hicks, Catherine M., Carol S. Lee, and Morgan Ramsey. 2024. {``Developer
{Thriving}: {Four Sociocognitive Factors That Create Resilient
Productivity} on {Software Teams}.''} \emph{IEEE Software} 41 (4):
68--77. \url{https://doi.org/10.1109/MS.2024.3382957}.

\bibitem[\citeproctext]{ref-hicksNewDeveloperAI2024}
Hicks, Catherine M., Carol Lee, and Kristen Foster-Marks. 2024. {``The
{New Developer}: {AI Skill Threat}, {Identity Change} \& {Developer
Thriving} in the {Transition} to {AI-Assisted Software Development}.''}
April 20, 2024. \url{https://doi.org/10.31234/osf.io/2gej5}.

\bibitem[\citeproctext]{ref-hicksDeveloperThrivingFour2023}
Hicks, Catherine, Carol S. Lee, and Morgan Ramsey. 2023. {``Developer
{Thriving}: {The} Four Factors That Drive {Software Developer
Productivity} Across {Industries}.''} Developer Success Lab at
Pluralsight.
\url{https://www.pluralsight.com/resource-center/guides/developer-thriving-research-paper}.

\bibitem[\citeproctext]{ref-kayTidybayesTidyData2023}
Kay, Matthew. 2023. \emph{{tidybayes}: {Tidy} Data and Geoms for
{Bayesian} Models}. Manual.
\url{https://doi.org/10.5281/zenodo.1308151}.

\bibitem[\citeproctext]{ref-kitchenhamEvidencebasedSoftwareEngineering2004}
Kitchenham, B. A., T. Dyba, and M. Jorgensen. 2004. {``Evidence-Based
Software Engineering.''} In \emph{Proceedings. 26th {International
Conference} on {Software Engineering}}, 273--81.
\url{https://doi.org/10.1109/ICSE.2004.1317449}.

\bibitem[\citeproctext]{ref-kruschkeRejectingAcceptingParameter2018}
Kruschke, John K. 2018. {``Rejecting or {Accepting Parameter Values} in
{Bayesian Estimation}.''} \emph{Advances in Methods and Practices in
Psychological Science} 1 (2): 270--80.
\url{https://doi.org/10.1177/2515245918771304}.

\bibitem[\citeproctext]{ref-kudrjavetsSmallCodeChanges2022}
Kudrjavets, Gunnar, Nachiappan Nagappan, and Ayushi Rastogi. 2022. {``Do
Small Code Changes Merge Faster?: A Multi-Language Empirical
Investigation.''} In \emph{Proceedings of the 19th {International
Conference} on {Mining Software Repositories}}, 537--48. Pittsburgh
Pennsylvania: ACM. \url{https://doi.org/10.1145/3524842.3528448}.

\bibitem[\citeproctext]{ref-lagiosExplainingNegativeConsequences2022}
Lagios, Constantin, Gaëtane Caesens, Nathan Nguyen, and Florence
Stinglhamber. 2022. {``Explaining the {Negative Consequences} of
{Organizational} {Dehumanization}.''} \emph{Journal of Personnel
Psychology} 21 (2): 86--93.
\url{https://doi.org/10.1027/1866-5888/a000286}.

\bibitem[\citeproctext]{ref-lawlessStatisticalModelsMethods2003}
Lawless, Jerald F. 2003. \emph{Statistical Models and Methods for
Lifetime Data}. 2nd ed. Wiley Series in Probability and Statistics.
Hoboken, N.J.: Wiley-Interscience.

\bibitem[\citeproctext]{ref-linesWhyEliteDev2023}
Lines, Dan. 2023. {``Why Elite Dev Teams Focus on Pull-Request
Metrics.''} LeadDev. May 11, 2023.
\url{https://leaddev.com/reporting/why-elite-dev-teams-focus-pull-request-metrics}.

\bibitem[\citeproctext]{ref-ludeckeExtractingComputingExploring2020}
Lüdecke, Daniel, Mattan S. Ben-Shachar, Indrajeet Patil, and Dominique
Makowski. 2020. {``Extracting, Computing and Exploring the Parameters of
Statistical Models Using {R}.''} \emph{Journal of Open Source Software}
5 (53): 2445. \url{https://doi.org/10.21105/joss.02445}.

\bibitem[\citeproctext]{ref-maxwellSoftwareDevelopmentProductivity1996}
Maxwell, K. D., L. Van Wassenhove, and S. Dutta. 1996. {``Software
Development Productivity of {European} Space, Military, and Industrial
Applications.''} \emph{IEEE Transactions on Software Engineering} 22
(10): 706--18. \url{https://doi.org/10.1109/32.544349}.

\bibitem[\citeproctext]{ref-mcelreathStatisticalRethinkingBayesian2020}
McElreath, Richard. 2020. \emph{Statistical {Rethinking}: {A Bayesian
Course} with {Examples} in {R} and {Stan}}. 2nd ed. Boca Raton: {Chapman
and Hall/CRC}. \url{https://doi.org/10.1201/9780429029608}.

\bibitem[\citeproctext]{ref-mettlerConnectedWorkplaceCharacteristics2024}
Mettler, Tobias. 2024. {``The Connected Workplace: {Characteristics} and
Social Consequences of Work Surveillance in the Age of Datification,
Sensorization, and Artificial Intelligence.''} \emph{Journal of
Information Technology} 39 (3): 547--67.
\url{https://doi.org/10.1177/02683962231202535}.

\bibitem[\citeproctext]{ref-meyerTodayWasGood2021}
Meyer, Andre N., Earl T. Barr, Christian Bird, and Thomas Zimmermann.
2021. {``Today {Was} a {Good Day}: {The Daily Life} of {Software
Developers}.''} \emph{IEEE Transactions on Software Engineering} 47 (5):
863--80. \url{https://doi.org/10.1109/TSE.2019.2904957}.

\bibitem[\citeproctext]{ref-meyerEnablingGoodWork2021}
Meyer, André N., Gail C. Murphy, Thomas Zimmermann, and Thomas Fritz.
2021. {``Enabling {Good Work Habits} in {Software Developers} Through
{Reflective Goal-Setting}.''} \emph{IEEE Transactions on Software
Engineering} 47 (9): 1872--85.
\url{https://doi.org/10.1109/TSE.2019.2938525}.

\bibitem[\citeproctext]{ref-murphy-hillWhatPredictsSoftware2021}
Murphy-Hill, Emerson, Ciera Jaspan, Caitlin Sadowski, David Shepherd,
Michael Phillips, Collin Winter, Andrea Knight, Edward Smith, and
Matthew Jorde. 2021. {``What {Predicts Software Developers}'
{Productivity}?''} \emph{IEEE Transactions on Software Engineering} 47
(3): 582--94. \url{https://doi.org/10.1109/TSE.2019.2900308}.

\bibitem[\citeproctext]{ref-nanImpactBudgetSchedule2009}
Nan, Ning, and Donald E. Harter. 2009. {``Impact of {Budget} and
{Schedule Pressure} on {Software Development Cycle Time} and
{Effort}.''} \emph{IEEE Transactions on Software Engineering} 35 (5):
624--37. \url{https://doi.org/10.1109/TSE.2009.18}.

\bibitem[\citeproctext]{ref-nelsonAppliedLifeData1982}
Nelson, Wayne. 1982. \emph{Applied Life Data Analysis}. Wiley Series in
Probability and Mathematical Statistics. {Applied} Probability and
Statistics. New York: Wiley.

\bibitem[\citeproctext]{ref-nicholsEndMythIndividual2019}
Nichols, William R. 2019. {``The {End} to the {Myth} of {Individual
Programmer Productivity}.''} \emph{IEEE Software} 36 (5): 71--75.
\url{https://doi.org/10.1109/MS.2019.2908576}.

\bibitem[\citeproctext]{ref-obstbaumOngoingResearchSoftware}
Obstbaum, Simon, and Yegor Denisov-Blanch. n.d. {``Ongoing {Research} on
{Software Engineering Productivity}.''}

\bibitem[\citeproctext]{ref-oroszMeasuringDeveloperProductivity2024a}
Orosz, Gergely. 2024a. {``Measuring Developer Productivity? {A} Response
to {McKinsey}.''} January 30, 2024.
\url{https://newsletter.pragmaticengineer.com/p/measuring-developer-productivity}.

\bibitem[\citeproctext]{ref-oroszMeasuringDeveloperProductivity2024}
---------. 2024b. {``Measuring Developer Productivity? {A} Response to
{McKinsey}, {Part} 2.''} October 22, 2024.
\url{https://newsletter.pragmaticengineer.com/p/measuring-developer-productivity-part-2}.

\bibitem[\citeproctext]{ref-paudelMeasuringImpactTechnical2024}
Paudel, Bhuwan, Javier Gonzalez-Huerta, Ehsan Zabardast, and Eriks
Klotins. 2024. {``Towards {Measuring} the {Impact} of {Technical Debt}
on {Lead Time}: {An Industrial Case Study}.''} June 3, 2024.
\url{https://doi.org/10.48550/arXiv.2406.01578}.

\bibitem[\citeproctext]{ref-pedersenPatchworkComposerPlots2024}
Pedersen, Thomas Lin. 2024. \emph{Patchwork: {The} Composer of Plots}.
Manual. \url{https://patchwork.data-imaginist.com}.

\bibitem[\citeproctext]{ref-pedersenScicoColourPalettes2025}
Pedersen, Thomas Lin, and Fabio Crameri. 2025. \emph{Scico: {Colour}
Palettes Based on the Scientific Colour-Maps}. Manual.
\url{https://github.com/thomasp85/scico}.

\bibitem[\citeproctext]{ref-qiuShowtextUsingFonts2024}
Qiu, Yixuan, and authors/contributors of the included software. See file
AUTHORS for details. 2024. \emph{Showtext: {Using} Fonts More Easily in
{R} Graphs}. Manual. \url{https://github.com/yixuan/showtext}.

\bibitem[\citeproctext]{ref-quadlinMarkWomansRecord2018}
Quadlin, Natasha. 2018. {``The {Mark} of a {Woman}'s {Record}: {Gender}
and {Academic Performance} in {Hiring}.''} \emph{American Sociological
Review} 83 (2): 331--60. \url{https://doi.org/10.1177/0003122418762291}.

\bibitem[\citeproctext]{ref-rcoreteamLanguageEnvironmentStatistical2023}
R Core Team. 2023. \emph{R: A Language and Environment for Statistical
Computing}. Manual. Vienna, Austria: R Foundation for Statistical
Computing. \url{https://www.R-project.org/}.

\bibitem[\citeproctext]{ref-ramirezMeasuringKnowledgeWorker2004}
Ramírez, Yuri W., and David A. Nembhard. 2004. {``Measuring Knowledge
Worker Productivity.''} \emph{Journal of Intellectual Capital} 5 (4):
602--28. \url{https://doi.org/10.1108/14691930410567040}.

\bibitem[\citeproctext]{ref-rigginsWhatMcKinseyGot2023}
Riggins, Jennifer. 2023. {``What {McKinsey} Got Wrong about Developer
Productivity.''} LeadDev. October 23, 2023.
\url{https://leaddev.com/career-development/what-mckinsey-got-wrong-about-developer-productivity}.

\bibitem[\citeproctext]{ref-riosaWrittenUnwrittenGuide2019}
Riosa, Blake. 2019. {``The (Written) Unwritten Guide to Pull
Requests.''} Work Life by Atlassian. July 25, 2019.
\url{https://www.atlassian.com/blog/git/written-unwritten-guide-pull-requests}.

\bibitem[\citeproctext]{ref-rosserSystemsPerspectiveTechnical2021}
Rosser, Larri Ann, and John H Norton. 2021. {``A {Systems Perspective}
on {Technical Debt}.''} In \emph{2021 {IEEE Aerospace Conference}
(50100)}, 1--10. \url{https://doi.org/10.1109/AERO50100.2021.9438359}.

\bibitem[\citeproctext]{ref-rummelAverageWeibullAnalysis2017}
Rummel, Bernard. 2017. {``Beyond {Average}: {Weibull Analysis} of {Task
Completion Times}''} 12 (2).

\bibitem[\citeproctext]{ref-ruvimovaExploratoryStudyProductivity2022}
Ruvimova, Anastasia, Alexander Lill, Jan Gugler, Lauren Howe, Elaine
Huang, Gail Murphy, and Thomas Fritz. 2022. {``An Exploratory Study of
Productivity Perceptions in Software Teams.''} In \emph{Proceedings of
the 44th {International Conference} on {Software Engineering}}, 99--111.
Pittsburgh Pennsylvania: ACM.
\url{https://doi.org/10.1145/3510003.3510081}.

\bibitem[\citeproctext]{ref-sackmanExploratoryExperimentalStudies1968}
Sackman, H., W. J. Erikson, and E. E. Grant. 1968. {``Exploratory
Experimental Studies Comparing Online and Offline Programming
Performance.''} \emph{Communications of the ACM} 11 (1): 3--11.
\url{https://doi.org/10.1145/362851.362858}.

\bibitem[\citeproctext]{ref-sadowskiSoftwareDevelopmentProductivity2019}
Sadowski, Caitlin, Margaret-Anne Storey, and Robert Feldt. 2019. {``A
{Software Development Productivity Framework}.''} In \emph{Rethinking
{Productivity} in {Software Engineering}}, edited by Caitlin Sadowski
and Thomas Zimmermann, 39--47. Berkeley, CA: Apress.
\url{https://doi.org/10.1007/978-1-4842-4221-6_5}.

\bibitem[\citeproctext]{ref-sadowskiRethinkingProductivitySoftware2019}
Sadowski, Caitlin, and Thomas Zimmermann, eds. 2019. \emph{Rethinking
{Productivity} in {Software Engineering}}. Springer Nature.
\url{https://doi.org/10.1007/978-1-4842-4221-6}.

\bibitem[\citeproctext]{ref-shrikanthAssessingPractitionerBeliefs2021}
Shrikanth, N. C., William Nichols, Fahmid Morshed Fahid, and Tim
Menzies. 2021. {``Assessing Practitioner Beliefs about Software
Engineering.''} \emph{Empirical Software Engineering} 26 (4): 73.
\url{https://doi.org/10.1007/s10664-021-09957-5}.

\bibitem[\citeproctext]{ref-sjobergFutureEmpiricalMethods2007}
Sjoberg, Dag I. K., Tore Dyba, and Magne Jorgensen. 2007. {``The
{Future} of {Empirical Methods} in {Software Engineering Research}.''}
In \emph{Future of {Software Engineering} ({FOSE} '07)}, 358--78.
\url{https://doi.org/10.1109/FOSE.2007.30}.

\bibitem[\citeproctext]{ref-standevelopmentteamStanHeadersHeadersInterface2020}
Stan Development Team. 2020. {``{StanHeaders}: {Headers} for the {R}
Interface to {Stan}.''} \url{https://mc-stan.org/}.

\bibitem[\citeproctext]{ref-storeyHowDevelopersManagers2022}
Storey, Margaret-Anne, Brian Houck, and Thomas Zimmermann. 2022a. {``How
{Developers} and {Managers Define} and {Trade Productivity} for
{Quality}.''} April 27, 2022.
\url{https://doi.org/10.1145/3528579.3529177}.

\bibitem[\citeproctext]{ref-storeyHowDevelopersManagers2022b}
---------. 2022b. {``How {Developers} and {Managers Define} and {Trade
Productivity} for {Quality}.''} In \emph{Proceedings of the 15th
{International Conference} on {Cooperative} and {Human Aspects} of
{Software Engineering}}, 26--35.
\url{https://doi.org/10.1145/3528579.3529177}.

\bibitem[\citeproctext]{ref-storeyTheorySoftwareDeveloper2021}
Storey, Margaret-Anne, Thomas Zimmermann, Christian Bird, Jacek
Czerwonka, Brendan Murphy, and Eirini Kalliamvakou. 2021. {``Towards a
{Theory} of {Software Developer Job Satisfaction} and {Perceived
Productivity}.''} \emph{IEEE Transactions on Software Engineering} 47
(10): 2125--42. \url{https://doi.org/10.1109/TSE.2019.2944354}.

\bibitem[\citeproctext]{ref-gelman}
Team, Stan Development. 2024. {``Stan {Modeling Language Users Guide}
and {Reference Manual}, 2.35.''} \url{https://mc-stan.org}.

\bibitem[\citeproctext]{ref-terhorst-northWorstProgrammerKnow2023}
Terhorst-North, Daniel. 2023a. {``The {Worst Programmer I Know}.''} Dan
North \& Associates Limited. September 2, 2023.
\url{https://dannorth.net/the-worst-programmer/}.

\bibitem[\citeproctext]{ref-terhorst-northMcKinseyDeveloperProductivity2023}
---------. 2023b. {``{McKinsey Developer Productivity Review}.''} Dan
North \& Associates Limited. October 4, 2023.
\url{https://dannorth.net/mckinsey-review/}.

\bibitem[\citeproctext]{ref-toxboeCycleTime2023}
Toxboe, Anders. 2023. {``Cycle {Time}.''} Learning Loop. February 10,
2023. \url{https://learningloop.io/glossary/cycle-time}.

\bibitem[\citeproctext]{ref-trendowiczChapter6Factors2009}
Trendowicz, Adam, and Jürgen Münch. 2009. {``Chapter 6 {Factors
Influencing Software Development Productivity}---{State}‐of‐the‐{Art}
and {Industrial Experiences}.''} In \emph{Advances in {Computers}},
77:185--241. Elsevier.
\url{https://doi.org/10.1016/S0065-2458(09)01206-6}.

\bibitem[\citeproctext]{ref-vazireCredibilityReplicabilityImproving2022}
Vazire, Simine, Sarah R. Schiavone, and Julia G. Bottesini. 2022.
{``Credibility {Beyond Replicability}: {Improving} the {Four Validities}
in {Psychological Science}.''} \emph{Current Directions in Psychological
Science} 31 (2): 162--68.
\url{https://doi.org/10.1177/09637214211067779}.

\bibitem[\citeproctext]{ref-vehtariRanknormalizationFoldingLocalization2021}
Vehtari, Aki, Andrew Gelman, Daniel Simpson, Bob Carpenter, and
Paul-Christian Bürkner. 2021. {``Rank-Normalization, Folding, and
Localization: {An} Improved {Rhat} for Assessing Convergence of {MCMC}
(with Discussion).''} \emph{Bayesian Analysis}.

\bibitem[\citeproctext]{ref-vonhippelFreeRevealingPrivatecollective2006}
Von Hippel, Eric, and Georg Von Krogh. 2006. {``Free Revealing and the
Private-Collective Model for Innovation Incentives.''} \emph{R and D
Management} 36 (3): 295--306.
\url{https://doi.org/10.1111/j.1467-9310.2006.00435.x}.

\bibitem[\citeproctext]{ref-walkerEverythingWrongDORA2023}
Walker, James. 2023a. {``Everything {Wrong} with {DORA Metrics}
\textbar{} {Aviator}.''} January 18, 2023.
\url{https://www.aviator.co/blog/everything-wrong-with-dora-metrics/}.

\bibitem[\citeproctext]{ref-walkerConsUsingSPACE2023}
---------. 2023b. {``Cons of {Using SPACE} to {Measure Productivity}
\textbar{} {Aviator}.''} February 27, 2023.
\url{https://www.aviator.co/blog/whats-wrong-with-using-space-to-measure-developer-productivity/}.

\bibitem[\citeproctext]{ref-wattsNewScienceNetworks2004}
Watts, Duncan J. 2004. {``The {`{New}'} {Science} of {Networks}.''}
\emph{Annual Review of Sociology} 30 (August): 243--70.
\url{https://doi.org/10.1146/annurev.soc.30.020404.104342}.

\bibitem[\citeproctext]{ref-waydevCycleTimeFormula2021}
Waydev. 2021. {``Cycle {Time Formula}: {How} to {Optimize} the {Key
Metric} to {Accelerate Software Delivery}.''} July 26, 2021.
\url{https://waydev.co/reduce-cycle-time/}.

\bibitem[\citeproctext]{ref-wickhamGgplot2ElegantGraphics2016}
Wickham, Hadley. 2016. \emph{Ggplot2: {Elegant} Graphics for Data
Analysis}. Springer-Verlag New York.
\url{https://ggplot2.tidyverse.org}.

\bibitem[\citeproctext]{ref-wickhamScalesScaleFunctions2023}
Wickham, Hadley, Thomas Lin Pedersen, and Dana Seidel. 2023.
\emph{Scales: {Scale} Functions for Visualization}. Manual.
\url{https://scales.r-lib.org}.

\bibitem[\citeproctext]{ref-woodThinplateRegressionSplines2003}
Wood, Simon N. 2003. {``Thin-Plate Regression Splines.''} \emph{Journal
of the Royal Statistical Society (B)} 65 (1): 95--114.

\bibitem[\citeproctext]{ref-woodStableEfficientMultiple2004}
---------. 2004. {``Stable and Efficient Multiple Smoothing Parameter
Estimation for Generalized Additive Models.''} \emph{Journal of the
American Statistical Association} 99 (467): 673--86.

\bibitem[\citeproctext]{ref-woodFastStableRestricted2011}
---------. 2011. {``Fast Stable Restricted Maximum Likelihood and
Marginal Likelihood Estimation of Semiparametric Generalized Linear
Models.''} \emph{Journal of the Royal Statistical Society (B)} 73 (1):
3--36.

\bibitem[\citeproctext]{ref-woodGeneralizedAdditiveModels2017}
---------. 2017. \emph{Generalized {Additive Models}: {An Introduction}
with {R}, {Second Edition}}. CRC Press.

\bibitem[\citeproctext]{ref-woodSmoothingParameterModel2016}
Wood, Simon N., N. Pya, and B. Saefken. 2016. {``Smoothing Parameter and
Model Selection for General Smooth Models (with Discussion).''}
\emph{Journal of the American Statistical Association} 111: 1548--75.

\bibitem[\citeproctext]{ref-xieKnitrComprehensiveTool2014}
Xie, Yihui. 2014. {``Knitr: A Comprehensive Tool for Reproducible
Research in {R}.''} In \emph{Implementing Reproducible Computational
Research}, edited by Victoria Stodden, Friedrich Leisch, and Roger D.
Peng. {Chapman and Hall/CRC}.

\bibitem[\citeproctext]{ref-xieDynamicDocumentsKnitr2015}
---------. 2015. \emph{Dynamic Documents with {R} and Knitr}. 2nd ed.
Boca Raton, Florida: {Chapman and Hall/CRC}.
\url{https://yihui.org/knitr/}.

\bibitem[\citeproctext]{ref-xieKnitrGeneralpurposePackage2024}
---------. 2024. \emph{Knitr: A General-Purpose Package for Dynamic
Report Generation in {R}}. Manual. \url{https://yihui.org/knitr/}.

\bibitem[\citeproctext]{ref-zhangPullRequestLatency2022}
Zhang, Xunhui, Yue Yu, Tao Wang, Ayushi Rastogi, and Huaimin Wang. 2022.
{``Pull Request Latency Explained: An Empirical Overview.''}
\emph{Empirical Software Engineering} 27 (6): 126.
\url{https://doi.org/10.1007/s10664-022-10143-4}.

\end{CSLReferences}
\end{document}